\documentclass[useAMS,usenatbib]{mnras}
\usepackage{graphicx}
\usepackage{amsmath, amssymb}
\usepackage[T1]{fontenc}
\usepackage[usenames]{color}
\usepackage[dvipsnames]{xcolor}
\usepackage{bm}
\usepackage[capitalize]{cleveref}
\usepackage{physics}
\usepackage{caption}
\usepackage{subcaption}
\usepackage{siunitx}
\usepackage{booktabs}
\usepackage{array}
\usepackage{multirow}
\usepackage{rotating}

\newcommand{\red}[1]{#1}

\DeclareRobustCommand{\VAN}[3]{#2}
\let\VANthebibliography\thebibliography
\def\thebibliography{\DeclareRobustCommand{\VAN}[3]{##3}\VANthebibliography}

\newcommand{\lx}{L_{\mathrm{X}}}
\newcommand{\lcdm}{$\Lambda$CDM\;}
\newcommand{\ysz}{Y_{\mathrm{SZ}}}

\newcommand{\bibnote}[2]{\global\@namedef{#1note}{#2}}
\newcommand{\biblink}[2]{\global\@namedef{#1link}{#2}}
\newcommand{\LT}{$LT$}

\newcommand{\YT}{$YT$}
\newcommand{\LTYT}{$LTYT$}

\newcommand{\zobs}{z^{\rm obs}}
\newcommand{\zpred}{z^{\rm pred}}

\newcommand{\zcosmo}{z_{\rm cosmo}}
\newcommand{\zpec}{z_{\rm pec}}

\newcommand{\Vpec}{V_{\rm pec}}
\newcommand{\Vext}{\bm{V}_{\rm ext}}

\newcommand{\Mpch}{\ensuremath{h^{-1}\,\mathrm{Mpc}}}
\newcommand{\kmsec}{\ensuremath{\mathrm{km}\,\mathrm{s}^{-1}}}
\newcommand{\kmsecMpc}{\ensuremath{\mathrm{km}\,\mathrm{s}^{-1}\,\mathrm{Mpc}^{-1}}}

\newcommand{\TWOMPP}{2M\texttt{++}}
\newcommand{\Manti}{\texttt{Manticore-Local}}

\newcommand{\Om}{\Omega_{\rm m}}

\defcitealias{migkasAnisotropyGalaxyCluster2018}{M18}
\defcitealias{migkasProbingCosmicIsotropy2020}{M20}
\defcitealias{carrickCosmologicalParametersComparison2015}{C15}
\defcitealias{VF_olympics}{S25}
\defcitealias{migkasCosmologicalImplicationsAnisotropy2021}{M21}

\title[]{Testing cosmic anisotropy with cluster scaling relations}
\author[T. Yasin et al.]{
Tariq Yasin$^{1}$\thanks{E-mail: tariq.yasin@physics.ox.ac.uk},
Richard Stiskalek$^{1}$,
Harry Desmond$^{2}$,
Sebastian von Hausegger$^{3}$,
Pedro G. Ferreira$^{1}$
\\
$^{1}$Astrophysics, University of Oxford, Denys Wilkinson Building, Keble Road, Oxford, OX1 3RH, UK\\
$^{2}$Institute of Cosmology and Gravitation, University of Portsmouth, Dennis Sciama Building, Burnaby Road, Portsmouth PO1 3FX, UK\\
$^{3}$Rudolf Peierls Centre for Theoretical Physics, University of Oxford, Parks Road, Oxford OX1 3PU, UK
}

\date{Accepted XXX. Received YYY; in original form ZZZ}

\begin{document}\label{firstpage}
\pagerange{\pageref{firstpage}--\pageref{lastpage}}
\maketitle

\begin{abstract}
We test claims of large-scale anisotropy in the local expansion rate using cluster scaling relations as distance indicators. Using a Bayesian forward model, we jointly fit the X-ray luminosity--temperature (\LT) and thermal Sunyaev–Zel'dovich--temperature (\YT) relations, marginalising over the latent cluster distances and modelling selection effects as well as peculiar velocities.
The latter are modelled using reconstructions of the local peculiar velocity field where we self-consistently account for possible anisotropic redshift--distance relations via an approximate scheme.
This treatment proves crucial to the inferred anisotropy and breaks the degeneracy between anisotropy in scaling relation normalisations and underlying cosmological anisotropy. We apply our method to 312 clusters at $z \lesssim 0.2$, testing dipolar, quadrupolar and general (pixelised) anisotropy models. Bayesian model selection finds no more than weak evidence for any anisotropic model. For dipole models, we obtain upper limits of $\delta H_0 / H_0 < 3.2\%$ and bulk flow magnitude $< 1300\,\mathrm{km\,s^{-1}}$. Our results contrast with previous claims of statistically significant anisotropy from the same data, which we attribute to our principled forward modelling of both redshifts and scaling relation observables through latent distances and our treatment of the impact of anisotropic redshift--distance relations when modelling the local peculiar velocity field. Our work highlights the importance of accurately modelling peculiar velocities when testing isotropy with distance indicators, and motivates the further development of reconstructions that self-consistently treat large-scale deviations from the Hubble flow.
\end{abstract}
\begin{keywords}
large-scale structure of Universe -- methods: statistical -- cosmic background radiation -- distance scale -- galaxies: clusters: general -- X-rays: galaxies: clusters
\end{keywords}

\section{Introduction}

\red{The cosmological principle (CP) asserts that the Universe is isotropic. In the Friedmann–Robertson–Walker metric this symmetry is exact, while \lcdm\ further requires that perturbations about this background be statistically isotropic. Departures from either the exact symmetry of the background or the statistical isotropy of fluctuations would challenge the foundations of standard cosmology. Modern datasets have been increasingly scrutinised for such evidence across multiple cosmic epochs}, including the Cosmic Microwave Background (CMB;~\citealt{jonesUniverseNotStatistically2023, gaztanagaFindingOriginsCMB2024}), distant radio galaxies, quasars~\citep{secrestChallengeStandardCosmological2022}, gamma-ray bursts \citep{mondalProbingCosmicIsotropy2026}, supernovae~\citep{rahmanNewConstraintsAnisotropic2022, kalbounehMultipoleExpansionLocal2023, huTestingCosmologicalPrinciple2024, sorrentiDipolePantheon+SH0ESData2023, sorrentiLowMultipolesPantheon+SH0ES2025,Verma2023, Verma2024, sahAnisotropyPantheonSupernovae2025, BarjouDelayre2025} and galaxy or cluster scaling relations used as distance tracers~\citep{saidJointAnalysis6dFGS2020, migkasCosmologicalImplicationsAnisotropy2021, rahmanNewConstraintsAnisotropic2022, watkinsAnalysingLargescaleBulk2023, boubelLargescaleMotionsGrowth2024, boubelTestingAnisotropicHubble2025, stiskalekNoEvidenceLocal2026, watkinsOriginsBulkFlow2025}. In the \lcdm\ framework, isotropy is a statistical statement: the cosmological principle requires only that anisotropies be consistent with the expected fluctuation amplitude on the scales probed.

Claimed departures from isotropy have taken several distinct forms, each corresponding to different physical mechanisms (for a review, see~\citealt{aluriObservableUniverseConsistent2023}). In the CMB, anisotropy may manifest as hemispherical power asymmetries or alignments among the lowest-order multipoles, sometimes interpreted as evidence for a preferred cosmic axis, large-scale mode coupling, or non-trivial topology~\citep{jonesUniverseNotStatistically2023,gaztanagaFindingOriginsCMB2024}. In the large-scale distribution of extragalactic sources such as quasars and radio galaxies, a dipole pattern is expected from our kinematic motion relative to the cosmic rest frame; however, recently detected excesses beyond the predicted kinematic dipole have been interpreted as evidence for either strong intrinsic anisotropy in the matter distribution (a \emph{clustering dipole}) or motion of the sources relative to the CMB, both of which would require physics beyond \lcdm \citep{secrestChallengeStandardCosmological2022,damTestingCosmologicalPrinciple2023,secrestColloquiumCosmicDipole2025, vH2026}.

At lower redshift, isotropy is often investigated using direct distance tracers. Here, two related but physically distinct phenomena are often discussed: large-scale coherent peculiar velocities (\emph{bulk flows}; e.g.~\citealt{Hoffman_2015,watkinsAnalysingLargescaleBulk2023}) and directional variation in the inferred Hubble expansion rate (e.g.~\citealt{luongoLargerH0Values2022,kalbounehMultipoleExpansionLocal2023}). A bulk flow corresponds to a uniform motion of sources within a volume, whereas an anisotropic $H_0$ corresponds to an anisotropic contribution to the recession velocity that grows linearly with distance. In practice, the limited depth and sky coverage of current data may make it difficult to cleanly distinguish between these phenomena, as both can produce statistically similar features in observed redshift-distance relations~\citep{maartensCovariantCosmography2024}.
Determining whether such features arise from genuine large-scale anisotropy, local flow structure, residual systematics or inadequate statistical methodology is an ongoing challenge.

In practice these questions are addressed using empirical distance indicators that allow peculiar velocities to be inferred when combined with redshift observations. The scaling relations of galaxies, in particular the Tully--Fisher relation~\citep{tullyNewMethodDetermining1977} and the Fundamental Plane~\citep{djorgovskiFundamentalPropertiesElliptical1987a}, have long been used to probe peculiar velocities and test isotropy in the local Universe~\citep{tullyCosmicflows2Data2013, magoulas6dFGalaxySurvey2012,howlettCosmologicalForecastsCombined2017,desmondSubtleStatisticsDistance2025,stiskalekNoEvidenceLocal2026}. They work by correlating distance-independent observables (e.g.~rotation velocity or velocity dispersion) with distance-dependent quantities derived from observables (e.g.~luminosity or physical size). This allows distances to be inferred independently of redshift.
Combined with the observed redshift, this allows one to separate the cosmological redshift from the Doppler shifts due to peculiar velocity. In this way, one can map the peculiar velocity field and test for anisotropies in the local expansion rate. Without additional information from the local density and/or velocity field, the normalisation of the scaling relation is exactly degenerate with $H_0$: a fractional change in $H_0$ produces the same effect on the inferred distances as a shift in the scaling relation zero point. Consequently, any apparent anisotropy in $H_0$ could equally well be attributed to directional variations in the scaling relation normalisation (arising from systematics such as anisotropic dust extinction or sky-varying selection effects). However, a central insight in this work is that incorporating local velocity/density information breaks this degeneracy: the reconstructed density and peculiar velocity field depend on the assumed $H_0$ through the redshift-to-distance conversion used to map them to real space, whereas the scaling relation zero point does not.

A powerful guard against systematics in the study of cosmological distances and peculiar velocities is the use of multiple observables and distinct astrophysical objects: for astrophysical systematics to masquerade as cosmological anisotropy, they would need to affect relations with distinct physical origins in precisely the same way. Consistency across independent scaling relations thus provides a strong test of whether any detected anisotropy is genuinely cosmological. This is analogous to checking for consistency between different distance indicators as a way of identifying potential systematics in the cosmic distance ladder~\citep{najeraConsistenciesInconsistenciesRedshiftindependent2025}.

More recently, cluster scaling relations have emerged as powerful distance indicators for studying the local expansion rate. Galaxy clusters, the most massive collapsed structures in the Universe, exhibit tight correlations between their observable properties and total mass. The \LT{} relation~\citep{kaiserEvolutionClusteringRich1986} connects the X-ray luminosity of a cluster to its temperature, while the \YT{} relation links the integrated thermal Sunyaev–Zel'dovich (tSZ) signal to temperature. Both relations arise from the physics of the intracluster medium (ICM): in \emph{self-similar} models~\citep{bryanStatisticalPropertiesXray1998}, luminosity and $\ysz$ trace the baryonic mass (assumed to be a fixed fraction of the total mass), while temperature is set by hydrostatic equilibrium within the gravitational potential well. Observational studies find that the \LT{} relation slope is broadly consistent with theoretical predictions, though with notable scatter~\citep{wuLXTLXsRelationships1999,giodiniScalingRelationsGalaxy2013}. Crucially, temperature is distance-independent (measured from the X-ray spectrum), whereas $L$ and $\ysz$ depend on distance through the luminosity and angular diameter distances respectively. Hence, by predicting the intrinsic luminosity or $\ysz$ from the observed temperature and comparing to the observed flux, one can infer distances independently of redshifts.

Anisotropy in cluster scaling relations has been studied in significant detail in a series of papers, beginning with~\citet{migkasAnisotropyGalaxyCluster2018} who found anisotropy in the X-ray luminosity-temperature relationship (henceforth \LT). This was extended to a larger sample in~\citet[][hereafter~\citetalias{migkasProbingCosmicIsotropy2020}]{migkasProbingCosmicIsotropy2020}.~\citet[][hereafter~\citetalias{migkasCosmologicalImplicationsAnisotropy2021}]{migkasCosmologicalImplicationsAnisotropy2021} subsequently analysed ten further cluster scaling relations, some dependent on cosmology and some independent, to further constrain the anisotropy and test for astrophysical systematics. They found the \YT{} relation, between the projected thermal Sunyaev–Zel'dovich signal ($Y$) and temperature, provided strong constraints on anisotropy in $H_0$, or alternatively on a bulk flow model, due to its low intrinsic scatter.~\citet{heCharacterisingGalaxyCluster2025} applied their method to mocks generated from the FLAMINGO hydrodynamical simulations, finding that the anisotropy of the real data was statistically higher than in the mocks.~\citet{pandyaExaminingLocalUniverse2024}, by studying relations containing the cluster velocity dispersion (another cosmologically independent property), found no evidence that temperature systematics were driving the anisotropy.

In the literature, ``$H_0$ variation'' typically refers to a phenomenological model in which $H_0$ varies with angular position in the distance--redshift relation, rather than a physical model in which the expansion rate of space is itself spatially varying. We therefore assume the universe remains homogeneous in real space (on large scales). An inhomogeneous universe would require significantly more complex treatment, with cluster formation physics (e.g.~gravitational collapse dynamics) varying with position. When studying $\ysz$, it is also assumed that the background CMB is isotropic in the sense that each cluster is close to the same CMB monopole temperature.

The aim of this paper is to revisit the claims of anisotropy in cluster scaling relations using a full Bayesian forward modelling approach that jointly fits the scaling relations while marginalising over cluster distances and accounting for peculiar velocities sourced by the local density field, as well as selection effects. This allows us to robustly test for anisotropies in the local expansion rate while properly accounting for uncertainties and correlations in the data. We compare different models of anisotropy, including dipole and quadrupole terms in either the external velocity field or an effective variation in $H_0$, using Bayesian evidence to assess their relative merits.

The principal novelty over previous work is threefold. First, we account for peculiar velocities through separate likelihoods for the cluster scaling relation observables and the observed CMB-frame redshift, using \red{peculiar velocity fields of} the local Universe to predict the expected redshifts of each cluster. Second, we fit the \LT\ and \YT\ relations jointly, properly accounting for the intrinsic correlation between them. Third, we model selection effects and Malmquist bias within our hierarchical framework. These advances enable us to distinguish genuine large-scale anisotropies from signals induced by unmodelled local structure. 

For the purpose of converting redshift to comoving distance, we adopt a fiducial \lcdm\ cosmology with $\Omega_m = 0.3$. All distances are quoted in comoving $\Mpch$ units (factorising out the mean $H_0$ dependence), and all logarithms are base--10.

\section{Data}\label{sec:data}

\begin{figure*}
  \centering
  \includegraphics[width=0.95\linewidth]{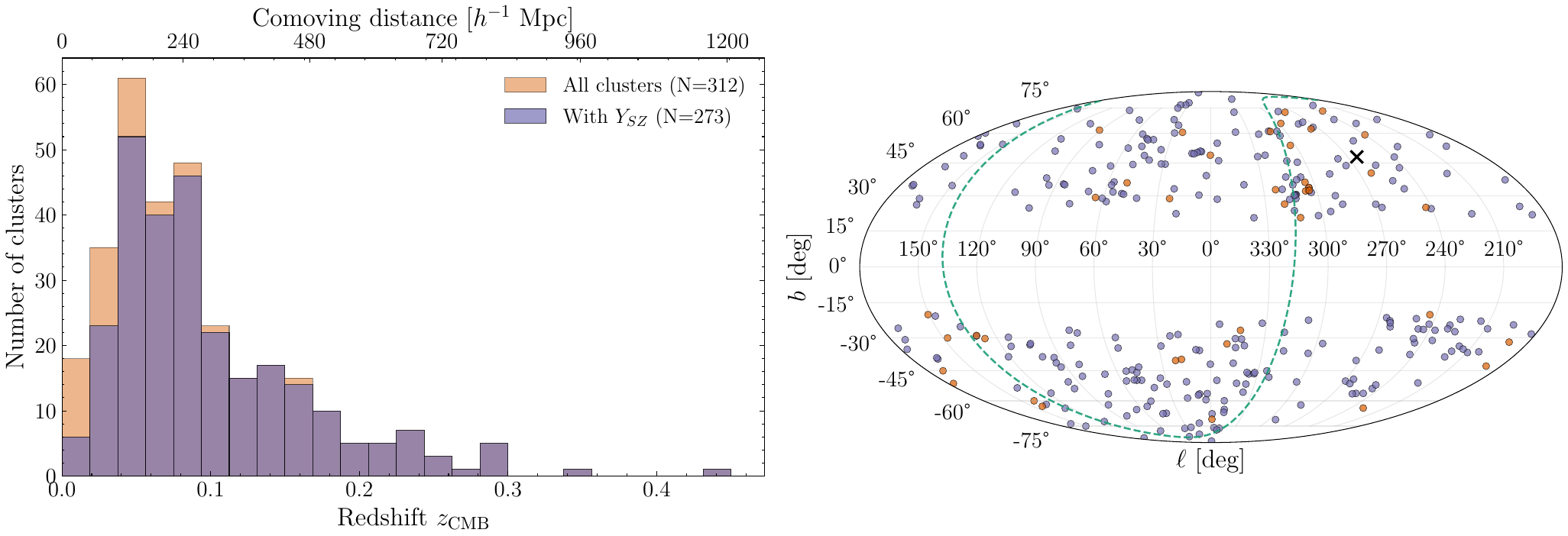}
\caption{
  \emph{Left:} A histogram of cluster CMB-frame redshifts, with clusters coloured by $\ysz$ availability. For reference, the upper axis shows comoving distance \red{in \Mpch} assuming $\zobs = \zcosmo$ (i.e.\ neglecting peculiar velocities). Only a small number of nearby clusters lack $\ysz$ measurements.
  \emph{Right:} Sky distribution in Galactic coordinates, with the same colour scheme. The supergalactic plane is shown as the dashed teal line and the CMB dipole direction as the black cross.
  }
\label{fig:redshift_dist}
\end{figure*}

Our analysis is based on the cluster catalogue compiled by~\citet[][hereafter~\citetalias{migkasCosmologicalImplicationsAnisotropy2021}]{migkasCosmologicalImplicationsAnisotropy2021}, which extends the earlier samples of~\citet{migkasAnisotropyGalaxyCluster2018} and~\citetalias{migkasProbingCosmicIsotropy2020} \red{by analysing additional scaling relations}. We only consider the \LT\ and \YT\ scaling relations in this work, as they are the ones that provide non-negligible constraints on anisotropy (\citetalias{migkasCosmologicalImplicationsAnisotropy2021}). The subsample with temperature $T$ available consists of 312 clusters from the eeHIFLUGCS \red{(extended HIghest X-ray FLUx Galaxy Cluster Sample)} flux-limited catalogue~\red{\citep{migkasProbingCosmicIsotropy2020}}, \red{with X-ray properties derived from XMM-Newton and Chandra spectroscopy}. Of these clusters, all 312 have X-ray luminosity $\lx$ measurements, and 273 of them have measurements of the integrated Compton-$y$ parameter $\ysz$ from the thermal Sunyaev–Zel'dovich effect. \red{The cluster sample used for the \LT\ and \YT\ relations by}~\citetalias{migkasCosmologicalImplicationsAnisotropy2021} is identical to that of~\citetalias{migkasProbingCosmicIsotropy2020}, except for one cluster which was removed due to the detection of foregrounds (\emph{K. Migkas}, private communication).

The \citetalias{migkasCosmologicalImplicationsAnisotropy2021} sample was restricted to clusters with Galactic latitude $|b|>20^{\circ}$ to minimise contamination and incompleteness near the Galactic plane.~\citetalias{migkasCosmologicalImplicationsAnisotropy2021} argue that the resulting sky distribution is approximately homogeneous across the celestial sphere, though residual large-scale anisotropies such as the Supergalactic Plane may contribute small spatial variations. The CMB-frame redshift and sky distributions of the clusters used in this work are shown in Figure~\ref{fig:redshift_dist}. Cluster redshifts are primarily obtained from the mean redshift of spectroscopically confirmed member galaxies. For a small subset of clusters, redshifts are derived from X-ray spectral fitting.

X-ray luminosities and temperatures are corrected for Galactic absorption using hydrogen column densities from~\citet{willingaleCalibrationXrayAbsorption2013}. The $K$-correction derived from the X-ray spectrum is applied consistently, with redshift uncertainties propagated to the final luminosity and temperature estimates. As these corrections depend on the same spectral fit, the resulting uncertainties are not fully independent; however, as the corrections are relatively small, any correlation in the uncertainties is likely negligible (\emph{K. Migkas}, private communication). As noted by~\citetalias{migkasCosmologicalImplicationsAnisotropy2021}, the X-ray luminosities reported in the catalogue were calculated assuming a fiducial cosmology with $H_0 = 70\,\kmsecMpc$ and $\Omega_{\mathrm{m}} = 0.3$, which we undo to obtain the flux ($F$) observable. Cluster temperatures correspond to the mean value measured in the 0.2–0.5\,$R_{500}$ annulus (core-excised), again from spectral fitting with the above absorption and $K$-corrections applied. The catalogue reports $T$, $T_{\min}$, and $T_{\mathrm{max}}$, corresponding to the $1\sigma$ confidence interval.
Following~\citetalias{migkasCosmologicalImplicationsAnisotropy2021}, uncertainties on $\log T$ are computed as
\begin{equation}
  \sigma_{\log x} = \log(e) \times \frac{x^{+} - x^{-}}{2x}
\end{equation}
where $x^{+}$ and $x^{-}$ denote the upper and lower bounds of the 68\% interval of parameter $x$. Flux uncertainties are provided directly in the catalogue.

The integrated Compton-$y$ parameter $Y_{\Omega} \equiv \int y \, \mathrm{d}\Omega$, where $y$ is the Compton-$y$ spectral distortion \red{and $\Omega$ denotes the solid angle of the integration region}, is extracted from \emph{Planck} HFI maps by~\citetalias{migkasCosmologicalImplicationsAnisotropy2021} using matched multi-filters. The integration region extends to an angular radius of $5\theta_{500}$ from the cluster centre, where $\theta_{500}$ is the angular size corresponding to $R_{500}$, the radius within which the mean density equals 500 times the critical density. Converting $\theta_{500}$ to a physical radius requires assuming a fiducial cosmology (the same as for luminosities above); however, because the thermal energy is concentrated at small radii, the integrated signal is insensitive to the precise choice of outer boundary. \red{Following~\citetalias{migkasCosmologicalImplicationsAnisotropy2021},} only clusters with signal-to-noise $>\!2$ in $Y$ are included in our analysis\red{; uncertainties on $Y_\Omega$ are derived from the matched multi-filter extraction}. The physical Sunyaev–Zel'dovich signal $\ysz = D_{\rm A}^2 Y_\Omega$, which represents the total thermal energy of the cluster, depends on the angular diameter distance $D_{\rm A}$ and hence on the assumed distance--redshift relation in our inference. 

The $Y_{\Omega}$ derived from a number of alternative methods from the \emph{Planck} maps were tabulated by~\citetalias{migkasCosmologicalImplicationsAnisotropy2021}. However, we found using these instead to make a negligible difference to our results, so we only consider the matched multi-filter estimates in this work.
  
\section{Methods}\label{sec:methods}

\subsection{Distances}\label{sec:methods_distances}

Our method follows~\citet{stiskalekVelocityFieldOlympics2026} in constructing a forward model for observed redshift and cluster observables, treating the distance to each cluster as a latent parameter that is marginalised over. The predicted redshift of a source, expressed in the CMB frame, can be written as
\begin{equation}\label{eq:redshift_composition}
  1 + \zpred = \left(1 + \zcosmo\right)\left(1 + \zpec\right),
\end{equation}
where $\zcosmo$ is the cosmological redshift arising from the Hubble expansion, while $\zpec \approx \Vpec /c$ is the Doppler shift produced by the cluster's peculiar velocity, $\Vpec$, projected along the line of sight.

In a spatially flat \lcdm cosmology with pressureless matter and a cosmological
constant, the relation between the cosmological redshift and the comoving
distance can be written as
\begin{equation}\label{eq:redshift_to_distance}
  r(\zcosmo) = \frac{c}{H_0} \int_{0}^{\zcosmo} \frac{\dd z^\prime}{E(z^\prime)},
\end{equation}
with
\begin{equation}
  E(z) =
  \sqrt{\Omega_{\rm m}(1+z)^3 + \Omega_{\Lambda}} .
\end{equation}
Throughout this work, the comoving distance $r$ is expressed in units of
$\Mpch$, where $h \equiv H_0/(100\,\kmsecMpc)$ denotes the dimensionless Hubble parameter, and the absolute Hubble scale is not inferred. Any global rescaling of distances is absorbed by the zero point (ZP) of the distance indicators, while direction-dependent rescalings of the inferred distance scale are introduced explicitly in our model and correspond to an anisotropy in $H_0$. We discuss our modelling of such anisotropies in \cref{sec:methods_flow_models}.

In this cosmology, the luminosity and angular diameter distances (which we also express in units of $\Mpch$) are
related to the comoving distance by
\begin{equation}\label{eq:distance_defs}
    D_{\rm L} = r (1+\zcosmo),
    \qquad
    D_{\rm A} = r (1+\zcosmo)^{-1}.
\end{equation}
We assume these distances depend only on the cosmological redshift $\zcosmo$, ignoring
relativistic corrections associated with peculiar velocities.

\red{In past work, reconstructions of the peculiar velocity field of the local volume, $\bm{v}(\bm{r})$, have been performed either from the redshift-space distribution of galaxies (e.g.~\citealt{Jasche_2013,carrickCosmologicalParametersComparison2015,Lilow_2021,Jasche_2019,Lilow_2024,mcalpineManticoreProjectDigital2025}) or directly from peculiar velocities (e.g.~\citealt{Hoffman_2015,Graziani_2019,Valade_2022,Courtois2023,Valade_2026}), with the former approach found to perform better in a recent comparison \citep{stiskalekVelocityFieldOlympics2026}.}

We adopt two redshift-space-based reconstructions:~\citetalias{carrickCosmologicalParametersComparison2015}, based on linear theory, and \Manti~\citep{mcalpineManticoreProjectDigital2025}, \red{a suite of digital twins (constrained simulations) run from initial conditions inferred by the} \textit{Bayesian Origin Reconstruction from Galaxies} algorithm (\texttt{BORG},~\citealt{Jasche_2013,Jasche_2015,Lavaux_2016,Leclercq_2017,Lavaux_2019,Porqueres_2019,Stopyra_2023}). Both~\citetalias{carrickCosmologicalParametersComparison2015} and \Manti\ are reconstructions based on the \TWOMPP\ galaxy redshift survey~\citep{Lavaux_2011}; further details are provided in \cref{app:reconstructions}. Under the single flow approximation, the line-of-sight component of the peculiar velocity of a cluster at position $\bm{r}$ is
\begin{equation}\label{eq:vpec}
  \Vpec = \left[\beta \bm{v}(\bm{r}) + \bm{V}_{\rm ext}\right] \cdot \hat{\bm{r}},
\end{equation}
with $\hat{\bm{r}}$ denoting the unit vector along the line of sight, and $\bm{V}_{\rm ext}$ representing contributions from structures outside the reconstruction volume, which we initially assume to be constant across the volume (a bulk flow), and later consider more complex models as outlined in Section~\ref{sec:methods_flow_models}.
The coefficient $\beta$ rescales the predicted velocity field from~\citetalias{carrickCosmologicalParametersComparison2015}, and is defined as $\beta = f(\Om)/b$, where $f(\Om) \approx \Om^{0.55}$ is the linear growth rate of structure~\citep{Bouchet_1995,Wang_1998} and $b$ is the linear galaxy bias factor of the \TWOMPP\ galaxy sample. For the~\citetalias{carrickCosmologicalParametersComparison2015} reconstruction, we infer $\beta$ with an informative Gaussian prior centred on the value from~\citetalias{carrickCosmologicalParametersComparison2015}, whereas for \Manti\ \red{we adopt a narrow Gaussian prior $\beta \sim \mathcal{N}(1, 0.04)$ (see Table~\ref{tab:parameters})} as the reconstruction assumes a fixed cosmology and the velocities are sourced self-consistently from the total matter distribution. If one neglects any flow sourced by the local potential, i.e. sets $\bm{v} = 0$, the motion reduces to a purely external contribution, $\bm{V}_{\rm ext}$.

Lastly, we introduce an additional inferred parameter, $\sigma_v$, to model random small-scale velocity dispersion not captured by the flow model, as well as any residual redshift uncertainty (subdominant for spectroscopic data and not provided for the datasets considered here). We assume these contributions are constant across all clusters and uncorrelated.\footnote{We do not attempt to separate redshift measurement errors from intrinsic small-scale velocities. For the five nearby clusters with direct distance measurements from NED (\url{https://ned.ipac.caltech.edu/}), we use the reported distances but neglect their uncertainties, as modelling them separately would require a more complex treatment for a negligible subset of the sample.}

\subsection{Cluster scaling relation}\label{sec:methods_cluster_relations}

\begin{figure*}
  \centering
  \includegraphics[width=0.95\linewidth]{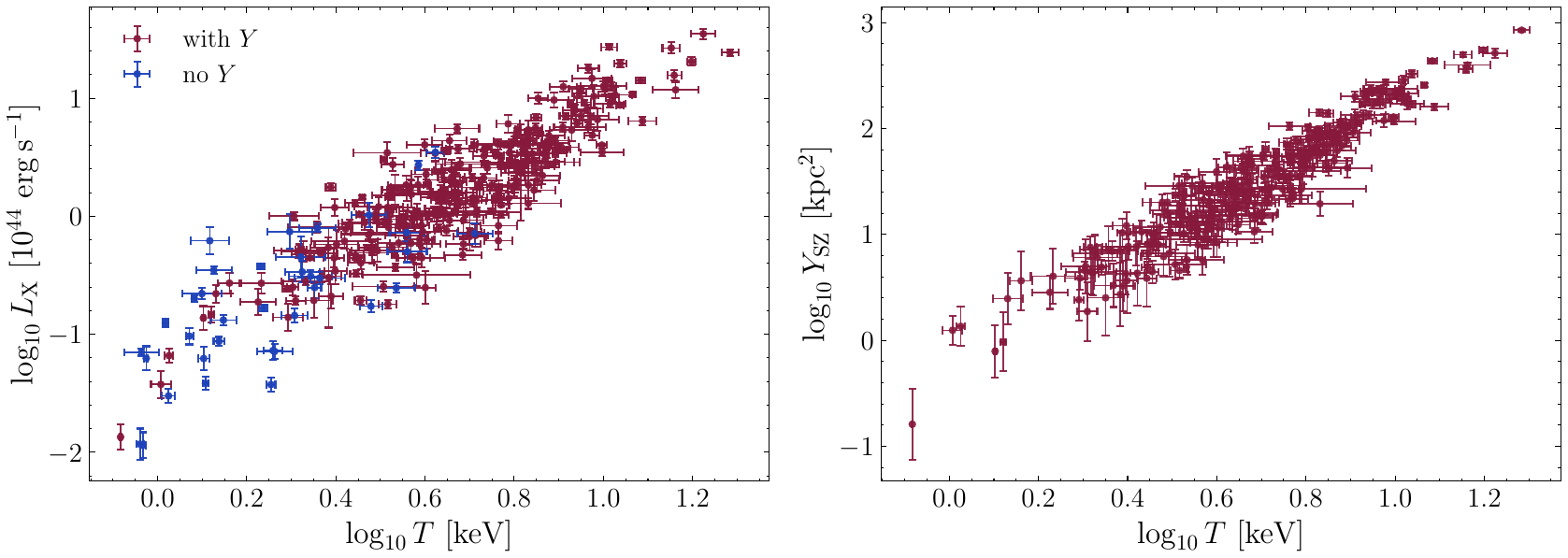}
    \caption{The two cluster scaling relations used in this work, assuming that the observed redshifts are purely cosmological (i.e.\ neglecting peculiar velocities), with $H_0 = 70\,\kmsecMpc$ and $\Om=0.3$.
    \emph{Left:} X-ray luminosity--temperature (\LT) relation.
    \emph{Right:} thermal Sunyaev–Zel'dovich–temperature (\YT) relation. Due to the SNR cut of~\citet{migkasCosmologicalImplicationsAnisotropy2021}, the \YT\ relation (right) contains preferentially hotter clusters.}
  \label{fig:relations}
\end{figure*}

The lowest-scatter cluster scaling relations for constraining peculiar velocities~\citep{giodiniScalingRelationsGalaxy2013} correlate the distance-independent temperature ($T$) with distance-dependent intrinsic cluster properties: the X-ray luminosity ($L$) and the projected thermal Sunyaev–Zel'dovich signal ($\ysz$).
$L$ is related to the observed X-ray flux $F$ by the luminosity distance
\begin{equation}
    L = 4 \pi D_{\rm L}^2 F,
\end{equation}
and $\ysz$ is related to the integrated Compton-$y$ parameter ($Y_{\Omega}$) on the sky as
\begin{equation}
    \ysz = D_{\rm A}^2 Y_{\Omega}.
\end{equation}
The two relations we study are the log-linear \LT\ and \YT\ relations
\begin{equation}\label{eq:cluster_relations}
\begin{aligned}
    \log L^{\rm pred} &= a_{L} + b_{L}\,\log T + \varepsilon_{L}, \\
    \log Y_{\rm SZ}^{\rm pred} &= a_{Y} + b_{Y}\,\log T + \varepsilon_{Y},
\end{aligned}
\end{equation}
where $a_{L,Y}$ and $b_{L,Y}$ are the intercepts and slopes, and $\varepsilon_{L,Y}$ is a noise term that accounts for the scatter in the relation. As $L^{\rm pred}$ and $Y_{\rm SZ}^{\rm pred}$ are correlated variables (both are driven by the amount of matter in the cluster and its thermodynamical state), the scatter in these two variables is expected to be correlated. We model it as coming from a multivariate normal distribution

\begin{equation}
\begin{bmatrix}
\varepsilon_{L} \\
\varepsilon_{Y}
\end{bmatrix}
\sim
\mathcal{N}\!\left(
\begin{bmatrix}
0 \\
0
\end{bmatrix},
\;
\bm{\Sigma}_{\rm int}
\right),
\end{equation}
where the intrinsic covariance $\bm{\Sigma}_{\rm int}$ is defined as
\begin{equation}
\bm{\Sigma}_{\rm int} =
\begin{bmatrix}
\sigma_{L}^{2} & \rho\,\sigma_{L}\sigma_{Y} \\
\rho\,\sigma_{L}\sigma_{Y} & \sigma_{Y}^{2}
\end{bmatrix}.
\end{equation}
Here $\sigma_L$ and $\sigma_Y$ are the 1D scatters of each relation, and $\rho \in [-1,1]$ is the (Pearson) correlation coefficient.

Our forward model predicts three observables for each cluster: the distance-independent temperature $T$, and the distance-dependent flux $F^{\rm pred}$ and integrated Compton-$y$ parameter $Y_{\Omega}^{\rm pred}$. The latter two depend on $T$ and the comoving distance $r$ via
\begin{equation}
\begin{split}
\red{\bm{x}^{\rm pred}_i}(T_i,r) &=
\begin{bmatrix}
\log F_i^{\rm pred}\\[2pt]
\log Y_{\Omega,i}^{\rm pred}
\end{bmatrix}\\
&=
\begin{bmatrix}
a_{L} + b_{L}\,\log T\red{_i} + \varepsilon_{L} - \log(4\pi) - 2\,\log D_{\rm L}\\[4pt]
a_{Y} + b_{Y}\,\log T\red{_i} + \varepsilon_{Y} - 2\,\log D_{\rm A}
\end{bmatrix}.
\end{split}
\end{equation}
The observables $\log F^{\rm obs}_{i}$ and $\log Y_{\Omega,i}^{\rm obs}$ for the $i^{\rm th}$ cluster are
\begin{equation}
\bm{x}^{\rm obs}_{i}=
\begin{bmatrix}
\log F^{\rm obs}_{i}\\[2pt]\log Y_{\Omega,i}^{\rm obs}
\end{bmatrix}.
\end{equation}
The likelihood for $\{\bm{x}^{\rm obs}\}$ is
\begin{equation}\label{eq:cluster_likelihood}
\begin{split}
\log \mathcal{L}&(\{\bm{x}^{\rm obs}\} \mid \{T\},\{r\}) =\\
&-\frac{1}{2}\sum_{i}\Bigg[
\left(\bm{x}^{\rm obs}_{i}-\bm{x}^{\rm pred}_{i}\right)^{\!\top}
\bm{\Sigma}_{{\rm tot},i}^{-1}\left(\bm{x}^{\rm obs}_{i}-\bm{x}^{\rm pred}_{i}\right)\\
&+ \log|2\pi\,\bm{\Sigma}_{{\rm tot},i}|
\Bigg],
\end{split}
\end{equation}
where the total covariance is
\begin{align}
\bm{\Sigma}_{{\rm tot},i}
&= \bm{\Sigma}_{\rm int} + \bm{\Sigma}_{{\rm meas},i},\\
\bm{\Sigma}_{{\rm meas},i} &=
\begin{bmatrix}
\delta F_{i}^{2} & 0\\
0 & \delta Y_{\Omega,i}^{2}
\end{bmatrix}.
\end{align}
The likelihood for $\{T^{\rm obs}\}$ is

\begin{equation}
\begin{split}
\log \mathcal{L}&\!\left(\{T^{\rm obs}\}\mid \{T\}\right)
=\\
&-\frac{1}{2}\sum_{i}
\left[
\frac{\big(\log T^{\rm obs}_{i} - \log T_{i}\big)^{2}}{\delta T_{i}^{2}}
+ \log\!\left(2\pi \delta T_{i}^{2}\right)
\right].
\end{split}
\end{equation}
Here $\delta F$, $\delta Y_{\Omega}$, and $\delta T$ are the measurement uncertainties on the logarithms of the cluster observables, i.e.\ $\delta F \equiv \sigma_{\log F}$. All measurement uncertainties are assumed to be Gaussian. We assign $T_i$ a Gaussian hyperprior,
\begin{equation}
\pi(T_i \mid \mu_T, w_T^2)
= \frac{1}{\sqrt{2\pi w_T^2}}
\exp\!\left(-\frac{(\log T_i - \mu_T)^2}{2w_T^2}\right).
\end{equation}
We adopt a uniform prior on $\mu_T$ and a Jeffreys prior on $w_T$, $\pi(w_T) \propto 1/w_T$.

We have framed our forward model for cluster observables as two correlated scaling relations where the distance-dependent parameters $L$ and $Y_{\Omega}$ are predicted from cluster temperature, which also recovers the intrinsic correlation between the two relations. An alternative approach would treat the relationship as three-dimensional, with $\log L = a_L + b_L\log T + c_L\log Y$ (or permutations thereof), which is akin to the fundamental plane but with only one distance-independent variable. However, using one distance-dependent quantity to predict another is problematic: $\log L$ and $\log Y$ depend on distance in nearly identical ways (differing only through $D_A$ versus $D_L$), leading to large degeneracies. We therefore do not consider this approach further.

\subsubsection{Relation evolution}

Theoretical predictions for the slopes of the $\lx$--$T$ and $\ysz$--$T$ relations follow from self-similar models in which both $\lx$ and $\ysz$ trace the baryonic mass---assumed to be a fixed fraction of the total mass---while temperature is set by hydrostatic and isothermal equilibrium~\citep{wuLXTLXsRelationships1999}. Observed clusters are broadly consistent with these assumptions, though with notable scatter~\citep{giodiniScalingRelationsGalaxy2013}. These self-similar models also predict a redshift evolution of the scaling relations~\citep[][Section 2.2]{bryanStatisticalPropertiesXray1998}:
\begin{equation}
\frac{L_{\text{X}}}{10^{44}\ \text{erg/s}}\ E(z)^{-1}=A \times \left(\frac{T}{4\ \text{keV}}\right)^B.
\label{eq1}
\end{equation}
Following~\citetalias{migkasProbingCosmicIsotropy2020}/\citetalias{migkasCosmologicalImplicationsAnisotropy2021}, we include this redshift evolution in our analysis by replacing $L^{\rm pred}$ with $L^{\rm pred} E(z)^{-1}$ and $Y_{\rm SZ}^{\rm pred}$ with $Y_{\rm SZ}^{\rm pred} E(z)$ in Equation~\eqref{eq:cluster_relations}, consistent with self-similar predictions. We find that omitting this evolution does not significantly affect our results.

\subsection{Inference model}\label{sec:methods_inference}

The total likelihood is the product of the likelihood for the cluster observables (equation~\ref{eq:cluster_likelihood}) with the likelihoods for the cluster redshifts, given by
\begin{equation}\label{eq:redshift_likelihood}
  \mathcal{L}(\zobs \mid r,\,\bm{\theta}) =
  \mathcal{N}\!\left(\zobs;\,\zpred,\,\sigma_v / c\right),
\end{equation}
where $\zpred$ is the model prediction for the CMB frame redshift of each cluster (Equation~\ref{eq:redshift_composition}), and $\sigma_v$ is the velocity dispersion introduced in \cref{sec:methods_distances}.

We adopt a prior on the source comoving distance (used to calculate $\zcosmo$) of the form
\begin{equation}\label{eq:empirical_prior_distance}
  \pi(r \mid \bm{\theta})
  = \frac{ f(r,\, \bm{\theta})}
      {\int \mathrm{d}r'\,  f(r',\, \bm{\theta})},
\end{equation}

where we introduce an empirical radial selection function
\begin{equation}
  f(r,\, \bm{\theta}) = r^p \exp\!\left[-\left(\frac{r}{R}\right)^n\right],
\end{equation}
with three additional free model parameters, that acts as an empirical model of selection effects~\citep{lavauxBayesian3DVelocity2016}. Choosing $p \simeq 2$ reproduces the homogeneous Malmquist scaling expected for a uniform spatial density ($n(r) \propto r^2$), $R$ sets the radial scale where incompleteness becomes significant, and $n$ controls the sharpness of the cutoff. The normalisation in \cref{eq:empirical_prior_distance} must be evaluated numerically since it depends on $\bm{\theta}$. \red{We note that this empirical treatment of selection effects is not derived from first principles; a more rigorous approach would forward-model the actual observational selection criteria \citep{kellyFlexibleMethodEstimating2008,desmondSubtleStatisticsDistance2025,stiskalek18CentMeasurement2026}. However, the precise selection function for this cluster sample is not well characterised, motivating our flexible empirical model.}

We do not apply inhomogeneous Malmquist bias (IMB) corrections in our analysis. The clusters in our sample already appear in the \TWOMPP\ catalogue, and their redshifts are typically derived from optical redshifts of constituent galaxies. Using the reconstructed density field to localise clusters via IMB would therefore double-count the redshift information. Instead, we use the \red{velocity fields} solely to provide peculiar velocity estimates at each cluster distance in our inference.

Since \Manti\ provides a full posterior over the density and velocity fields rather than a single point estimate, we marginalise over reconstruction uncertainty by drawing $N_s = 30$ samples from the posterior chain. At each step of our inference, we evaluate the likelihood for each sample and average, i.e.\ $\mathcal{L} = N_s^{-1} \sum_{s=1}^{N_s} \mathcal{L}(D \mid \bm{\theta}, \bm{v}_s)$, following the approach of~\citet{stiskalekVelocityFieldOlympics2026}.

A summary of the universal auxiliary model parameters and their priors is given in \cref{tab:parameters}, and we introduce the specific anisotropy models in the next section.
\begin{table*}
\centering
\caption{Model parameters and their priors. Parameters related to anisotropic flow and $H_0$ variation models are described separately in \cref{sec:methods_flow_models}. Priors are denoted as follows: uniform ($\mathcal{U}(a,b)$) on $[a,b]$; normal ($\mathcal{N}(\mu,\sigma)$) with mean $\mu$ and standard deviation $\sigma$; truncated normal ($\mathcal{N}^+$) with lower bound at zero; and $\mathrm{Jeffreys}(a,b)$, the scale-invariant prior $\pi(x) \propto 1/x$ on $[a,b]$. Uniform priors are chosen to fully encompass the posterior. For $\beta$, we use $\mathcal{N}(0.43, 0.02)$ following~\citetalias{carrickCosmologicalParametersComparison2015} and $\mathcal{N}(1, 0.04)$ for \Manti.}
\label{tab:parameters}
\renewcommand{\arraystretch}{1.3}
\begin{tabular}{@{}llllc@{}}
\toprule
Parameter & Units & Description & Prior & Section \\
\midrule
\multicolumn{5}{l}{\textit{Scaling relation parameters}} \\
$a_L$ & dex & \LT{} relation intercept & $\mathcal{U}(-5,\, 5)$ & \ref{sec:methods_cluster_relations} \\
$b_L$ & -- & \LT{} relation slope & $\mathcal{U}(-5,\, 5)$ & \ref{sec:methods_cluster_relations} \\
$a_Y$ & dex & \YT{} relation intercept & $\mathcal{U}(-5,\, 5)$ & \ref{sec:methods_cluster_relations} \\
$b_Y$ & -- & \YT{} relation slope & $\mathcal{U}(-5,\, 5)$ & \ref{sec:methods_cluster_relations} \\
$\sigma_L$ & dex & Intrinsic scatter on \LT{} & Jeffreys$(0,\, 1)$ & \ref{sec:methods_cluster_relations} \\
$\sigma_Y$ & dex & Intrinsic scatter on \YT{} & Jeffreys$(0,\, 1)$ & \ref{sec:methods_cluster_relations} \\
$\rho$ & -- & Correlation between $\varepsilon_L$ and $\varepsilon_Y$ & $\mathcal{U}(-1,\, 1)$ & \ref{sec:methods_cluster_relations} \\
$\mu_T$ & dex & Mean of $\log T$ hyperprior & $\mathcal{U}(-5,\, 5)$ & \ref{sec:methods_cluster_relations} \\
$w_T$ & dex & Width of $\log T$ hyperprior & Jeffreys$(0,\, 1)$ & \ref{sec:methods_cluster_relations} \\
\midrule
\multicolumn{5}{l}{\textit{Velocity field parameters}} \\
$\beta$ & -- & Velocity field scaling factor & See caption & \ref{sec:methods_distances} \\
$\sigma_v$ & \kmsec & Velocity dispersion / redshift scatter & Jeffreys$(150,\, 3000)$ & \ref{sec:methods_distances} \\
\midrule
\multicolumn{5}{l}{\textit{Selection function parameters}} \\
$p$ & -- & Distance prior power-law index & $\mathcal{N}^+(2,\, 0.1)$ & \ref{sec:methods_inference} \\
$R$ & \Mpch & Radial scale for incompleteness & $\mathcal{U}(1,\, 500)$ & \ref{sec:methods_inference} \\
$n$ & -- & Selection cutoff sharpness & $\mathcal{U}(0,\, 10)$ & \ref{sec:methods_inference} \\
\bottomrule
\end{tabular}

\end{table*}

\subsection{$H_0$ variation and flow models}\label{sec:methods_flow_models}

\subsubsection{Overview and the application of reconstructions}\label{sec:anisotropic_models_overview}

\red{We test various models of anisotropy in the external velocity field $\Vext$, $H_0$ (parameterised as its fractional variation $\delta H_0 / H_0$) and the zero point (ZP). Specifically, we consider the following:}
\begin{itemize}
  \item No anisotropy (base model)
  \item Dipole models:
    \begin{itemize}
      \item Constant $\Vext$
      \item Dipole in $\delta H_0 / H_0$
      \item Dipole in the ZP
    \end{itemize}
  \item Quadrupole models:
  \begin{itemize}
      \item Dipole + quadrupole in $\Vext$ projection (tidal/shear flow)
      \item Dipole + quadrupole in $\delta H_0 / H_0$
      \item Dipole + quadrupole in the ZP
  \end{itemize}
  \item Pixelised models:
  \begin{itemize}
      \item Pixelised $\Vext$ projection
      \item Pixelised $\delta H_0 / H_0$
      \item Pixelised ZP
  \end{itemize}
  \item Combined models:
  \begin{itemize}
      \item Constant $\Vext$ + dipole in $\delta H_0 / H_0$
      \item Constant $\Vext$ + dipole in ZP
  \end{itemize}
  \item Radially varying models:
  \begin{itemize}
      \item $\Vext$ with radially varying magnitude, fixed direction
      \item $\Vext$ with radially varying magnitude and direction
  \end{itemize}
\end{itemize}
\red{A constant $\Vext$ produces a dipolar pattern in the line-of-sight velocity projection. A quadrupole in the velocity field corresponds to a tidal or shear flow, which can be produced by large-scale structures such as superclusters or voids.}

Within the scaling relations, allowing ZPs to vary as
$a \rightarrow a + \Delta_{\rm ZP}$ is exactly degenerate with a rescaling of distances, since the observables enter as $-2\log D$. To allow for direction-dependent rescaling, we parameterise the fractional anisotropy in $H_0$ as
$\delta H_0(\hat{\bm n})/H_0$, where $\hat{\bm n}$ is the unit direction vector
on the sky. This is related to a directional ZP shift by
\begin{equation}\label{eq:Deltaa_DeltaH0_exact}
\Delta_{\rm ZP}(\hat{\bm n}) = 2\log\!\left[1+\frac{\delta H_0(\hat{\bm n})}{H_0}\right].
\end{equation}
For $|\delta H_0/H_0|\ll 1$, this reduces to
\begin{equation}\label{eq:Deltaa_DeltaH0_linear}
\frac{\delta H_0(\hat{\bm n})}{H_0} \simeq \frac{\ln 10}{2}\,\Delta_{\rm ZP}(\hat{\bm n})
\;\simeq\; 1.15\,\Delta_{\rm ZP}(\hat{\bm n}) .
\end{equation}
In the absence of a density or velocity reconstruction, both parameterisations enter the likelihood only through the scaling relations and are therefore observationally indistinguishable.

When a reconstructed density and/or velocity field  is included self-consistently, the degeneracy between $H_0$ anisotropy and the scaling relation zero point variation is broken. This is because the reconstructed fields are derived from galaxy redshift surveys, and hence should transform under the same redshift--distance mapping assumed in the inference (but are independent of scaling relation parameters). In past studies the reconstruction has been treated as fixed to the real space positions derived using a fiducial cosmology \red{\citep{saidJointAnalysis6dFGS2020, rahmanNewConstraintsAnisotropic2022, stiskalekNoEvidenceLocal2026, boubelLargescaleMotionsGrowth2024, boubelTestingAnisotropicHubble2025, watkinsOriginsBulkFlow2025}}, but we go beyond this approximation.

\begin{figure}
 \centering
 \includegraphics[width=1.0\linewidth]{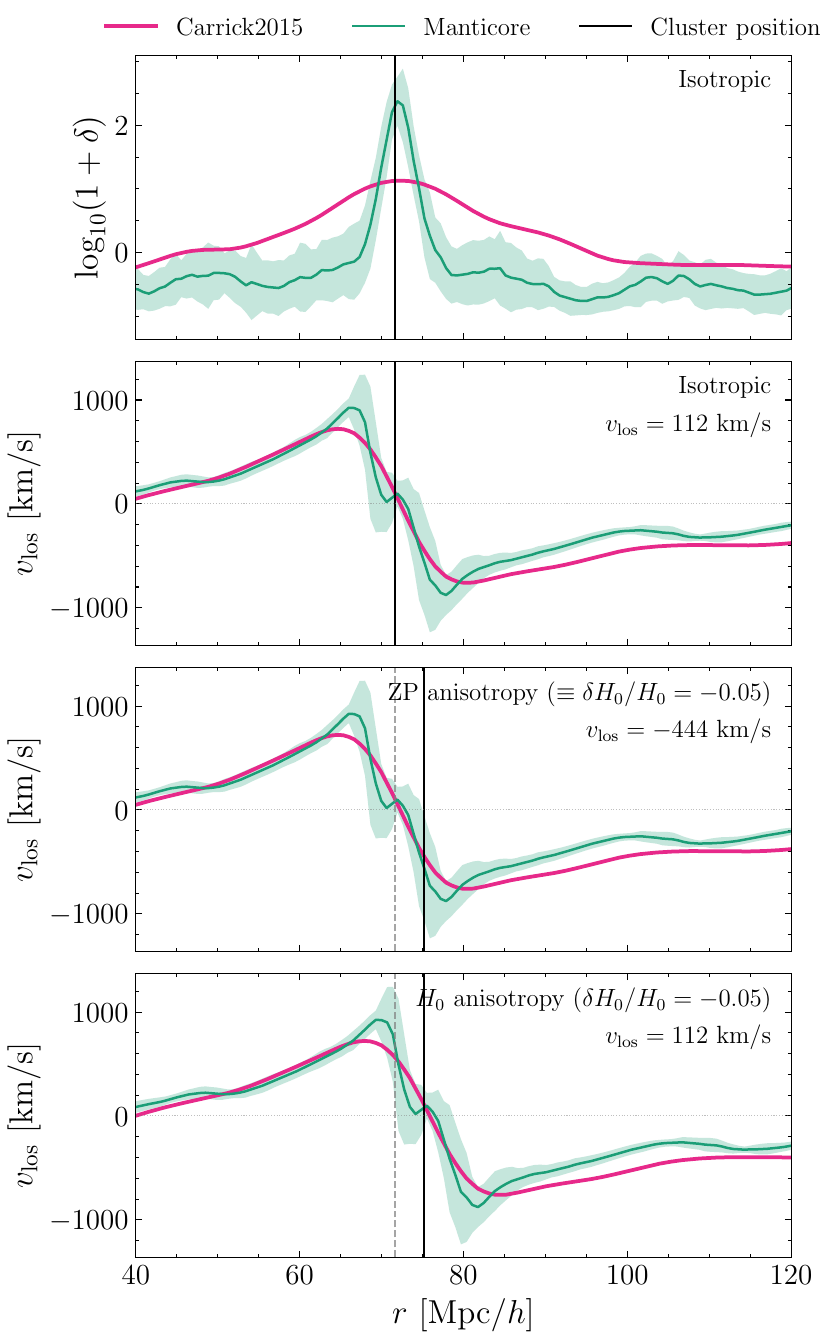}
 \caption{\red{The top two panels show the line-of-sight overdensity and velocity profiles toward the Coma cluster from the~\citetalias{carrickCosmologicalParametersComparison2015} (pink) and \Manti\ (green) reconstructions, with the cluster position for an illustrative set of scaling relation parameters in a no-anisotropy model shown in black. $v_{\rm los}$ is the line of sight velocity at the cluster radius. The bottom two panels illustrate the effect of a 10\% zeropoint anisotropy (scaling relation-inferred cluster position shifts but velocity profile unchanged) versus a 10\% $H_0$ anisotropy (both cluster and velocity profile shift together, preserving the velocity at the cluster position). The dashed grey line shows the original cluster position for comparison.}}
 \label{fig:rho_los}
\end{figure}

To treat a density/velocity field entirely self-consistently, one would need to re-derive the reconstruction for each proposed anisotropic $H_0$ and/or bulk flow model. Even though technically possible for \citetalias{carrickCosmologicalParametersComparison2015}, \red{it is extremely computationally expensive due to the iterative procedure used to remove redshift-space distortions from the density field.}

\red{We therefore adopt an approximate approach that captures the leading-order effects of anisotropic $H_0$ on the predicted peculiar velocities whilst retaining the original reconstruction. For each cluster line of sight, we first convert the reconstructed velocity profile $v_{\rm LOS}(r) \equiv \bm{v}(\bm{r}) \cdot \hat{\bm{r}}$ (see Equation~\ref{eq:vpec}) from distance to redshift using the fiducial isotropic cosmology of \citetalias{carrickCosmologicalParametersComparison2015}/\Manti\footnote{As noted previously, we removed the external velocity \citetalias{carrickCosmologicalParametersComparison2015} added to the field.}. This gives the velocity as a function of redshift, $v_{\rm LOS}(z)$, where we interpret $z$ as the CMB-frame redshift with the peculiar velocity contribution from local structures removed. At each step of the inference, we then map from redshift back to comoving distance $r$ using the assumed anisotropic cosmology, by inverting
\begin{equation}\label{eq:reconstruction_inversion}
(1 + z) = (1 + z_{\rm cosmo})(1 + \bm{V}_{\rm ext} \cdot \hat{\bm{r}} / c),
\end{equation} and applying Equation~\ref{eq:redshift_to_distance}—to obtain the $v_{\rm LOS}(r)$ used in Equation~\ref{eq:vpec}.}

This procedure assumes the linearity of distance with respect to local peculiar velocity and the bulk flow contribution, which approximately holds in the low-redshift regime covered by the reconstructions ($\lesssim 200$ Mpc).
It captures the effect on the predicted peculiar velocities of structures within the reconstructed volume, \red{under the assumption that changes in the density field due to the anisotropic redshift--distance mapping are small compared to the density fluctuations themselves. We discuss this approximation and its limitations in detail in \cref{sec:methods_reconstructions}.}

\red{For \citetalias{carrickCosmologicalParametersComparison2015} we could have applied the anisotropic redshift--distance mapping directly to the reconstructed density field, and then re-derived the velocity field using linear theory. However, in testing we find that for the anisotropies considered here this makes only a small difference to the peculiar velocities ($<3\%$ for $\delta H_0/H_0 = 0.05$), which has a negligible impact on the posteriors. Therefore we remap the velocity field directly for computational efficiency.}

\red{For the radially varying $\Vext(r)$ models there is an additional complexity. We need to evaluate the velocity field at the real space position corresponding to each $z$ in Equation~\ref{eq:reconstruction_inversion}, but the mapping itself depends on $\Vext(r)$. We therefore begin by assuming $z = z_{\rm cosmo}$ to evaluate $\Vext(r)$, and then iterate. This converges rapidly, producing sub-percent level error in $r$ after only 3 iterations.}

\red{We illustrate the contrasting effects ZP and $H_0$ anisotropy in \cref{fig:rho_los}, which shows the line-of-sight overdensity and velocity profiles from the~\citetalias{carrickCosmologicalParametersComparison2015} and \Manti\ reconstructions toward the Coma cluster, for their fiducial isotropic cosmologies. In the top two panels we show the inferred position of the cluster for some fiducial scaling relation parameters (we assume zero scatter, so the cluster is perfectly localised) and no anisotropy. The cluster sits near the peak of the \TWOMPP derived overdensity profiles, which gives rise to a peculiar velocity of $\sim 100$ km/s at its radius.}

\red{In the third panel we introduce an anisotropy in the ZP towards Coma of $\sim 0.085$ dex (which is equivalent to a $\sim10$\% $H_0$ anisotropy by Equation~\ref{eq:Deltaa_DeltaH0_linear}). This shifts the inferred cluster position, but leaves \citetalias{carrickCosmologicalParametersComparison2015}/\Manti\ velocity profiles unchanged, meaning in this model the cluster no longer sits at the overdensity peak, and the peculiar velocity at the new position is $\sim -700\,\kmsec$. In the bottom panel, we show the effect of an equivalent anisotropy in $H_0$. This remaps the positions of both the cluster and the velocity fields coherently, leaving the assumed peculiar velocity of the cluster unchanged from the isotropic case. }

\red{Hence, in the presence of a reconstructed density/velocity field, anisotropy in $H_0$ and anisotropy in the scaling relation normalisation correspond to physically distinct effects, and so can be constrained as two independent models. We summarise the reconstruction models used in this work in \cref{tab:reconstructions}.}

\begin{table*}
\centering
\caption{Summary of the three peculiar velocity models used in this work. \red{The velocity field column indicates whether the reconstruction's velocity predictions are used in the likelihood (see Section~\ref{sec:anisotropic_models_overview}).}}
\label{tab:reconstructions}
\renewcommand{\arraystretch}{1.3}
\begin{tabular}{@{}lccc@{}}
\toprule
Model & Survey & Velocity field & Isotropy assumptions \\
\midrule
No velocity field & --- & No & None \\
\citetalias{carrickCosmologicalParametersComparison2015} & \TWOMPP & Yes & Linear perturbation theory, $z$--$r$ for weighting/selection \\
\Manti & \TWOMPP & Yes & Power spectrum, $z$--$r$ for luminosity/selection \\
\bottomrule
\end{tabular}
\end{table*}

\subsubsection{Model definitions}

We define the dipole and quadrupole contributions at each cluster position as
\begin{equation}
g(\hat{\bm{n}}) \;=\;
 D \, \hat{\bm{d}}\cdot\hat{\bm{n}}
\;+\; Q \left[ (\hat{\bm{q}}_1\cdot\hat{\bm{n}})
       (\hat{\bm{q}}_2\cdot\hat{\bm{n}})
       - \frac{1}{3}\,\hat{\bm{q}}_1\cdot\hat{\bm{q}}_2 \right].
\label{eq:multipoles}
\end{equation}
where $g(\hat{\bm{n}})$ denotes the field evaluated along the cluster direction $\hat{\bm{n}}$, $D$ and $Q$ are the dipole and quadrupole amplitudes, $\hat{\bm{d}}$ is the unit dipole axis, and $\hat{\bm{q}}_1$ and $\hat{\bm{q}}_2$ are the unit quadrupole axes.

For the pixelised models, we divide the sky into $N_{\rm pix} = 12$ equal-area regions using the HEALPix tessellation scheme~\citep{gorskiHEALPixFrameworkHighResolution2005} at resolution $N_{\rm side} = 1$, the coarsest possible equal-area tessellation. Each pixel is assigned an independent parameter representing the fractional $H_0$ variation, ZP shift, or line-of-sight velocity component for clusters in that direction. This parameterisation allows for arbitrary angular structure without imposing a particular multipole decomposition.

For the radially varying bulk flow models, we parameterise the velocity field at a set of radial knots $\{r_k\}$. In the ``radial magnitude'' model, the flow direction $\hat{\bm{d}}$ is assumed constant across all radii, and only the magnitude $|\Vext|(r)$ varies; the magnitude is constrained to be non-negative. Interpolation of the magnitude between knots uses cubic splines. In the ``radial'' model, both the direction and velocity magnitude are inferred at each knot. To interpolate the direction smoothly between knots, we employ spherical linear interpolation \red{(SLERP; \citealt{shoemakeAnimatingRotationQuaternion1985})}, which traces the shortest arc on the unit sphere between consecutive direction vectors. The magnitude is interpolated separately using cubic splines.

At low redshift (where $c\,\zcosmo\simeq100\,r$), a dipole in the
distance--redshift relation, parameterised as $D(\hat{\bm d}\cdot\hat{\bm n})$, is observationally equivalent to (see Appendix~\ref{app:H0_to_flow}) a dipolar peculiar velocity field with
\begin{equation}
V_{\rm pec}(r,\hat{\bm n})
\simeq
\frac{100\,r}{1 + 100r/c}\,
D(\hat{\bm d}\cdot\hat{\bm n}) ,
\label{eq:Vpec_r_main}
\end{equation}
which, for $r \ll c/100$, reduces to
\begin{equation}
V_{\rm pec}(r,\hat{\bm n})
\simeq
100\,r\,D\,(\hat{\bm d}\cdot\hat{\bm n})
\end{equation}
\red{i.e.\ a dipolar velocity field with amplitude increasing linearly with distance.}

\subsection{Inference procedure}\label{sec:inference_procedure}

Posterior sampling is carried out with the No-U-Turn Sampler~\citep[NUTS;][]{hoffmanNoUTurnSamplerAdaptively2011},
a variant of Hamiltonian Monte Carlo, as implemented in the
\texttt{numpyro}\footnote{\url{https://num.pyro.ai/en/latest/}} framework~\citep{phanComposableEffectsFlexible2019, binghamPyroDeepUniversal2019}.
We ensure convergence by requiring the Gelman–Rubin statistic $\hat{R}-1 \leq 0.01$
for all parameters~\citep{gelmanInferenceIterativeSimulation1992}.
For model comparison, we compute the Bayesian evidence, defined as
\begin{equation}
\mathcal{Z} \equiv p(D\mid M) = \int \dd\bm{\theta} \:
\mathcal{L}(D\mid\bm{\theta}, M)\,\pi(\bm{\theta}),
\end{equation}
where $\bm{\theta}$ denotes the free parameters, $\mathcal{L}$ the likelihood,
and $\pi$ the prior.
Evidence is estimated using the \texttt{harmonic} package\footnote{\url{https://github.com/astro-informatics/harmonic}},
which fits a normalising flow to posterior samples and applies the harmonic estimator~\citep{polanskaLearnedHarmonicMean2024}.

We numerically marginalise over the latent variables ($r$ and $T$) within each MCMC iteration, producing a lower-dimensional posterior well-suited for evidence calculation.
For $r$, we use a fine grid extending to $1400\,\Mpch$, ensuring all clusters lie within the boundary even under strong anisotropies. The grid spacing is $0.8\,\Mpch$ for~\citetalias{carrickCosmologicalParametersComparison2015} and the no velocity field case, and $0.65\,\Mpch$ for \Manti; \red{these spacings oversample the original reconstruction grids by a factor of $\sim 2$ to avoid interpolation artefacts.}
\red{For $T$, we marginalise over an adaptive grid for each cluster, spanning $\pm 5\sigma$ around the observed temperature (where $\sigma$ is the measurement uncertainty) with 31 uniformly spaced points.} Competing models $M_1$ and $M_2$ are compared using the Bayes factor, \red{assuming equal prior model probabilities},
\begin{equation}
\mathcal{B} \equiv \frac{p(D\mid M_1)}{p(D\mid M_2)},
\end{equation}
which provides a quantitative measure of relative support. We interpret the strength of evidence using the Jeffreys scale,
which classifies $|\ln\mathcal{B}| < 1$ as inconclusive, $1 < |\ln\mathcal{B}| < 2.5$ as weak, $2.5 < |\ln\mathcal{B}| < 5$ as moderate, and $|\ln\mathcal{B}| > 5$ as strong evidence~\citep{jeffreysTheoryProbability1939}.

\subsection{Prior choices}\label{sec:priors}

For all dipole and quadrupole models ($H_0$, ZP, and $\Vext$), the dipole direction $\hat{\bm{d}}$ and quadrupole eigenvectors $\hat{\bm{q}}_1$, $\hat{\bm{q}}_2$ are assigned isotropic priors on the sphere. For $H_0$ and ZP anisotropy, we adopt a uniform prior on the amplitude $|\delta H_0/H_0|$ from 0 to 0.15, motivated by the ${\sim}9\%$ variations reported by~\citetalias{migkasCosmologicalImplicationsAnisotropy2021}. When fitting for ZP anisotropy, we place the prior on $\Delta_{\rm ZP}$ to be uniform in the equivalent $\delta H_0/H_0$ via Equation~\eqref{eq:Deltaa_DeltaH0_linear}. For the pixelised $H_0$ and ZP models, we assign independent uniform priors on each HEALPix pixel value, with $\delta H_0/H_0 \in [-0.15, 0.15]$.

\red{For dipole and quadrupole $\Vext$ models we adopt a uniform prior on the amplitude} from 0 to $5{,}000\,\kmsec$, which is approximately equivalent to the $H_0$ dipole prior at $r \sim 300\,\Mpch$ via Equation~\eqref{eq:Vpec_r_main}. The same amplitude prior applies to each radial knot in the radially varying models.

For the pixelised $\Vext$ model, we impose a sum-to-zero constraint $\sum_i V_i = 0$ to remove the unphysical monopole (uniform radial inflow) degree of freedom. \red{Without this constraint, we find that uniform priors on each pixel 
lead to a strong preference for a ${\sim}4{,}000\,\kmsec$ global inflowing 
monopole, which pushes clusters to larger inferred distances. A sufficiently large monopole renders the reconstruction's peculiar 
velocities negligible, removing their localising effect and allowing 
the $r^2$ volume factor of the distance prior 
(Equation~\ref{eq:empirical_prior_distance}) to push clusters to larger 
distances. Although not exactly degenerate with the scaling relation 
normalisation, the inflow partially compensates for the larger distances. 
This pathology does not occur for the pixelised $H_0$ model, whose radial 
scaling ($V_{\rm pec} \sim r$) prevents it from dominating the peculiar velocity 
field for nearby clusters.} We implement the constraint by sampling $N_{\rm pix}-1$ independent standard normal variates $\bm{u}$ and projecting them onto the zero-sum subspace via $\bm{V} = \sigma_{\rm pix} Q\,\bm{u}$, where $Q$ is a $N_{\rm pix} \times (N_{\rm pix}-1)$ matrix whose columns span the space orthogonal to the all-ones vector. The amplitude $\sigma_{\rm pix}$ is assigned a uniform prior from 0 to $5{,}000\,\kmsec$.

For the pixelised ZP and no velocity field $H_0$ models, a uniform monopole in the pixels is exactly degenerate with the scaling relation normalisation $a$, which is already a free parameter; we therefore impose the same sum-to-zero constraint to remove the degeneracy. For the pixelised $H_0$ model \emph{with} reconstruction, the perfect degeneracy is broken because $H_0$ also affects the redshift-to-distance conversion used to query the velocity field; we therefore use independent uniform priors on each pixel without a sum-to-zero constraint.

Other model parameters are assigned broad uniform priors, except for the velocity dispersion $\sigma_v$ and intrinsic scatters $\sigma_L$ and $\sigma_Y$, for which we adopt scale-invariant (Jeffreys) priors of the form $\pi(x)\propto 1/x$. For a uniform prior where the posterior is well enclosed within the prior range (which is case for all our anisotropy parameter magnitudes), the evidence can be rescaled to a different prior width by adding $\ln(\Delta_{\rm old}/\Delta_{\rm new})$ to $\ln \mathcal{Z}$, where $\Delta_{\rm old,new}$ is the old/new prior width; the reader can thus convert our evidences to their preferred priors for nearly all our flow models. 

\section{Results}\label{sec:results}

\red{We now present constraints on our models of anisotropy in the external velocity field ($\Vext$), $H_0$ and the scaling relation zero point (ZP). We present full posterior distributions for our base model (no anisotropy) and the $H_0$ / $\Vext$ dipole models for the combined \LTYT\ relation in \cref{app:full_posterior}. The key model parameters exhibit mild degeneracies: the scaling relation intercepts ($a_L$, $a_Y$) are correlated with the radial selection parameters ($p$, $R$, $n$), as both affect the inferred distance distribution. Similarly, the velocity dispersion $\sigma_v$ is partially degenerate with the intrinsic scatter parameters ($\sigma_L$, $\sigma_Y$), since both contribute to the total scatter in the redshift likelihood. Despite these correlations, the anisotropy parameters (dipole amplitude and direction) are well-constrained independently. The only strong degeneracy is between the selection parameters $R$ and $n$.}

\begin{figure}
  \centering
  \includegraphics[width=\columnwidth]{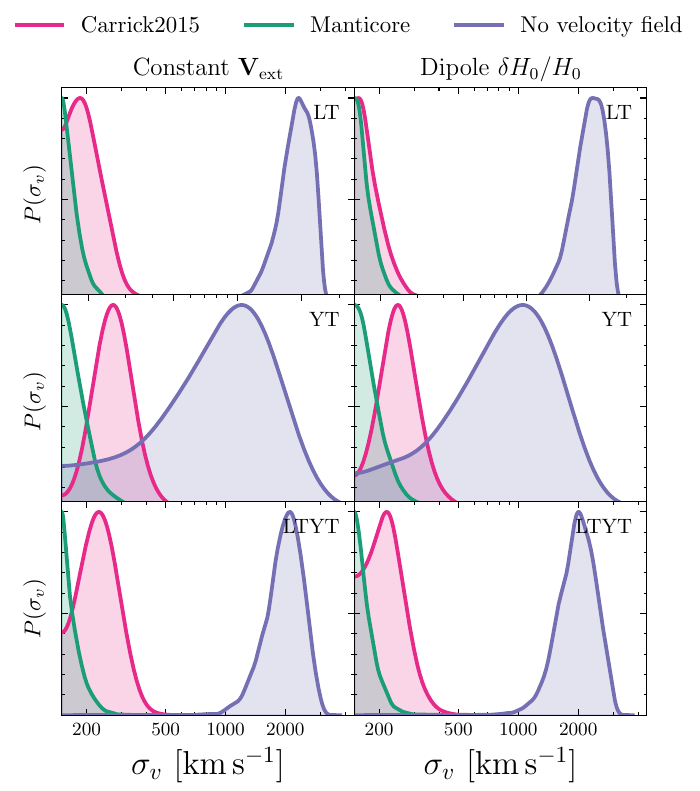}
  \caption{The posterior on $\log \sigma_v$ (its prior is flat in this space) for different reconstructions for the constant $\Vext$ and $H_0$ dipole models for \LT, \YT\ and \LTYT\ relations. Both~\citetalias{carrickCosmologicalParametersComparison2015} and \Manti\ find $\sigma_v \sim 200$--$250\,\kmsec$ for \LT\ and \LTYT, consistent with small-scale peculiar velocity dispersion. For \YT, $\sigma_v$ is less constrained. The no velocity field models prefer high values ($\sigma_v \sim 1500$--$2500\,\kmsec$), showing they cannot accurately predict cluster redshifts without the velocity field information. The plot is made using reflection KDE with Scott's rule for smoothing width.}
\label{fig:sigma_v}
\end{figure}

We first verify whether our models can accurately predict cluster redshifts by examining the inferred velocity dispersion parameter $\sigma_v$. \Cref{fig:sigma_v} shows the posterior distribution of $\sigma_v$, which captures residual scatter from unmodelled peculiar velocities, redshift measurement errors, and potentially any remaining systematic effects. The~\citetalias{carrickCosmologicalParametersComparison2015} reconstruction yields $\sigma_v \approx 200$--$250\,\kmsec$ for the \LT\ and \LTYT\ relations, consistent with expectations from small-scale peculiar velocity dispersion, while \Manti\ yields $\sigma_v < 200$ but with the posterior peaking at the lower bound of the prior. For \YT, $\sigma_v$ is less constrained due to the lower sensitivity of the \YT\ relation to distance errors. In contrast, models without a velocity field require much larger values ($\sigma_v \sim 1500$--$2500\,\kmsec$), indicating substantial unmodelled velocity scatter. The reduced $\sigma_v$ when reconstructions are applied demonstrates that these models are necessary to predict the velocity field traced by our cluster sample.

Comparing the Bayesian evidence across reconstruction methods, we find \Manti\ is strongly favoured over both~\citetalias{carrickCosmologicalParametersComparison2015} ($\Delta\ln\mathcal{Z} \sim 20$) and no velocity field ($\Delta\ln\mathcal{Z} \sim 40$). This suggests that \Manti's more accurate modelling of non-linear structure formation successfully captures the velocity field traced by our cluster sample, consistent with~\citet{stiskalekVelocityFieldOlympics2026}, who found that \texttt{BORG}-based reconstructions outperformed~\citetalias{carrickCosmologicalParametersComparison2015} in explaining peculiar velocities. \red{However, both reconstructions assume isotropy in their derivation: \Manti\ through its prior on the $\Lambda$CDM power spectrum and fixed cosmological parameters, and~\citetalias{carrickCosmologicalParametersComparison2015} through the assumption of linear perturbation theory. For~\citetalias{carrickCosmologicalParametersComparison2015}, we approximately account for potential anisotropy by remapping the velocity field self-consistently (Section~\ref{sec:anisotropic_models_overview}), whereas \Manti\ remains tied to its fiducial isotropic cosmology.} We present results for both reconstructions throughout to assess the robustness of our findings, as well as results without a velocity field to illustrate the impact of not including peculiar velocity information. All evidences in the tables are reported relative to the base model within each reconstruction.

\subsection{Dipole models}\label{sec:results_dipoles}

\begin{figure*}
 \centering
 \begin{subfigure}[t]{0.48\textwidth}
  \centering
  \includegraphics[width=\linewidth]{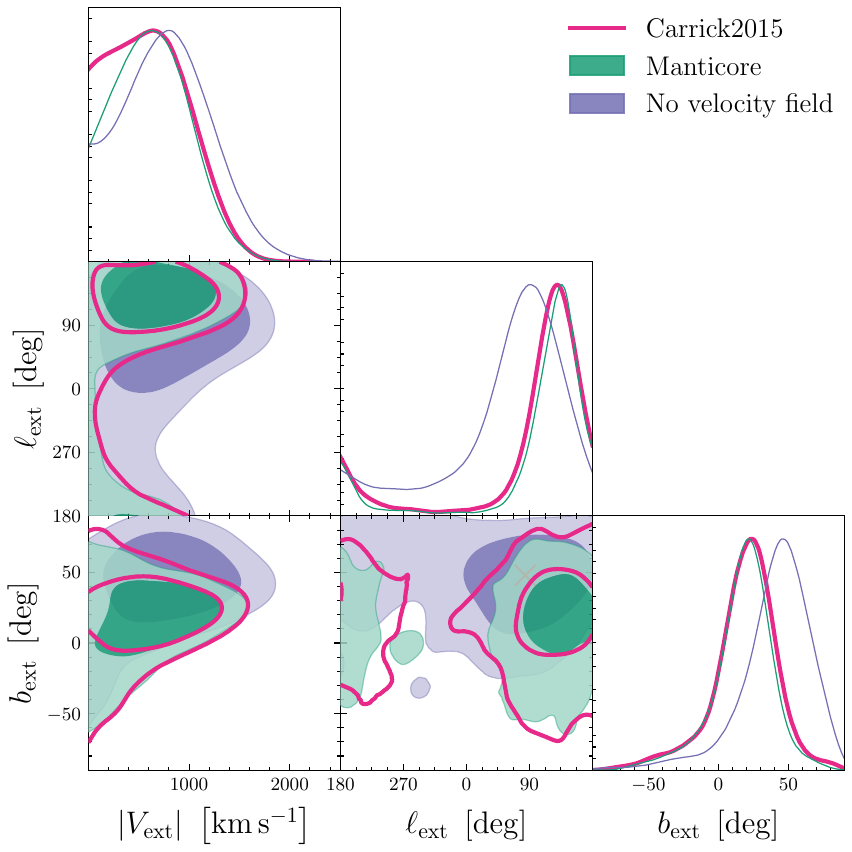}
 \end{subfigure}

 \vspace{0.6em}
 \begin{subfigure}[t]{0.48\textwidth}
  \centering
  \includegraphics[width=\linewidth]{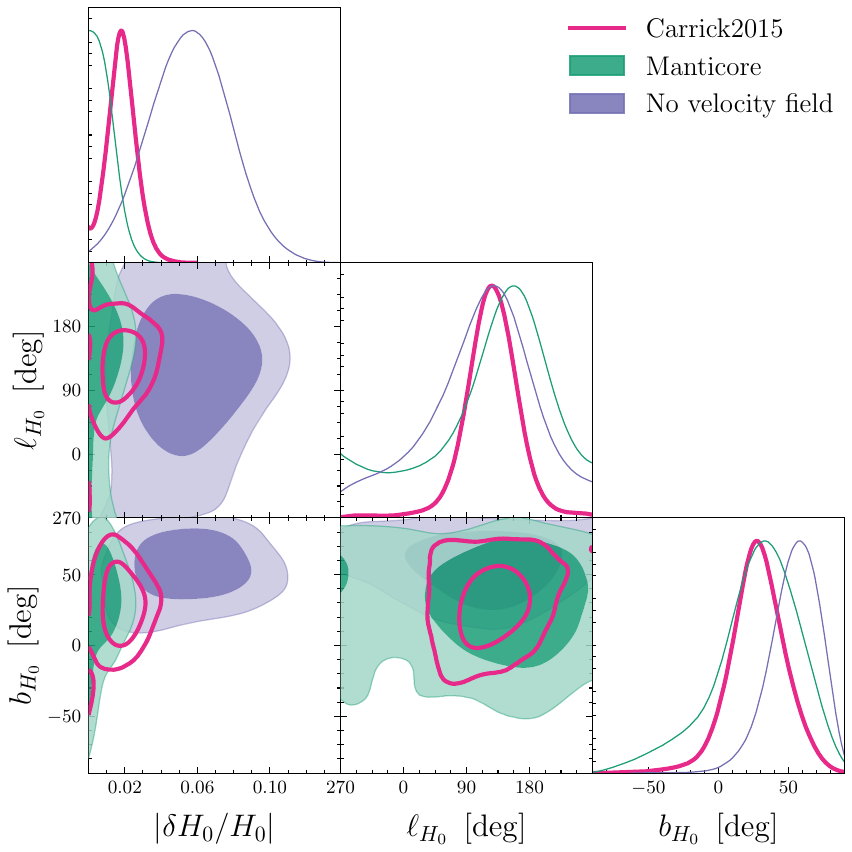}
 \end{subfigure}\hfill
 \begin{subfigure}[t]{0.48\textwidth}
  \centering
  \includegraphics[width=\linewidth]{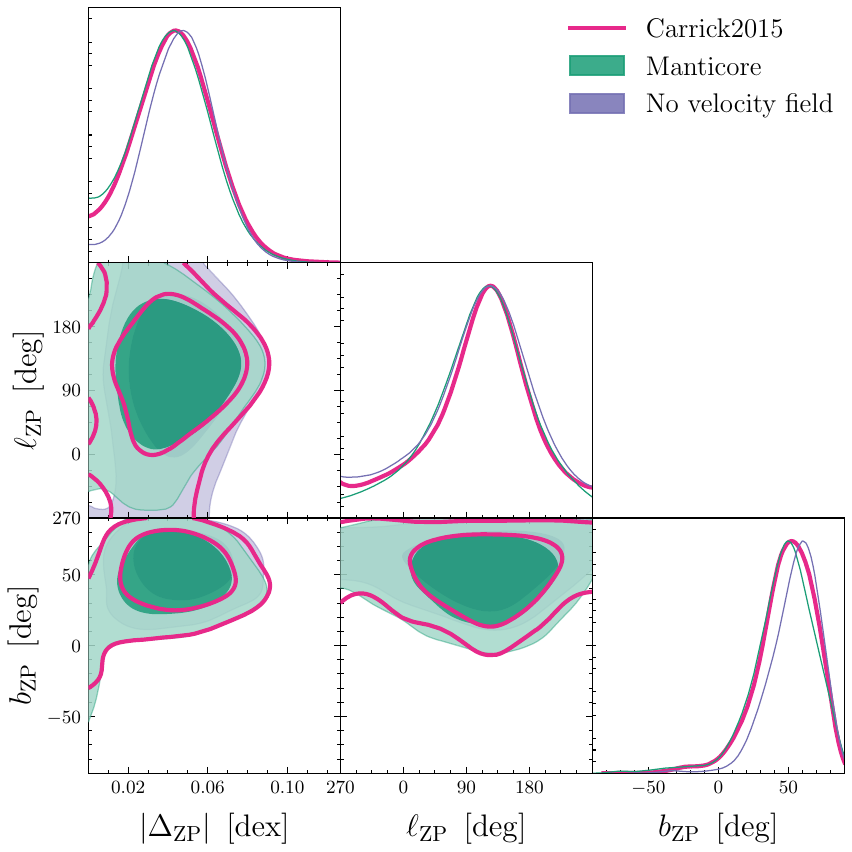}
 \end{subfigure}
 \caption{Constraints on the dipolar bulk flow (top), $H_0$ anisotropy (bottom left), and ZP dipole (bottom right) for different peculiar velocity field models.}
 \label{fig:reconstruction_comparison}
\end{figure*}

We present constraints \red{on the constant $\Vext$} and $H_0$/ZP dipole anisotropy models in \cref{tab:dipole_results} for the joint \LTYT\ relation across all three reconstruction scenarios. \Cref{fig:reconstruction_comparison} shows the corresponding posterior distributions.

\red{For the $\Vext$ dipole model}, both reconstructions yield upper limits consistent with zero:~\citetalias{carrickCosmologicalParametersComparison2015} gives $<1276\,\kmsec$ at 95\% confidence ($\Delta\ln\mathcal{Z} = -0.93 \pm 0.09$), while \Manti\ gives $628^{+376}_{-369}\,\kmsec$ ($\Delta\ln\mathcal{Z} = -1.02 \pm 0.07$). The inferred directions are poorly constrained, consistent with the posteriors being compatible with zero amplitude.

For the $H_0$ dipole,~\citetalias{carrickCosmologicalParametersComparison2015} yields $\delta H_0/H_0 = 1.8^{+0.8}_{-0.8}\%$ with negligible evidence ($\Delta\ln\mathcal{Z} = 0.01 \pm 0.08$), while \Manti\ yields only an upper limit $\delta H_0/H_0 < 2.1\%$ with negative evidence ($\Delta\ln\mathcal{Z} = -1.93 \pm 0.09$), indicating the additional complexity is not warranted by the data.

Without a velocity field, we find a bulk flow amplitude of $785^{+440}_{-440}\,\kmsec$ directed towards $(\ell, b) = (124^\circ \pm 83^\circ, 41^\circ \pm 26^\circ)$, with negligible evidence ($\Delta\ln\mathcal{Z} = -0.55 \pm 0.07$) relative to the base model. The $H_0$ dipole shows $\delta H_0/H_0 = 5.6^{+2.2}_{-2.2}\%$ towards $(\ell, b) = (144^\circ, 54^\circ)$, with weak positive evidence ($\Delta\ln\mathcal{Z} = 1.34 \pm 0.07$).

Comparing across reconstructions, the choice of velocity field has a substantial impact on the inferred $H_0$ anisotropy: the amplitude decreases from $\sim 6\%$ without a velocity field to $\sim 2\%$ with~\citetalias{carrickCosmologicalParametersComparison2015}, and to an upper limit with \Manti. In contrast, the $\Vext$ dipole amplitude is less sensitive to the reconstruction choice, with all three scenarios yielding consistent (and statistically insignificant) values. The ZP dipole (which can arise from systematic calibration errors or astrophysical effects) shows a persistent signal of $\sim 4$--$5\%$ amplitude towards $(\ell, b) \approx (145^\circ, 50^\circ)$ across all reconstruction scenarios, with only weak or negligible Bayesian evidence ($\Delta\ln\mathcal{Z} = 0.04$--$0.90$). 

Comparing \LT, \YT, and joint constraints (\cref{fig:relation_comparison}), we find consistent results for the $H_0$ dipole across all three relations. For the $\Vext$ dipole, however, both \LT\ and \YT\ individually show a moderate dipole signal in the posterior, but pointing in slightly different directions. When fitted jointly, this tension leads to a significantly reduced amplitude in the \LTYT\ posterior, which settles on a compromise direction between the two with much lower significance. For the ZP dipole, \LT\ prefers a dipole, but \YT\ does not, and the joint fit is close to the YT result (due to its stronger constraining power, but with reduced significance). 

We infer a high intrinsic correlation ($0.85 < \rho < 0.9$) between the scatter of the \LT\ and \YT\ relations for all models, indicating limited independent information between them. These results demonstrate the importance of jointly fitting both relations in a consistent model rather than combining constraints post hoc. 

\subsection{Extended models}\label{sec:results_beyond_dipoles}

We extend our analysis beyond simple dipoles to test for more complex anisotropy phenomenologies. \Cref{tab:beyond_dipoles} summarises the Bayesian evidence for all extended models relative to the base models for the \LTYT\ relation. For models with reconstructions, \red{we find at most negligible evidence for any model other} than ZP pixelised for \citetalias{carrickCosmologicalParametersComparison2015} and \Manti\, $\Vext$ pixelised for \citetalias{carrickCosmologicalParametersComparison2015} only, and radial $\Vext$ with free direction for \Manti.

\begin{figure*}
 \centering
 \begin{subfigure}[t]{0.48\linewidth}
  \centering
  \includegraphics[width=\linewidth]{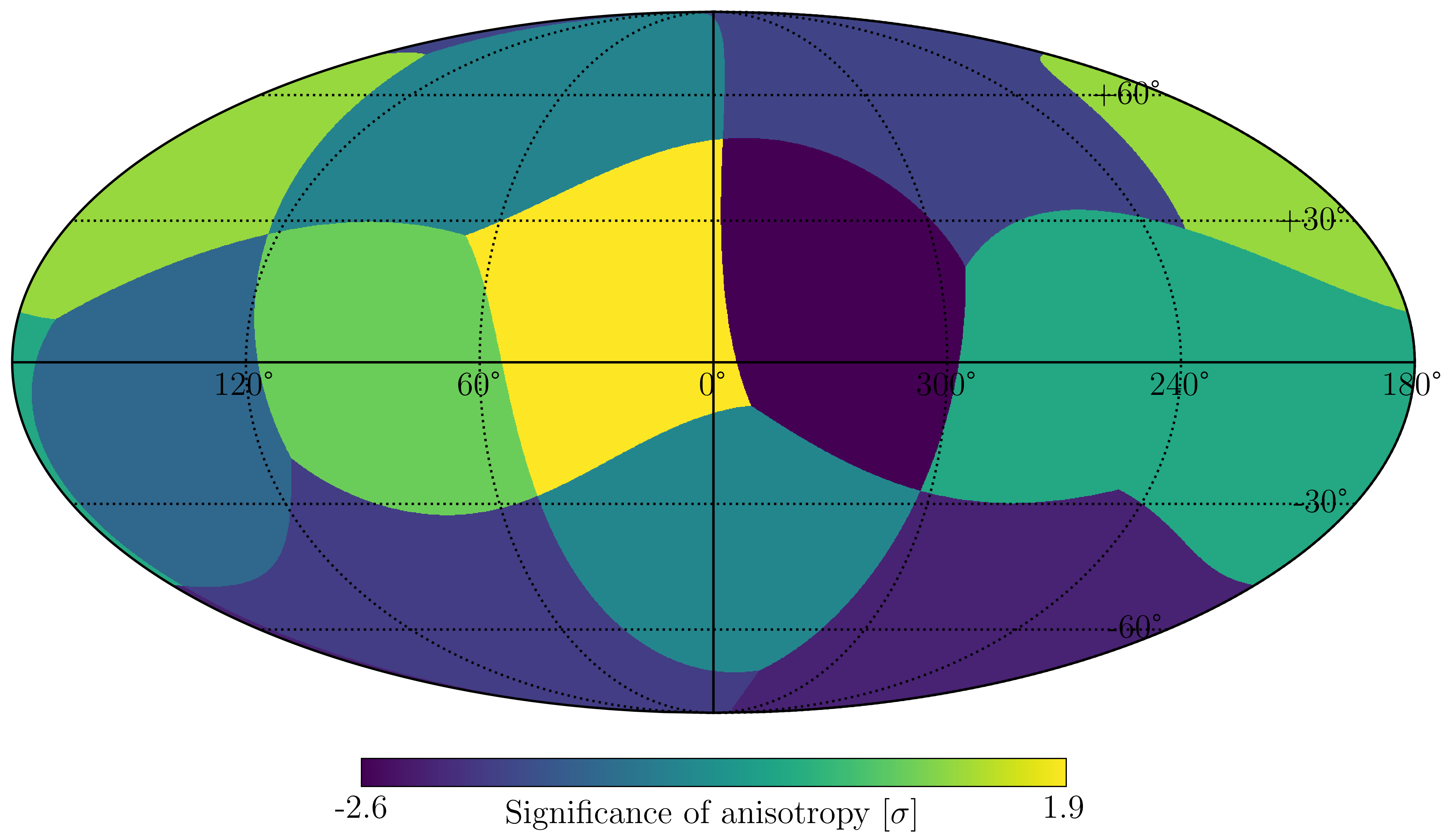}
 \end{subfigure}\hfill
 \begin{subfigure}[t]{0.48\linewidth}
  \centering
  \includegraphics[width=\linewidth]{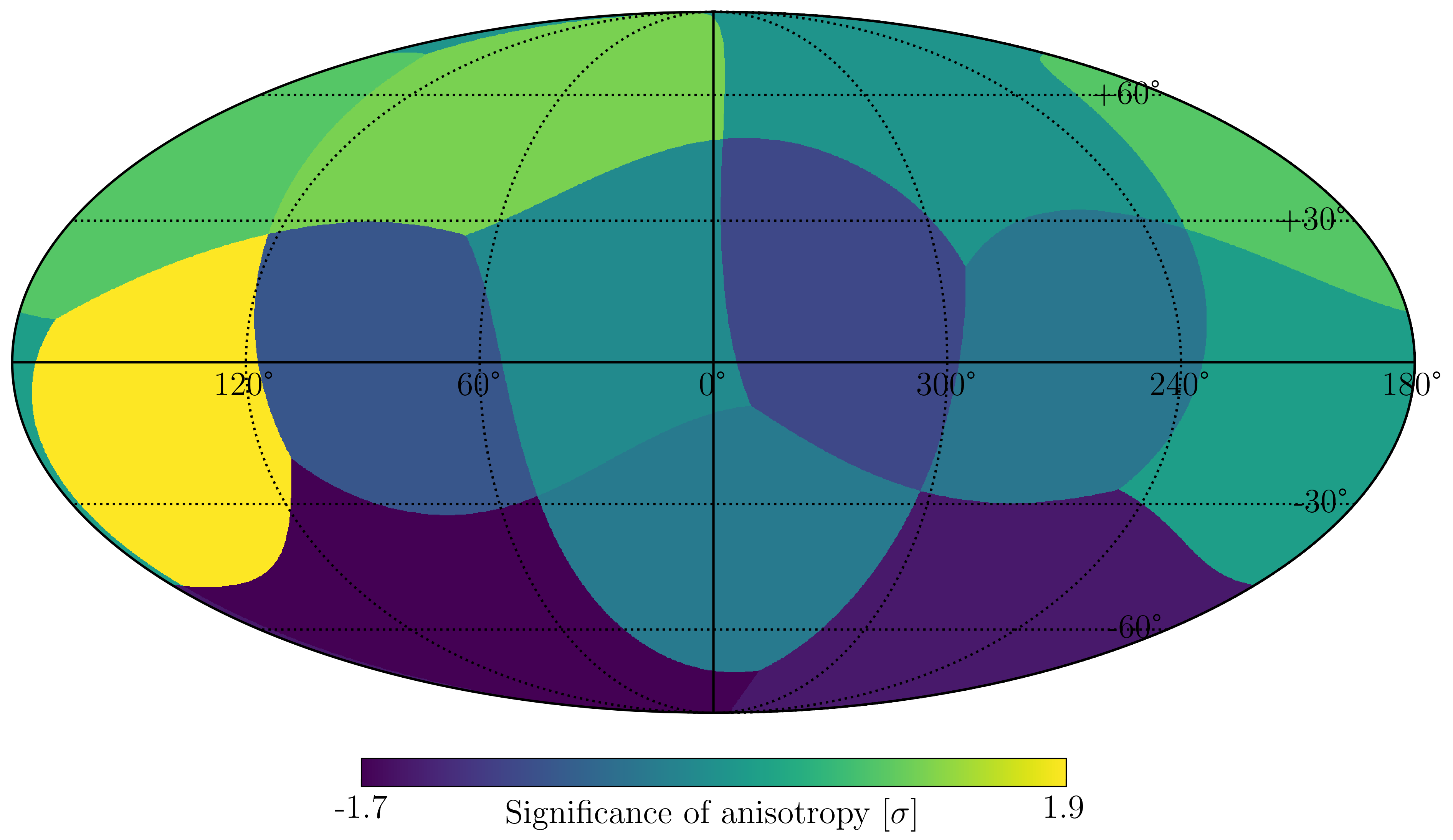}
 \end{subfigure}
 \caption{Significance of anisotropy in HEALPix pixels for the~\citetalias{carrickCosmologicalParametersComparison2015} reconstruction with the joint \LTYT\ relation. Left: $V_{\mathrm{ext}}$ pixelised model. Right: $H_0$ pixelised model. The colourbar shows the significance in units of $\sigma$.}
 \label{fig:ltyt_pix_dipoles}
\end{figure*}

We parameterise direction-dependent variations using 12 HEALPix pixels ($N_{\rm side}=1$), allowing independent values for $\Vext$, ZP, or $H_0$ in each sky direction (\cref{fig:ltyt_pix_dipoles}). For the ZP pixelised model, we find weak evidence with both~\citetalias{carrickCosmologicalParametersComparison2015} and \Manti. For the $\Vext$ pixelised model, we find weak evidence with~\citetalias{carrickCosmologicalParametersComparison2015} only ($\Delta\ln\mathcal{Z} = 2.08 \pm 0.19$). \Cref{fig:ltyt_pix_dipoles} shows the pixelised $\Vext$ and $H_0$ maps for \LTYT\ with~\citetalias{carrickCosmologicalParametersComparison2015}. The most outlying pixel shows a $2.2\sigma$ deviation from zero for $\Vext$, corresponding to a few hundred $\kmsec$ outflow. The largest flow amplitudes ($-4000$ and $+1600\,\kmsec$) appear in adjacent pixels covering the Galactic plane, but with only $\sim 1.5\sigma$ significance due to large uncertainties. We attribute these to chance fluctuations arising from the lower number of clusters around the Galactic plane. Given the weak evidence, the lack of statistically significant deviations from zero, the absence of a coherent flow pattern, and the large number of models tested, we interpret $\Vext$ pixelised model's weak evidence as likely noise.

In contrast, the $H_0$ pixelised model is strongly disfavoured when reconstructions are applied ($\Delta\ln\mathcal{Z} = -7.75$ for~\citetalias{carrickCosmologicalParametersComparison2015}, $-11.55$ for \Manti), indicating that this model's additional complexity is not justified by the data.

\begin{figure*}
 \centering
 \includegraphics[width=\linewidth]{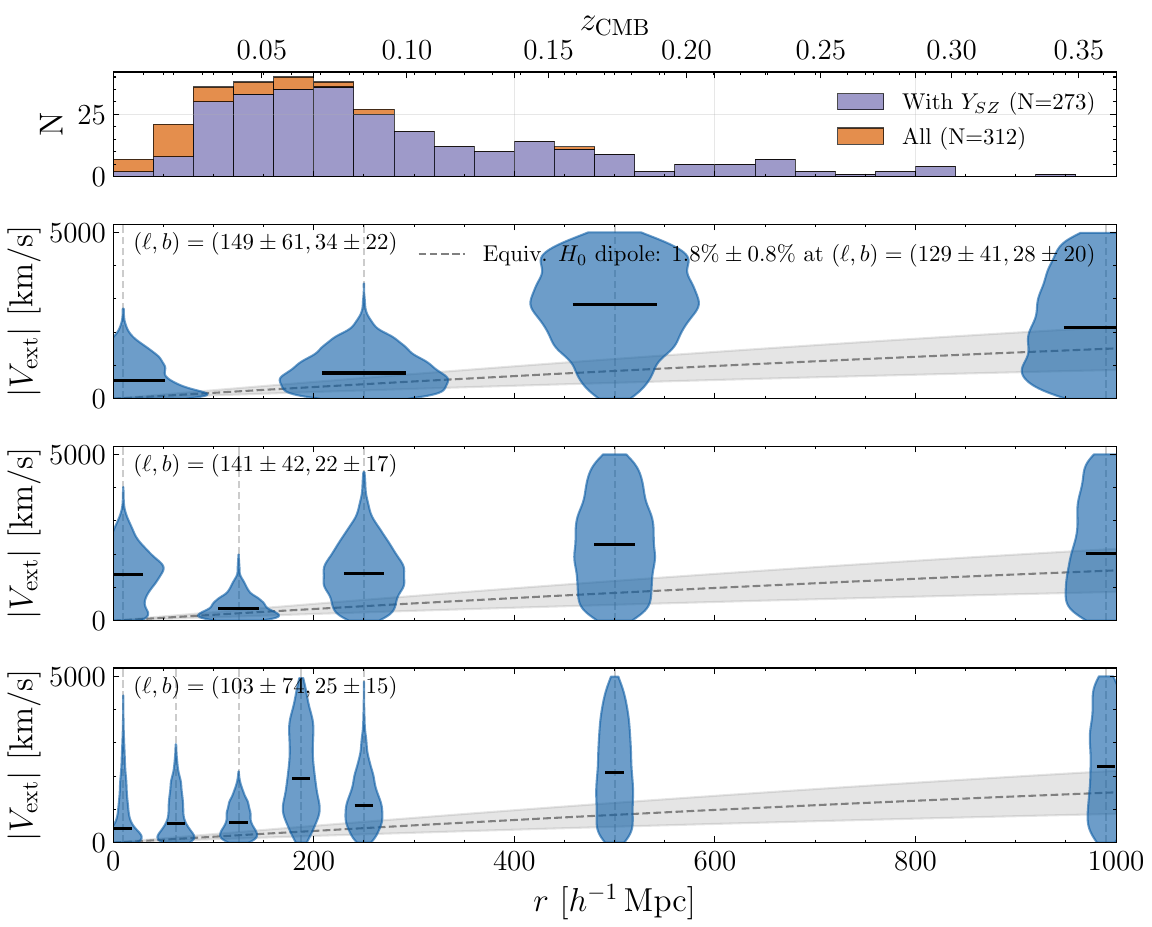}
 \caption{Posterior distributions of the radially-varying bulk flow amplitude using the joint \LTYT\ sample with the~\citetalias{carrickCosmologicalParametersComparison2015} reconstruction. Each panel shows violin plots of the full posterior at each radial knot position, with increasing radial resolution from top to bottom. The shaded horizontal band shows the $H_0$ dipole equivalent bulk flow magnitude inferred from an independent $H_0$ anisotropy model (dipH0), converted to velocity units. The inferred dipole direction $(\ell, b)$ is indicated in each panel. The top panel shows the CMB-frame redshifts of the clusters corresponding to the distance $r$ on the other panels according to the fiducial isotropic distance--redshift relation. All radial bins are consistent with zero residual flow, and the Bayesian evidence disfavours this model compared to the simpler dipole models.}
 \label{fig:radially_binned}
\end{figure*}

We test whether the bulk flow amplitude varies with distance by fitting independent velocity amplitudes at four radial knots. In the fixed-direction model, all knots share a common direction while amplitudes vary independently (\cref{fig:radially_binned}; parameters in \cref{tab:radial_params}). For the \LTYT\ relation, this model yields $\Delta\ln\mathcal{Z} = -1.30 \pm 0.12$ with~\citetalias{carrickCosmologicalParametersComparison2015} and $\Delta\ln\mathcal{Z} = -1.30 \pm 0.09$ with \Manti, disfavouring any radial dependence of the bulk flow. The inferred amplitudes at each radial knot show large uncertainties, with several bins consistent with zero. We also test a free-direction model where each radial bin has an independent direction (\cref{tab:radial_free_dir_params}). The free-direction model shows weak positive evidence with \Manti\ ($\Delta\ln\mathcal{Z} = 1.25 \pm 0.10$), while~\citetalias{carrickCosmologicalParametersComparison2015} is neutral ($\Delta\ln\mathcal{Z} = 0.36 \pm 0.66$). We note that using five radial knots instead of four increases the evidence further ($\Delta\ln\mathcal{Z} \sim 2$), but the preference disappears with a sixth knot (both knots added at low radius). As this result is sensitive to the binning choice, we do not consider it robust.

Adding a quadrupole component to the dipole (parameterised by two additional axes and an amplitude) does not improve the fit. For the $\Vext$ dipole+quadrupole model, we find $\Delta\ln\mathcal{Z} = -3.24 \pm 0.12$ with~\citetalias{carrickCosmologicalParametersComparison2015} and $\Delta\ln\mathcal{Z} = -3.68 \pm 0.09$ with \Manti, strongly disfavouring the additional complexity. The quadrupole amplitude is constrained to $<593\,\kmsec$ at 95\% confidence. Similar results hold for the ZP and $H_0$ quadrupole models (see \cref{tab:quad_params}).

We also test whether simultaneous $\Vext$ and ZP/$H_0$ dipoles improve the fit. For both reconstructions, the ZP+$\Vext$ model yields negligible or negative evidence ($\Delta\ln\mathcal{Z} = -0.55 \pm 0.12$ for~\citetalias{carrickCosmologicalParametersComparison2015}, $-0.63 \pm 0.10$ for \Manti), as does the $H_0$+$\Vext$ model ($\Delta\ln\mathcal{Z} = -1.20 \pm 0.07$ and $-3.32 \pm 0.07$ respectively). Neither combination is favoured over the simpler single-dipole models. Full parameters are given in \cref{tab:mixed_dipoles}.

For the no velocity field case, the only extended models with positive evidence are the pixelised ZP and $H_0$ models, with $\Delta\ln\mathcal{Z} = 1.78 \pm 0.08$ and $2.08 \pm 0.08$ respectively (\cref{tab:beyond_dipoles}). 

\begin{table*}
\centering
\caption{Dipole model parameter constraints and Bayesian evidence for the joint \LTYT\ relation with all reconstructions. For each model, we report the dipole amplitude, direction in Galactic coordinates $(\ell, b)$, and the log-evidence difference $\Delta\ln\mathcal{Z}$ relative to the base model (no anisotropy). Amplitudes reported as ``$<X$'' indicate 95\% upper limits where the posterior is consistent with zero. For the ZP dipole, the amplitude is expressed as a fractional magnitude variation equivalent to $\delta H_0/H_0$ via \cref{eq:Deltaa_DeltaH0_linear}, enabling direct comparison with the $H_0$ dipole. \red{The ``Preference'' column indicates the Jeffreys scale interpretation of the evidence: ``Disfavoured'' ($\Delta\ln\mathcal{Z} < 0$), ``Negligible'' ($0 \leq \Delta\ln\mathcal{Z} < 1$), ``Weak'' ($1 \leq \Delta\ln\mathcal{Z} < 2.5$), ``Moderate'' ($2.5 \leq \Delta\ln\mathcal{Z} < 5$), or ``Strong'' ($\Delta\ln\mathcal{Z} \geq 5$).} No model shows more than negligible evidence for dipolar anisotropy, and with reconstructions applied the $\Vext$ dipole amplitude is consistent with zero.}
\label{tab:dipole_results}
\begin{tabular}{|c|c|l|c|c|c|c|c|}
\hline\hline
Relation & Velocity field & Model & Amplitude & $\ell$ [$^\circ$] & $b$ [$^\circ$] & $\Delta\ln\mathcal{Z}$ & Preference \\
\hline\hline
\multirow{9}{*}{LTYT} & \multirow{3}{*}{C15} & $\mathbf{V}_{\rm ext}$ & $< 1276$\,km/s & $135 \pm 48$ & $19 \pm 24$ & $-0.93 \pm 0.09$ & Disfavoured \\
 &  & $H_0$ & $1.8^{+0.8}_{-0.8}$\% & $129 \pm 41$ & $28 \pm 20$ & $0.01 \pm 0.08$ & Negligible \\
 &  & ZP & $4.4^{+1.9}_{-2.0}$\% & $143 \pm 80$ & $49 \pm 21$ & $0.26 \pm 0.07$ & Negligible \\
\cline{2-8}
 & \multirow{3}{*}{Manticore} & $\mathbf{V}_{\rm ext}$ & $628^{+376}_{-369}$\,km/s & $138 \pm 44$ & $17 \pm 22$ & $-1.02 \pm 0.07$ & Disfavoured \\
 &  & $H_0$ & $< 2.1$\% & $166 \pm 78$ & $27 \pm 31$ & $-1.93 \pm 0.09$ & Disfavoured \\
 &  & ZP & $4.2^{+1.9}_{-2.0}$\% & $138 \pm 78$ & $46 \pm 23$ & $0.04 \pm 0.07$ & Negligible \\
\cline{2-8}
 & \multirow{3}{*}{None} & $\mathbf{V}_{\rm ext}$ & $785^{+440}_{-440}$\,km/s & $124 \pm 83$ & $41 \pm 26$ & $-0.55 \pm 0.07$ & Disfavoured \\
 &  & $H_0$ & $5.6^{+2.2}_{-2.2}$\% & $144 \pm 81$ & $54 \pm 18$ & $1.34 \pm 0.07$ & Weak \\
 &  & ZP & $4.7^{+1.6}_{-1.7}$\% & $147 \pm 83$ & $54 \pm 18$ & $0.90 \pm 0.08$ & Negligible \\
\hline
\end{tabular}
\end{table*}

\begin{table*}
\centering
\caption{Bayesian evidence comparison for extended anisotropy models for the joint \LTYT\ relation. $\Delta\ln\mathcal{Z}$ is relative to the base model (no anisotropy) for each velocity field (C15/Manticore/None). Models include: pixelised variations (12 HEALPix pixels), radially varying bulk flow with free direction (rad.) or fixed direction (rad. fix.\ dir.), dipole+quadrupole combinations, and simultaneous dipole models. With reconstructions applied, most models show negative or negligible evidence, indicating the data do not require anisotropy beyond that captured by the \red{peculiar velocity field}. The $\Vext$ pixelised model with~\citetalias{carrickCosmologicalParametersComparison2015} shows weak positive evidence ($\Delta\ln\mathcal{Z} = 2.08$), though this is not statistically significant.}
\label{tab:beyond_dipoles}
\begin{tabular}{|l|c|c|c|}
\hline\hline
Model & C15 & Manticore & None \\
\hline\hline
$\mathbf{V}_{\rm ext}$ pix. & $2.08 \pm 0.19$ & $0.44 \pm 0.20$ & $-0.23 \pm 0.07$ \\
ZP pix. & $1.30 \pm 0.14$ & $1.23 \pm 0.09$ & $1.78 \pm 0.08$ \\
$H_0$ pix. & $-7.75 \pm 0.15$ & $-11.55 \pm 0.09$ & $2.08 \pm 0.08$ \\
$\mathbf{V}_{\rm ext}$ rad. & $0.36 \pm 0.66$ & $1.25 \pm 0.10$ & $0.20 \pm 0.07$ \\
$\mathbf{V}_{\rm ext}$ rad. (fix. dir.) & $-1.30 \pm 0.12$ & $-1.30 \pm 0.09$ & $-0.28 \pm 0.07$ \\
$\mathbf{V}_{\rm ext}$ dip.+quad. & $-3.24 \pm 0.12$ & $-3.68 \pm 0.09$ & $0.54 \pm 0.10$ \\
ZP dip.+quad. & $-1.36 \pm 0.14$ & $-1.47 \pm 0.10$ & $-1.18 \pm 0.22$ \\
$H_0$ dip.+quad. & $-1.25 \pm 0.10$ & $-3.58 \pm 0.14$ & $0.80 \pm 0.16$ \\
ZP dip. + $\mathbf{V}_{\rm ext}$ dip. & $-0.55 \pm 0.12$ & $-0.63 \pm 0.10$ & $-0.66 \pm 0.08$ \\
$H_0$ dip. + $\mathbf{V}_{\rm ext}$ dip. & $-1.20 \pm 0.07$ & $-3.32 \pm 0.07$ & $-0.05 \pm 0.08$ \\
\hline
\end{tabular}
\end{table*}

\begin{figure*}
 \centering
 \includegraphics[width=0.7\linewidth]{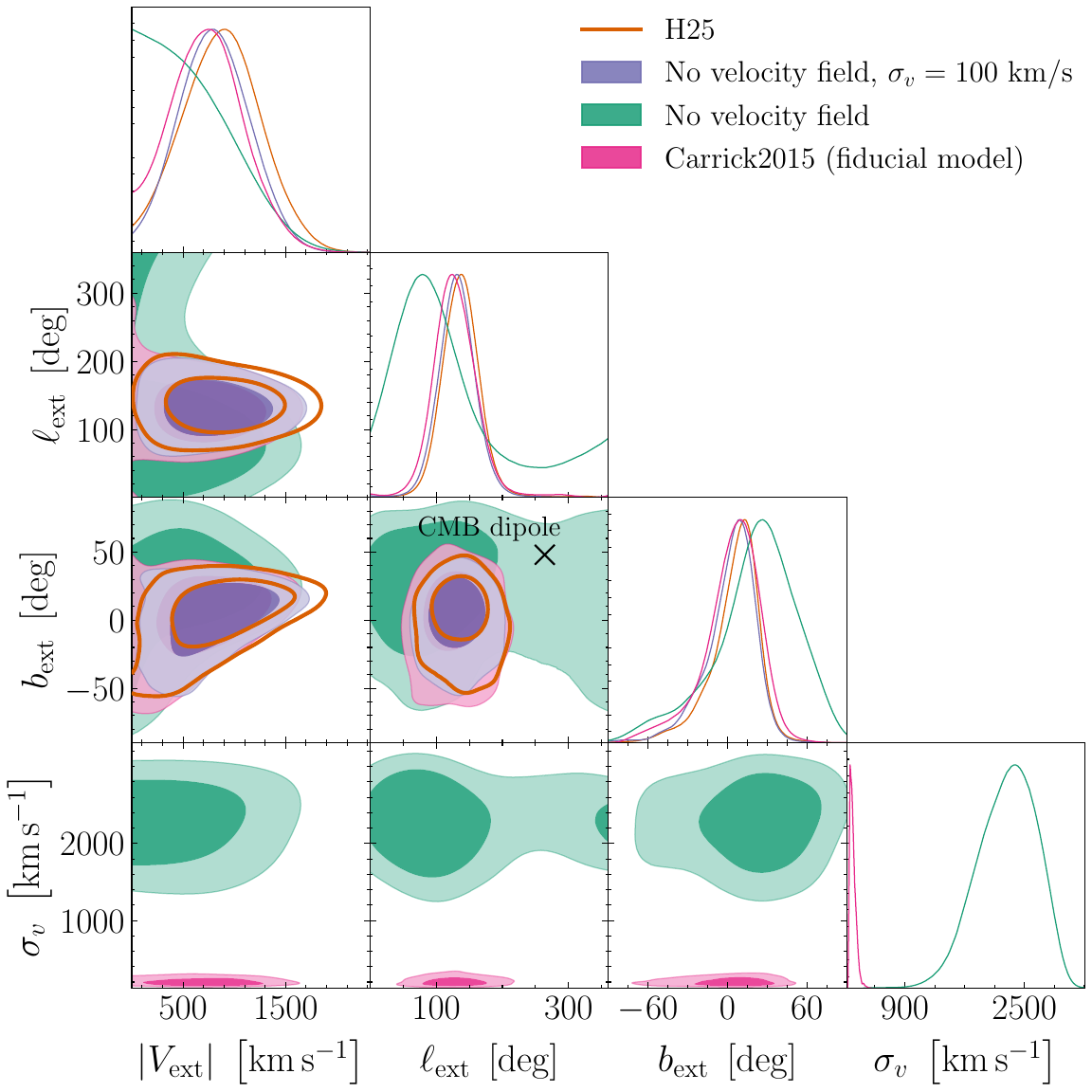}
 \caption{\red{The dipolar bulk flow and $\sigma_v$ parameters for the \LT\ relation with our fiducial model using the~\citetalias{carrickCosmologicalParametersComparison2015} reconstruction (pink), no velocity field (green), no velocity field with fixed $\sigma_v = 100\,\kmsec$ (purple), and the~\citet{heCharacterisingGalaxyCluster2025} method (orange; labelled H25). Our model without a velocity field finds a bulk flow similar to H25, but it is not statistically significant. With the reconstruction, we find a small dipole consistent with \lcdm\ expectations. The cross marks the CMB dipole direction.}}
 \label{fig:migkas-comparison}
\end{figure*}

\section{Discussion}

\subsection{Overview and broader implications}

We have presented a Bayesian hierarchical model to jointly constrain cluster scaling relations and models for bulk flows and $H_0$ anisotropy, accounting for selection effects, unmodelled redshift uncertainties and peculiar velocities, and Malmquist bias. We have applied this model to the~\citetalias{migkasCosmologicalImplicationsAnisotropy2021} sample of 312 X-ray selected clusters with measurements of their X-ray flux and temperature, and integrated Compton-$y$ parameter from Planck. We have used two different reconstructions of the local velocity field to model the peculiar velocities from local structure: the linear theory reconstruction of~\citetalias{carrickCosmologicalParametersComparison2015} and the non-linear \Manti\ reconstruction. We also present results without any peculiar velocity field.

We find that without a velocity field, a large scatter on redshifts of $\sim 2000\,\kmsec$ is necessary to explain the observed cluster redshifts. With \citetalias{carrickCosmologicalParametersComparison2015}, this is reduced to $200 \lesssim \sigma_v \lesssim 300\,\kmsec$, which is broadly in line with expectations from small scale unmodelled peculiar velocities. For \Manti\ $\sigma_v$ has an upper limit of $200\,\kmsec$, with the posterior peaking at the lower bound of the prior, which may indicate that some of the redshift scatter is being absorbed by the intrinsic scatter in the scaling relations, with which it is moderately degenerate. Both reconstructions are strongly favoured by Bayesian evidence over the no velocity field case. Together, the evidence and $\sigma_v$ constraints show that \red{peculiar velocity fields are} necessary to accurately model the peculiar velocities traced by our cluster sample. 

We consider both reconstructions on equal footing, given their complementary strengths and weaknesses. \Manti\ models are strongly favoured by Bayesian evidence over \citetalias{carrickCosmologicalParametersComparison2015} \red{for all our models} \red{(noting that \Manti\ also performed better at predicting peculiar velocities on independent datasets)}. However, the \texttt{BORG} algorithm puts a $\Lambda$CDM prior on the initial density field and evolves it to present day using a full structure formation model. In contrast,~\citetalias{carrickCosmologicalParametersComparison2015} simply takes the observed galaxy density field and applies linear theory to compute the velocity field, making it less model-dependent. Thus while \Manti\ seems to provide a more accurate velocity field, there are stronger assumptions of standard cosmology embedded in its construction, which is undesirable when aiming to test for beyond-$\Lambda$CDM anisotropy. \red{We note that both reconstructions assume the cosmological principle in their construction.}


We begin by fitting our simplest models for each class of anisotropy: a constant $\Vext$, a dipole in $\delta H_0/H_0$, and a dipole in the scaling relation zero points (ZPs). By properly treating the redshift-distance conversion assumed when constructing the velocity field model \red{(remapping the reconstruction to the anisotropic distance--redshift relation at each step of the inference, as described in \cref{sec:anisotropic_models_overview})}, we separate anisotropy in $H_0$ from anisotropy in the ZP. In particular, we find that the ZP dipole persists across all reconstruction choices (although not strongly favoured by evidence). However, modelling the peculiar velocity field reduces the preferred $H_0$ dipole amplitude from $\sim$6\% to $\sim$2\% with the Carrick2015 reconstruction, and down to an upper limit with \Manti. This demonstrates the importance of modelling the peculiar velocity field when inferring $H_0$ anisotropy from cluster scaling relations.

\red{In our model, the degeneracy between $H_0$ and ZP anisotropy is broken. We find the $H_0$ anisotropy has an upper limit of 3.2\%. The zero point dipole must therefore be driven by systematics (such as anisotropic dust foregrounds affecting the Sunyaev-Zel'dovich measurements, temperature measurements, or residual angular selection effects relating to the \YT\ sample construction) rather than cosmology.}

Our $\Vext$ dipole constraints are less sensitive to reconstruction choice. Using individual scaling relations, they show weak preference for a dipole with the posterior peaked at $\sim 800\,\kmsec$. We find that \LT\ and \YT\ each individually produce dipole constraints pointing in slightly different directions. When fitted jointly, this directional tension causes a much reduced signal, with the \LTYT\ posterior settling on an intermediate direction with lower amplitude. \red{The directional disagreement between the two relations may point to unmodelled systematics affecting one or both observables.} This demonstrates the importance of jointly fitting both relations rather than combining constraints separately. We also infer a high intrinsic correlation \red{($0.85 < \rho < 0.9$)} between the \LT\ and \YT\ relations, indicating limited independent information between them. This correlation coefficient may be useful for future cluster studies requiring correlated mock observations.

We then fit a broad suite of extended anisotropic models in $H_0$, $\Vext$, and ZP: pixelised variations, radially varying bulk flows, dipole+quadrupole combinations, and mixed dipole models. With reconstructions applied, we find no significant evidence for any of these extended models beyond the base no-anisotropy model, indicating that the data do not require additional anisotropy beyond that captured by the \red{peculiar velocity field}. The only exception is the $\Vext$ pixelised model, which shows weak evidence with C15 ($\Delta\ln\mathcal{Z} = 2.08$). Given the large number of models tested and the lack of coherent flow patterns, we conclude that this weak evidence is unlikely to be of physical significance but more likely to arise from the look-elsewhere effect.

\subsection{Comparison with cluster anisotropy literature}

Anisotropies were studied in the same cluster sample in \citetalias{migkasProbingCosmicIsotropy2020} and~\citetalias{migkasCosmologicalImplicationsAnisotropy2021}. Methodologically, the primary differences with our work are: (i) our use of velocity field reconstructions to model peculiar velocities; (ii) our Bayesian hierarchical modelling framework, which jointly fits multiple scaling relations and anisotropy parameters while modelling selection effects and Malmquist bias, with a separate likelihood for cluster scaling relation observables and redshifts; and (iii) our separation of $H_0$ and ZP anisotropies through the treatment of the redshift-distance conversion in the presence of peculiar velocities.

\citetalias{migkasProbingCosmicIsotropy2020} and \citetalias{migkasCosmologicalImplicationsAnisotropy2021} consider two anisotropy models: a dipole in $\Vext$, and a non-parametric $H_0$ anisotropy method. Our results are most directly comparable to their bulk flow, as we fit an identical flow model. While~\citetalias{migkasProbingCosmicIsotropy2020} and~\citetalias{migkasCosmologicalImplicationsAnisotropy2021} use ``minimal residuals'' and ``minimum anisotropies'' approaches to fit their bulk flow, \citet{heCharacterisingGalaxyCluster2025} present a Bayesian method which obtains \red{similar} results and fits the normalisation, slope and scatter of the scaling relation simultaneously with the bulk flow. This is similar to our no-velocity-field model fitted to individual scaling relations, with $\sigma_v \to 0$ (which removes the redshift likelihood, then completely determined by the cluster scaling relation), and selection removed.

They only apply this method to the \LT\ relation, and we show our reimplementation of their method in Fig.~\ref{fig:migkas-comparison}\red{, which recovers their contours}. We see that for our no-velocity-field model with \red{$\sigma_v$ fixed to} 100\,\kmsec\ (limited by our grid resolution), we obtain similar results to the~\citet{heCharacterisingGalaxyCluster2025} MCMC method. With $\sigma_v$ free to vary, we obtain a large value to absorb the unmodelled small scale peculiar velocities, which destroys the constraining power. With reconstructions, $\sigma_v$ is constrained to lower values, and we recover a similar posterior to theirs, although with more posterior mass at zero. However, as discussed in \cref{sec:results_dipoles}, when jointly fitting both scaling relations, the tension between their preferred dipole directions causes the joint posterior to settle on a compromise direction with almost no preference for a non-zero value. In conclusion, whilst the reconstruction and our \red{comprehensive Bayesian forward modelling} just lowers the significance for individual scaling relations, \red{jointly fitting both relations} causes the significance to collapse for the dipole model. These statements are not generally applicable to all flow models e.g. as highlighted in \cref{sec:results_beyond_dipoles}, the $\Vext$ pixelised model shows weak evidence when modelling the peculiar velocity field and jointly fitting both relations (but none without a velocity field compared to the base model).

The second model \citetalias{migkasProbingCosmicIsotropy2020} and \citetalias{migkasCosmologicalImplicationsAnisotropy2021} consider is a non-parametric $H_0$ anisotropy method. The sky is scanned with overlapping cones, fitting an $H_0$ value to each, with cluster observables weighted by a distance-based kernel from the cone centre. The significance is assessed using frequentist mocks. The direction of \citetalias{migkasCosmologicalImplicationsAnisotropy2021}'s claimed $H_0$ anisotropy broadly aligns with the direction of our preferred dipoles for the three velocity field models. The model itself is most similar to our pixelised $H_0$ model, although our model fits $H_0$ in 12 independent \texttt{HEALpix} pixels in order to avoid having overlapping pixels where each cluster's observables are used in multiple likelihoods. With our no-velocity-field method we find weak evidence for our pixel model, but with reconstructions we find strong evidence against it, with no obvious dipole pattern, likely due to the large pixels and low cluster counts in each. Loss of evidence for anisotropy when the velocity field is included is generally true for other $H_0$ models: we find once the reconstruction is added the significance of $H_0$ anisotropy drops considerably.

Overall, even without a velocity field, our method finds much weaker significance for anisotropy than \citetalias{migkasProbingCosmicIsotropy2020} and \citetalias{migkasCosmologicalImplicationsAnisotropy2021}. With reconstructions added, the significance drops further for $H_0$ models.









\subsection{Comparison with anisotropy studies with other tracers}

Our analysis is most directly comparable to recent studies of $H_0$ anisotropy using other distance tracers, particularly the Tully--Fisher relation (TFR) and Type Ia supernovae. \citet{boubelTestingAnisotropicHubble2025} analyse CosmicFlows-4 TFR distances using the \citetalias{carrickCosmologicalParametersComparison2015} reconstruction, reporting a dipole variation in $H_0$ of approximately 3 per cent ($0.063 \pm 0.016$\,mag) towards $(\ell, b) = (142^\circ \pm 30^\circ, 52^\circ \pm 10^\circ)$ at $3.9\sigma$ significance. \citet{stiskalekNoEvidenceLocal2026} re-analysed the same data, as well as 2MTF, SFI++, and Pantheon+ samples, using a forward-modelling approach \red{with peculiar velocity fields}. They find a somewhat larger dipole of $\sim$4 per cent in CosmicFlows-4 towards $(\ell, b) = (127^\circ \pm 11^\circ, 10^\circ \pm 8^\circ)$, attributing the difference to methodological choices including velocity assignment in real versus redshift space. Crucially, by allowing for a \red{radially varying $\Vext$}, they show that the anisotropic model captures local flow features rather than an actual linearly growing effective bulk flow, consistent with \lcdm expectations.

Our $H_0$ dipole with \citetalias{carrickCosmologicalParametersComparison2015} points towards $(\ell, b) = (129^\circ \pm 41^\circ, 28^\circ \pm 20^\circ)$, broadly consistent in Galactic longitude with both \citet{boubelTestingAnisotropicHubble2025} ($\ell = 142^\circ \pm 30^\circ$) and \citet{stiskalekNoEvidenceLocal2026} ($\ell = 127^\circ \pm 11^\circ$). However, the Galactic latitude shows significant scatter: our $b = 28^\circ \pm 20^\circ$ is intermediate between \citet{boubelTestingAnisotropicHubble2025} ($b = 52^\circ \pm 10^\circ$) and \citet{stiskalekNoEvidenceLocal2026} ($b = 10^\circ \pm 8^\circ$), with ranges that barely overlap.

Our forward-modelling framework is based on the approach of \citet{stiskalekNoEvidenceLocal2026}: we jointly calibrate the distance relation, marginalise over cluster distances, and account for peculiar velocities using both linear-theory (\citetalias{carrickCosmologicalParametersComparison2015}) and non-linear (\Manti) reconstructions. A key extension is that we rescale the position at which the velocity field is queried to account for the anisotropic redshift--distance relation and possible bulk flows, ensuring consistency between the assumed cosmography and the peculiar velocity correction. Additionally, we do not include an inhomogeneous Malmquist bias term, as the clusters are composed of galaxies that appear in \TWOMPP\ and are (likely) used to derive the cluster redshifts, potentially introducing correlations with the reconstruction. We also employ a more comprehensive set of flow models than previous studies, which is appropriate given weak priors on the form any anisotropy should take. Applying this framework to cluster scaling relations rather than galaxy scaling relations or supernovae probes a somewhat different volume and relies on distinct astrophysical systematics: the \LT\ and \YT\ relations depend on \red{intracluster medium} physics rather than stellar populations, providing an independent check on anisotropy claims.

Bulk flow studies not incorporating velocity field reconstructions have generally reported larger amplitudes. \citet{watkinsAnalysingLargescaleBulk2023}, using CosmicFlows-4, claim a bulk flow of $\sim 400$\,\kmsec\ towards $(\ell, b) = (298^\circ \pm 5^\circ, -8^\circ \pm 4^\circ)$ at $250\,h^{-1}$\,Mpc, in apparent tension with \lcdm at $\sim 5\sigma$. This direction is nearly opposite to our \Manti\ $\Vext$ dipole towards $(\ell, b) = (138^\circ \pm 44^\circ, 17^\circ \pm 22^\circ)$, differing by $\sim 160^\circ$ in Galactic longitude. Earlier work by \citet{hudsonLargeScaleBulkFlow1999} found a bulk flow of $630 \pm 200$\,\kmsec\ from Fundamental Plane distances to 56 galaxy clusters. By contrast, studies that include velocity reconstructions---including ours, \citet{stiskalekNoEvidenceLocal2026}, and \citet{rahmanNewConstraintsAnisotropic2022} who found no evidence for anisotropic expansion in \red{Joint Light-curve Analysis (JLA) supernovae~\citep{betouleSNLSSupernovaeLegacy2014}} using the \citetalias{carrickCosmologicalParametersComparison2015} reconstruction---find bulk flows consistent with \lcdm expectations. This suggests that apparent excess bulk flows may arise from local flow features that are captured by reconstruction, rather than genuine departures from isotropy on large scales\red{~\citep[as argued by][]{stiskalekNoEvidenceLocal2026}}.

Our results thus align with the emerging picture that apparent anisotropies in distance-indicator analyses largely reflect local structure rather than genuine departures from the cosmological principle. The consistency across different tracers---clusters, TFR galaxies, and supernovae---\red{when peculiar velocity fields are modelled} underscores the importance of \red{properly accounting for} peculiar velocities in tests of cosmic isotropy.

Relatedly, \citet{stiskalekTestingLocalSupervoid2025} deployed a similar framework to study homogeneity of the universe rather than isotropy, testing claims of a large-scale local underdensity. Their preferred model constrained the void to be less than 10 per cent the size claimed by previous studies of photometric galaxy catalogues.

\subsection{Reconstructions in anisotropic models}\label{sec:methods_reconstructions}

\red{Our work demonstrates the critical importance of} reconstructed peculiar velocity fields of the local universe in studies of cosmological anisotropy. However, when testing for anisotropies in $H_0$ or the bulk flow, one must carefully consider how isotropy assumptions embedded in the reconstruction methodology might affect the results.

We employ two reconstructions in this work, both based on the \TWOMPP\ redshift survey (introduced in \cref{sec:methods_distances}). The~\citetalias{carrickCosmologicalParametersComparison2015} reconstruction \red{(iteratively) derives} the density field directly from the observed galaxy distribution, then derives the velocity field using linear perturbation theory. This approach is relatively model-agnostic: it requires only the assumption that structure growth is well-described by linear theory on the scales of interest. \red{Then the continuity equation is used to derive velocities from densities (which assumes the cosmological principle).}
In contrast, \Manti\ employs the \texttt{BORG} algorithm to forward-model the full particle trajectories from Gaussian initial conditions, constrained by the observed galaxy distribution. While this captures non-linear dynamics, it necessarily assumes the $\Lambda$CDM power spectrum and expansion history, making it more model-dependent.

To infer anisotropies self-consistently, the reconstruction should ideally be generated under the same anisotropic assumptions as the inference model. For linear-theory reconstructions like~\citetalias{carrickCosmologicalParametersComparison2015}, this would require constructing the density field from the redshift survey using the anisotropic redshift--distance relation assumed in a given likelihood evaluation. However, the iterative procedure used by~\citetalias{carrickCosmologicalParametersComparison2015} to remove redshift-space distortions is computationally expensive, making it impractical to regenerate the reconstruction at each step of the inference. We therefore adopt an approximation (fully described in Section~{\ref{sec:methods_flow_models}}): as both~\citetalias{carrickCosmologicalParametersComparison2015} and \Manti\ provide their overdensity $\delta(\bm{r})$ and velocity $\bm{v}(\bm{r})$ fields in real space, we first convert these back to redshift space $\delta(z)$ and $\bm{v}(z)$ along the line of sight using their assumed isotropic redshift--distance relations. We also remove the~\citetalias{carrickCosmologicalParametersComparison2015} $\Vext$ at this stage, so we can replace it with our own $\Vext$ model. Then, at each step of the inference, we convert the reconstructions back to real space by first correcting the redshifts with our $\Vext$ model and then applying the anisotropic redshift--distance relation. At low redshift, where the linear approximation holds, the bulk flow and local peculiar velocities are approximately separable---they contribute linearly to the observed redshift---so this remapping procedure is expected to be accurate. Note that the reconstructed velocities are only used when predicting the redshifts of the clusters; we do not use the density field for inhomogeneous Malmquist bias corrections.

Both reconstructions contain embedded isotropy assumptions that cannot be removed by \red{the velocity field remapping procedure described in \cref{sec:methods_flow_models}}. For~\citetalias{carrickCosmologicalParametersComparison2015}, this enters in three ways. First, the luminosity of each galaxy in the redshift survey is calculated using an isotropic redshift--distance relation. Second, these luminosities are used both to weight the density field directly and to compute radial selection corrections. Third, the velocity field is derived from the density field via the continuity equation, assuming isotropic linear perturbation theory; here the isotropy assumptions in calculating the density field may also propagate, particularly through the strong impact of long-wavelength modes in the Fourier transform to velocities. Since the velocity field depends on the absolute density field (not just local variations), and is sensitive to long-wavelength modes via the Fourier transform, the isotropy assumptions are more important for velocity predictions than for density-based corrections. 

\red{In this work we only use reconstructions to model the peculiar velocity field, and don't direct use the density field to inform the distance prior (i.e. no inhomogenous Malmquist correction, see \cref{sec:methods_inference})}. However, since the inhomogeneous Malmquist term is primarily sensitive to \emph{variations} in the density field in the vicinity of each object rather than absolute values, the effects discussed above would be expected to have a smaller impact on the inhomogeneous Malmquist correction as long as the anisotropy varies slowly compared to the reconstruction smoothing scale ($\sim 4\,\Mpch$).

For \Manti, the isotropy assumptions are stronger: both the power spectrum prior used in the reconstruction and the redshift--distance relation used for luminosities and selection corrections assume isotropy. Since the prior is on the $\Lambda$CDM power spectrum, departures from statistical isotropy are disfavoured, which may bias inferences with respect to anisotropic models. To fully account for these effects would require modifying the \texttt{BORG} algorithm to incorporate a fully specified anisotropic cosmological model. However, the theoretical development of such models---capable of accommodating large-scale bulk flows or $H_0$ anisotropies in a self-consistent manner---is only in its early stages~\citep[e.g.][]{Krishnan2023, Ebrahimian2024, Constantin2023, Martin2025, Tsagas2022}


\subsection{Future work}\label{sec:future_work}

\subsubsection{Methodology}\label{sec:future_methodology}

As discussed in \cref{sec:methods_reconstructions}, our approximate treatment of anisotropy in existing reconstructions---rescaling distances but not accounting for anisotropy in galaxy luminosities and completeness corrections---highlights the need for methodological improvements. Future work should focus on building all anisotropy dependence directly into the reconstructions from the ground up in their construction. For \Manti\ this would require building anisotropy parameters into BORG, which is a longer-term endeavour. However, it should be feasible for linear-theory methods such as those of~\citetalias{carrickCosmologicalParametersComparison2015}, if one is prepared to assume the density to velocity transformation from linear perturbation theory approximately holds in anisotropic cosmologies. A brute force, but costly procedure to achieve this would be to repeat the entire \citetalias{carrickCosmologicalParametersComparison2015} construction pipeline at each step of the inference.

However, another inadequacy of \citetalias{carrickCosmologicalParametersComparison2015} is its lack of a posterior on the density field. Recent advances in Bayesian reconstruction methods~\citep[e.g.][]{Jasche_2013,Jasche_2015,Lavaux_2016,mcalpineManticoreProjectDigital2025} \red{now enable sampling from the posterior distribution of density and velocity fields given the galaxy data in a $\Lambda$CDM context. These methodologies could be adapted to linear theory reconstructions, with joint sampling of the density and velocity fields at the present day along with the assumed anisotropic redshift--distance relation and any additional constraining direct distance data.} Such an approach would be computationally intensive \red{(although relatively cheap compared to BORG)}, but would provide a fully self-consistent treatment of anisotropy in peculiar velocity modelling.

A further refinement to our overall modelling procedure would be to forward model the raw observational data more directly, rather than relying on derived quantities such as $T$, $L$, and $Y$. For X-ray clusters, one could in principle fit the X-ray spectrum directly within the inference framework, extracting temperature and luminosity as latent parameters with their full covariance structure preserved.
\red{This is analogous to the approach taken by the BayeSN framework for Type Ia supernovae~\citep{mandelHierarchicalBayesianSED2022}.}
Similarly, for the tSZ signal, one could forward model the \emph{Planck} frequency maps to infer $Y$ jointly with the anisotropy parameters, accounting for the matched multi-filter extraction process and its associated uncertainties. Such an approach would eliminate potential biases introduced by the two-stage procedure of first extracting cluster properties and then fitting scaling relations, and would naturally propagate all observational uncertainties into the final constraints. While computationally demanding, this hierarchical approach represents the most principled, fully forward-modelled treatment of the data.

\subsubsection{Data}\label{sec:future_data}

For the purposes of comparison to previous literature, we restricted ourselves to the~\citetalias{migkasCosmologicalImplicationsAnisotropy2021} cluster sample. However, future work should apply our methodology to upcoming larger cluster samples with well characterised selection functions. In particular, the all-sky eROSITA survey~\citep{predehleROSITAMissionOverview2021} is expected to detect on the order of $10^5$ galaxy clusters~\citep{pillepichSimulatingeROSITAXray2012}, with a significant fraction having measured temperatures from eROSITA or follow-up observations. The catalogues from the eROSITA Final Equatorial Depth Survey (eFEDS)~\citep{brunnerSRGEROSITAFinalEquatorial2022} and the first all-sky data release~\citep{merloniSRGEROSITAAllSky2024} are already available, with larger catalogues forthcoming from future data releases~\citep{baharEROSITAFinalEquatorialDepth2022,migkasSRGEROSITAAllSky2024,ramos-cejaSRGEROSITAAllsky2025}. Applying our methodology to these expanded samples will enable more precise constraints on anisotropies and better characterisation of any radial dependence. \red{Cluster scaling relations probe peculiar velocities at larger distances than Tully--Fisher or fundamental plane samples, making them complementary to supernova studies.}

Beyond X-ray surveys, large cluster samples are being assembled from CMB observations via the thermal Sunyaev--Zel'dovich effect. Current and upcoming surveys such as the Simons Observatory~\citep{adeSimonsObservatoryScience2019} and CMB-S4~\citep{abrahamoCMBS4ScienceCase2022} will detect tens of thousands of clusters across the full sky, providing complementary tSZ-selected samples with different selection functions and systematics compared to X-ray samples. The combination of X-ray and tSZ-selected catalogues will help disentangle genuine cosmological signals from selection-dependent biases.

In addition to expanding cluster samples, our methodology can be extended to incorporate additional cluster observables. Galaxy velocity dispersion, as an alternative distance-independent mass proxy to X-ray temperature, has been used by~\citet{pandyaExaminingLocalUniverse2024} to test for isotropy with results consistent with our findings. Weak lensing masses and optical richness measurements offer further independent mass proxies that could be combined with X-ray and tSZ observables in a multi-observable ``cluster fundamental plane'' approach, analogous to the galaxy fundamental plane, which should be capable of reducing scatter in the distance estimates.

More broadly, the forward-modelling framework and peculiar velocity treatment developed here can be applied to other distance tracers used in peculiar velocity studies~\citep{stiskalekVelocityFieldOlympics2026,stiskalekNoEvidenceLocal2026}. \red{Future Tully--Fisher samples from instruments such as the Widefield ASKAP L-band Legacy All-sky Blind surveY (WALLABY)~\citep{koribalskiWALLABYSKAPathfinder2020}
will provide denser sampling of the local velocity field}, while the Dark Energy Spectroscopic Instrument (DESI) is delivering both Tully--Fisher~\citep{douglassDESIDR1Peculiar2025} and fundamental plane~\citep{saidDESIPeculiarVelocity2025} distance measurements. For Type Ia supernovae, the Vera C.\ Rubin Observatory Legacy Survey of Space and Time (LSST)~\citep{ivezicLSSTScienceDrivers2019} and the Nancy Grace Roman Space Telescope~\citep{hounsellSimulationsNancyGrace2018} will dramatically increase the number of low-redshift supernovae available for peculiar velocity analyses. Joint analyses combining these diverse tracers with cluster scaling relations will provide important cross-checks on any claimed anisotropies.

\section{Conclusion}

We present a Bayesian forward-modelling analysis of cluster scaling relations to test for large-scale anisotropy in the local expansion rate, jointly fitting the \LT\ and \YT\ relations for 312 clusters at $z \lesssim 0.2$ while marginalising over cluster distances and accounting for peculiar velocities using either \red{linear theory (\citetalias{carrickCosmologicalParametersComparison2015}) or non-linear modelling with the BORG algorithm (\Manti).}
Our conclusions are as follows:
\begin{itemize}
    \item Without accounting for peculiar velocities, a large unmodelled velocity dispersion of $\sigma_v \sim 2000\,\kmsec$ is required to explain the observed cluster redshifts. We find weak evidence for an $H_0$ dipole with amplitude $\delta H_0/H_0 = 5.6^{+2.2}_{-2.2}\%$ and a bulk flow of $785^{+440}_{-440}\,\kmsec$. However, this model cannot accurately predict cluster distances and is strongly disfavoured by Bayesian evidence ($\Delta\ln\mathcal{Z} \sim 40$) relative to models including local peculiar velocities.
    \item \red{Modelling the peculiar velocity field} dramatically reduces the inferred anisotropy. With the~\citetalias{carrickCosmologicalParametersComparison2015} reconstruction, the $H_0$ dipole amplitude decreases to $1.8^{+0.8}_{-0.8}\%$ (with negligible Bayesian evidence over the fiducial model), the bulk flow is constrained to $< 1300\,\kmsec$ at 95\% confidence, and $\sigma_v$ reduces to $200$--$250\,\kmsec$, consistent with small-scale peculiar velocity dispersion. With \Manti, the $H_0$ dipole is constrained to $\delta H_0/H_0 < 2.1\%$ (95\% confidence) with negative evidence, the bulk flow to $628^{+376}_{-369}\,\kmsec$, and $\sigma_v < 200\,\kmsec$. Both reconstructions are strongly preferred over no velocity field by the Bayesian evidence.
    \item \red{We stress that the above decrease in preferred anisotropy compared to literature results is not because we have simply ``absorbed'' the anisotropy into the peculiar velocity field: rather, neglecting peculiar velocities leads to spurious inferred anisotropy signals much larger than the modest average peculiar velocities in the reconstructions.}
    \item Crucially, proper treatment of the anisotropic redshift--distance relation \red{when modelling the peculiar velocity field} breaks the degeneracy between $H_0$ anisotropy and scaling relation zero point (ZP) anisotropy. The ZP dipole persists weakly across all reconstruction choices ($\sim 4$--$5\%$ amplitude) but is not favoured by evidence ($\Delta\ln\mathcal{Z} = 0.04$--$0.90$). In contrast, the $H_0$ dipole signal is substantially suppressed when reconstructions are properly applied, demonstrating that previous claims of $H_0$ anisotropy from cluster scaling relations may largely reflect unmodelled peculiar velocity structure.
    \item Extended models including quadrupoles, pixelised sky variations, and radially-varying bulk flows show no more than weak evidence relative to the base model when reconstructions are applied. Given the suite of models tested and the lack of more than weak evidence, we conclude that none of the anisotropy detected is likely to be of physical significance.
\end{itemize}

\red{Our work demonstrates that principled forward modelling of redshifts through latent distances, as well as accounting for small scale peculiar velocities is essential when using distance indicators to test for cosmological anisotropy. The approximate scheme we develop for treating anisotropic distance--redshift relations in reconstructed density/peculiar velocity fields, while computationally efficient, highlights the need for future work to build anisotropy directly into reconstruction algorithms in a fully self-consistent manner.} Upcoming large cluster samples from eROSITA and other surveys, combined with improved reconstruction methods and joint analyses with complementary tracers such as supernovae and the Tully--Fisher relation, will enable increasingly stringent tests of the cosmological principle in the local Universe.

\section*{Data availability}

The \citetalias{carrickCosmologicalParametersComparison2015} reconstruction\footnote{\url{http://cosmicflows.iap.fr/Cosmicflows-3.php}} and the \Manti\ reconstruction\footnote{\url{https://digitaltwin.fysik.su.se}} are publicly available. The rest of the data underlying this article will be shared on reasonable request to the corresponding author.

\section*{Acknowledgements}

We thank Mike Hudson and Guilhem Lavaux for useful discussions and comments, and Konstantinos Migkas for providing the cluster catalogue and for useful discussions.

TY acknowledges support from UKRI Frontiers Research Grant [EP/X026639/1], which was selected by the ERC.
RS acknowledges financial support from STFC Grant No. ST/X508664/1 and the Snell Exhibition of Balliol College, Oxford.
HD is supported by a Royal Society University Research Fellowship (grant no. 211046). SvH is supported by a Leverhulme Trust
International Professorship Grant to S. Sondhi (No. LIP-2020-014).

We thank Jonathan Patterson for smoothly running the Glamdring Cluster hosted by the University of Oxford, where the data processing was performed.

\bibliographystyle{mnras}
\bibliography{main,new}

\appendix

\section{Density and velocity field reconstructions}\label{app:reconstructions}

All reconstructions used in this work are derived from the \TWOMPP\ galaxy redshift catalogue~\citep{Lavaux_2011}, a whole-sky compilation of approximately 70\,000 galaxies assembled from 2MASS photometry~\citep{skrutskie2006TwoMicronAll} combined with redshifts from 2MRS~\citep{huchraKilometreSecondRedshift2012}, 6dF~\citep{jonesDegreefieldGalaxySurvey2009}, and SDSS DR7~\citep{abazajianSeventhDataRelease2009}. The catalogue is magnitude-limited to $K < 11.5$ in the 2MRS region and $K < 12.5$ in the 6dF and SDSS regions, with apparent magnitudes corrected for Galactic extinction, $k$-corrections, evolution, and surface brightness dimming.

\subsection{Carrick et al. (2015)}\label{app:carrick}

The~\citetalias{carrickCosmologicalParametersComparison2015} reconstruction employs the iterative method of~\citet{yahilCatalogInfraredObservations1991} to infer the luminosity-weighted galaxy density field from the redshift-space positions of galaxies in \TWOMPP.

To account for the magnitude-limited nature of the survey, each galaxy is assigned a luminosity weight $w^L(r)$ equal to the ratio of the total expected luminosity (integrating the Schechter luminosity function above a minimum luminosity $L_{\min}$) to that observable at distance $r$ given the survey flux limit. Galaxies from 2MRS with distances exceeding $125\,\Mpch$ are assigned zero weight, as the shallower magnitude limit ($K < 11.5$) leads to severe incompleteness at these distances. The Zone of Avoidance ($|b| < 5^\circ$, or $|b| < 10^\circ$ near the Galactic centre) is filled by cloning galaxies from adjacent strips at higher latitudes.

The weighted galaxy luminosities are placed on a $256^3$ grid covering $(400\,\Mpch)^3$ to compute the luminosity-density field. The galaxy density contrast $\delta_\mathrm{g}$ is then normalised to a common effective bias $b^*$ (that of an $L^*$ galaxy) to account for the fact that more luminous---and hence more biased---galaxies are preferentially observed at larger distances in a magnitude-limited survey. This correction uses a luminosity-dependent bias relation, yielding a normalised density contrast $\delta_\mathrm{g}^* = \delta_\mathrm{g}/\psi^L(r)$ where $\psi^L(r)$ is the effective luminosity-weighted bias at distance $r$ (see equation~8 of~\citetalias{carrickCosmologicalParametersComparison2015}).

The algorithm iteratively solves for galaxy distances by assuming the density field sources a velocity field via linear perturbation theory. In the linear regime where density fluctuations are small ($\delta \lesssim 1$), the continuity equation relates the peculiar velocity field to the matter overdensity through
\begin{equation}\label{eq:linear_velocity}
\bm{v}(\bm{r}) = \frac{f(\Om)}{4\pi} \int \dd^3\bm{r}' \, \delta(\bm{r}') \frac{\bm{r}' - \bm{r}}{|\bm{r}' - \bm{r}|^3},
\end{equation}
where $f(\Om) \approx \Om^{0.55}$ is the linear growth rate of density perturbations~\citep{Wang_1998}. Starting from redshift-space positions, the reconstruction ``adiabatically turns on gravity'' by incrementing $\beta \equiv f(\Om)/b^*$ from 0 to 1.0 in steps of 0.01. At each iteration, predicted line-of-sight velocities are used to update galaxy distances, with the previous five distance estimates averaged to suppress oscillations in triple-valued regions near clusters. The density field is smoothed with a $4\,\Mpch$ Gaussian kernel at each step to ensure linearity. The optimal value of $\beta$ is determined by comparing the predicted peculiar velocities to an independent Tully--Fisher peculiar velocity dataset, yielding $\beta = 0.43 \pm 0.021$~\citep{carrickCosmologicalParametersComparison2015}. In this work, we treat $\beta$ as a free parameter to scale the velocity field amplitude when applying the reconstruction to cluster data, with the \citetalias{carrickCosmologicalParametersComparison2015} value used as a prior.

\subsection{Manticore}\label{app:manticore}

\Manti~\citep{mcalpineManticoreProjectDigital2025} represents a more advanced approach, using the \texttt{BORG} algorithm~\citep{Jasche_2013,Jasche_2015,Lavaux_2016} to perform full field-level Bayesian inference of the initial conditions of the local Universe.

The \texttt{BORG} algorithm infers the posterior distribution of the primordial Gaussian white noise field by forward-modelling the gravitational evolution of structure and comparing to the observed galaxy distribution. The \TWOMPP\ catalogue is divided into 32 subcatalogues binned by $K$-band magnitude and redshift, each with associated angular and radial selection functions accounting for survey completeness. The forward model employs the COLA (COmoving Lagrangian Acceleration) approximate gravity solver during inference for computational efficiency, coupled with a sigmoid-truncated double power-law galaxy bias model that captures scale and density-dependent relationships between galaxies and dark matter.

The inference is performed on a $256^3$ grid covering a parent volume of $(1000\,\mathrm{Mpc})^3$, yielding a spatial resolution of approximately $4\,\Mpch$ in the initial white noise phases. From the posterior, 50 independent samples of the initial conditions are drawn and resimulated to $z = 0$ using the high-accuracy \texttt{SWIFT} $N$-body code~\citep{2024MNRAS.530.2378S}, providing physically consistent realisations of the density and velocity fields. The reconstruction is constrained within a radius of $\sim200\,\Mpch$ ($z \lesssim 0.05$), with unconstrained small-scale and large-scale modes drawn from the prior.

A fixed $\Lambda$CDM cosmology is adopted from the Dark Energy Survey Year 3 analysis~\citep{abbottDarkEnergySurvey2022}, with $\Om = 0.306$, $h = 0.681$, and $\sigma_8 = 0.807$. Because \Manti\ self-consistently evolves the matter distribution under gravity, inferring the full dynamical state of the local Universe, the velocity field amplitude does not require an additional scaling parameter. The peculiar velocities emerge directly from the $N$-body dynamics of the inferred initial conditions, and hence the velocities do not require a rescaling parameter as in \citetalias{carrickCosmologicalParametersComparison2015}. However, to capture possible residual uncertainties on the velocity field we put an equivalent parameter on $\beta = 1.0 \pm 0.04$ analogous to \citetalias{carrickCosmologicalParametersComparison2015}.

\section{Full posteriors}\label{app:full_posterior}

\begin{figure*}
  \centering
  \includegraphics[width=0.7\linewidth]{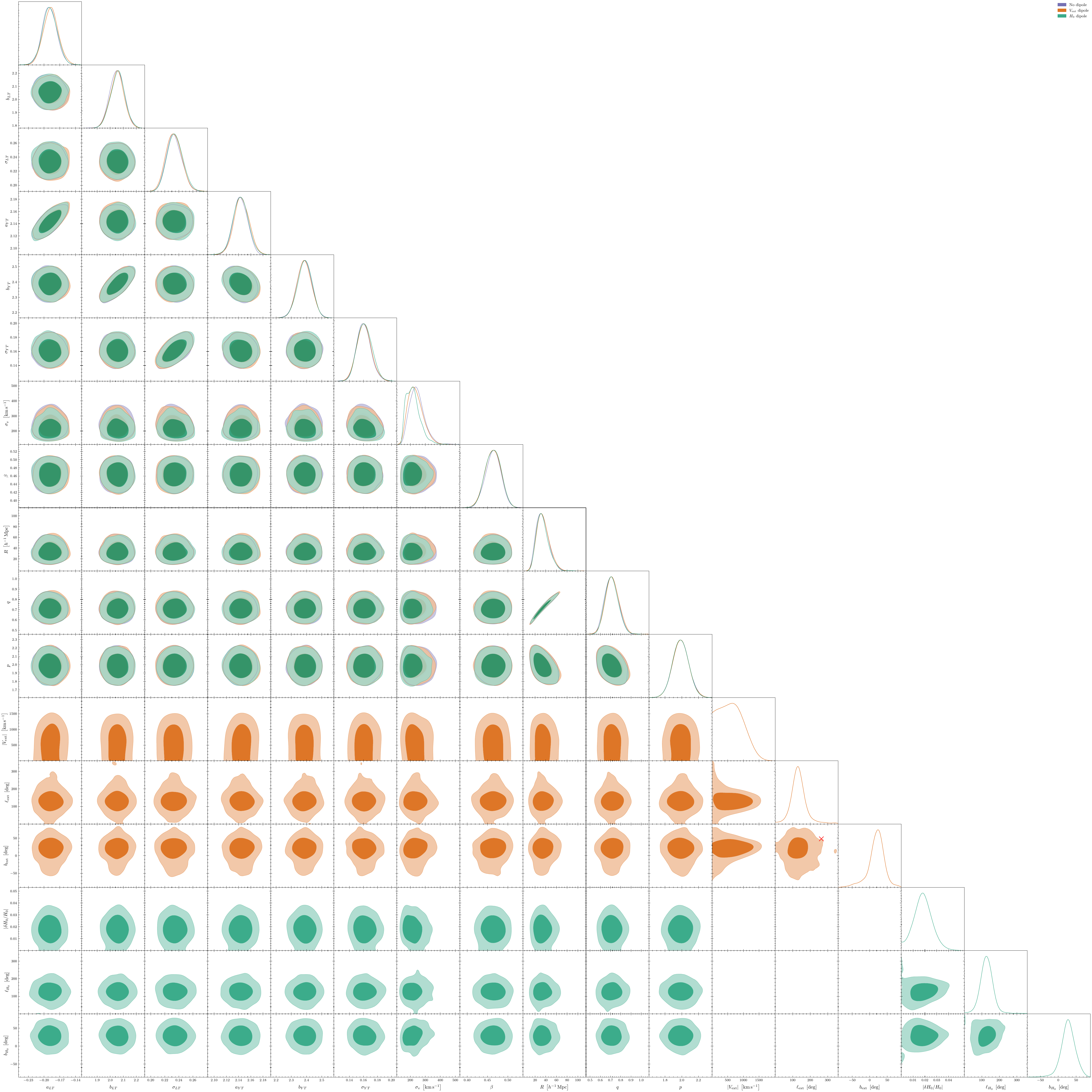}
  \caption{Posterior distributions for the model parameters for the base no flow/$H_0$ variation model, and the dipole flow and $H_0$ variation for the \Manti\ reconstruction. }
  \label{fig:full_posterior}
\end{figure*}

\begin{figure*}
 \centering
 \begin{subfigure}[t]{0.48\textwidth}
  \centering
  \includegraphics[width=\linewidth]{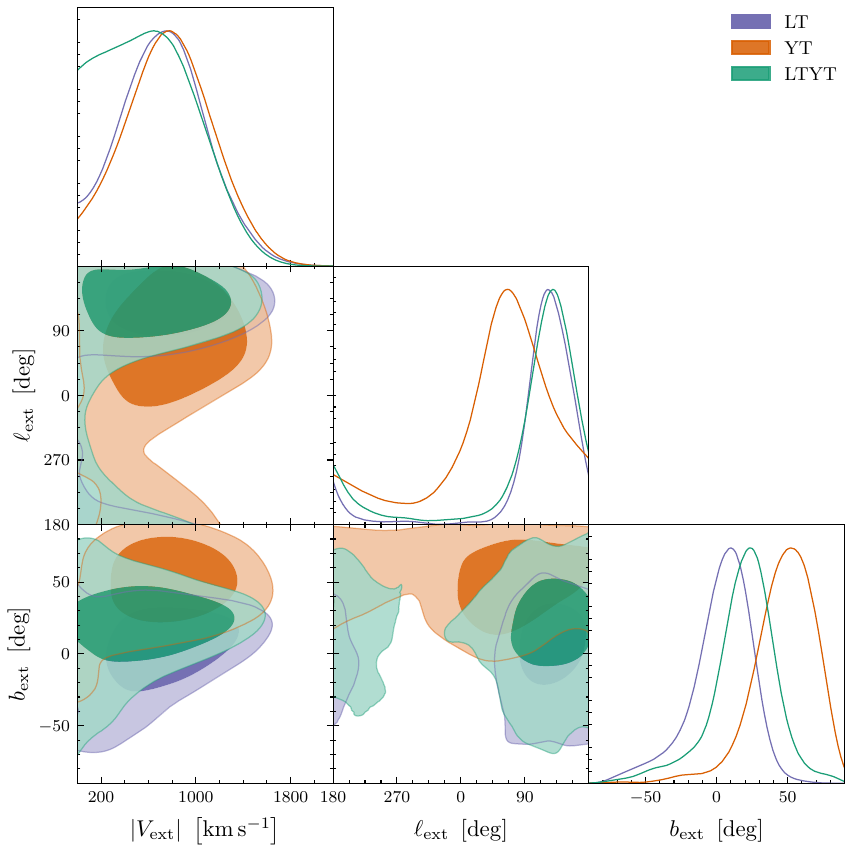}
 \end{subfigure}

 \vspace{0.6em}
 \begin{subfigure}[t]{0.48\textwidth}
  \centering
  \includegraphics[width=\linewidth]{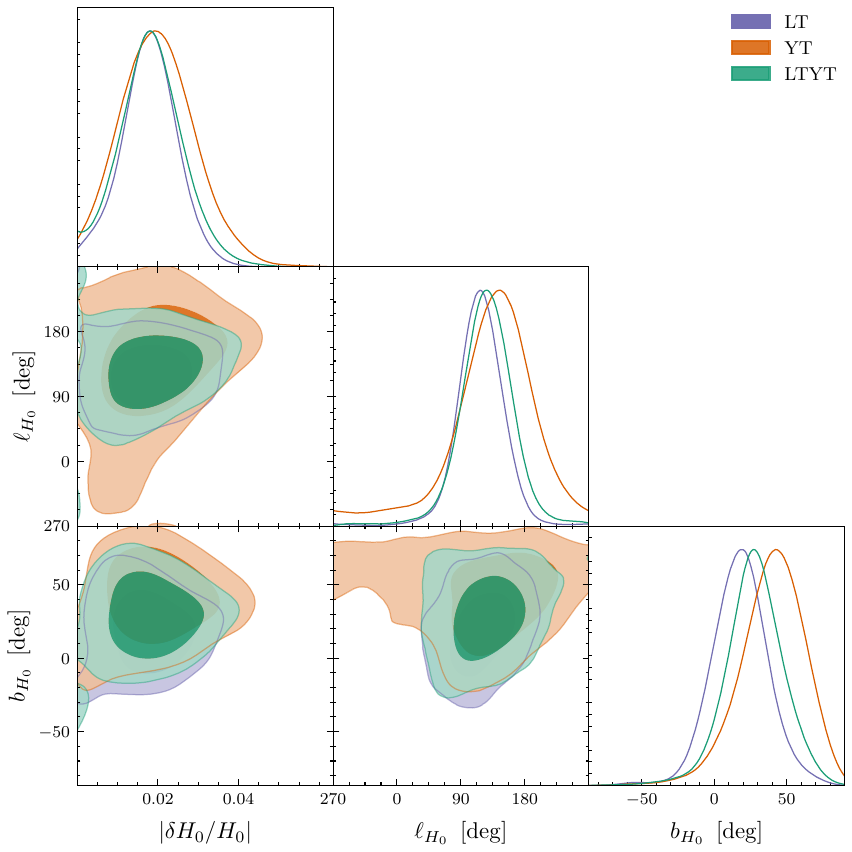}
 \end{subfigure}\hfill
 \begin{subfigure}[t]{0.48\textwidth}
  \centering
  \includegraphics[width=\linewidth]{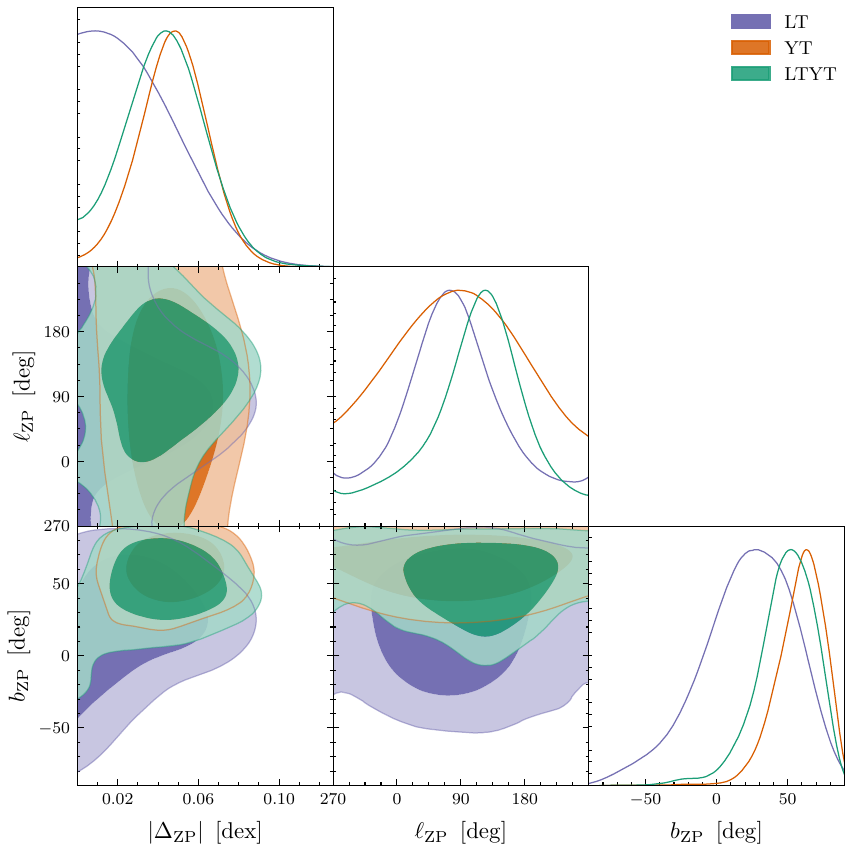}
 \end{subfigure}
 \caption{Constraints on the dipolar bulk flow (top), $H_0$ anisotropy (bottom left), and ZP dipole (bottom right) from the \LT\ relation (purple), the \YT\ relation (orange), and both relations jointly (green), using the~\citetalias{carrickCosmologicalParametersComparison2015} reconstruction. The constraints are consistent between relations, with the dipole signal driven primarily by \YT\ due to its lower intrinsic scatter.}
 \label{fig:relation_comparison}
\end{figure*}

This appendix presents the full posterior distributions for the model parameters in \Cref{fig:full_posterior} and the joint constraints on scaling relation parameters for the base model and dipole variants in \Cref{fig:relation_comparison}.

\section{Comprehensive tables}\label{app:tables}

This appendix provides detailed parameter constraints for all model variants tested in this work. In most tables we report results for each scaling relation (\LT, \YT, and \LTYT) and each velocity field reconstruction (Carrick2015, \Manti, and no velocity field). The tables are for dipole models (\cref{tab:dipole_results_full}), radially-varying bulk flow models (\cref{tab:radial_params,tab:radial_free_dir_params}), pixelised anisotropy models (\cref{tab:pixel_params}), quadrupole models (\cref{tab:quad_params}), and mixed dipole models (\cref{tab:mixed_dipoles}).

\begin{table*}
\centering
\caption{Dipole model constraints for different scaling relations and reconstructions. $\Delta\ln\mathcal{Z}$ is relative to the base model for each reconstruction. Upper limits (95\%) are reported when the posterior is consistent with zero. \red{The ``Preference'' column indicates the Jeffreys scale interpretation of the evidence (see Table~\ref{tab:dipole_results} caption for scale).}}
\begin{tabular}{|c|c|l|c|c|c|c|c|}
\hline\hline
Relation & Velocity field & Model & Amplitude & $\ell$ [$^\circ$] & $b$ [$^\circ$] & $\Delta\ln\mathcal{Z}$ & Preference \\
\hline\hline
\multirow{9}{*}{LT} & \multirow{3}{*}{C15} & $\mathbf{V}_{\rm ext}$ & $731^{+347}_{-360}$\,km/s & $130 \pm 36$ & $3 \pm 22$ & $0.07 \pm 0.12$ & Negligible \\
 &  & $H_0$ & $1.8^{+0.7}_{-0.7}$\% & $120 \pm 35$ & $18 \pm 20$ & $0.49 \pm 0.06$ & Negligible \\
 &  & ZP & $< 7.3$\% & $130 \pm 96$ & $21 \pm 33$ & $-1.35 \pm 0.07$ & Disfavoured \\
\cline{2-8}
 & \multirow{3}{*}{Manticore} & $\mathbf{V}_{\rm ext}$ & $730^{+398}_{-379}$\,km/s & $132 \pm 42$ & $3 \pm 22$ & $-0.31 \pm 0.15$ & Disfavoured \\
 &  & $H_0$ & $< 1.8$\% & $173 \pm 84$ & $20 \pm 32$ & $-2.35 \pm 0.05$ & Disfavoured \\
 &  & ZP & $< 7.2$\% & $132 \pm 98$ & $21 \pm 33$ & $-1.40 \pm 0.07$ & Disfavoured \\
\cline{2-8}
 & \multirow{3}{*}{None} & $\mathbf{V}_{\rm ext}$ & $< 1465$\,km/s & $122 \pm 89$ & $22 \pm 30$ & $-1.38 \pm 0.05$ & Disfavoured \\
 &  & $H_0$ & $4.3^{+2.9}_{-2.6}$\% & $141 \pm 115$ & $35 \pm 29$ & $-0.36 \pm 0.07$ & Disfavoured \\
 &  & ZP & $3.5^{+2.5}_{-2.2}$\% & $138 \pm 110$ & $33 \pm 29$ & $-0.77 \pm 0.05$ & Disfavoured \\
\hline\hline
\multirow{9}{*}{YT} & \multirow{3}{*}{C15} & $\mathbf{V}_{\rm ext}$ & $779^{+366}_{-372}$\,km/s & $119 \pm 89$ & $47 \pm 22$ & $-0.12 \pm 0.05$ & Disfavoured \\
 &  & $H_0$ & $2.0^{+0.9}_{-0.9}$\% & $149 \pm 61$ & $40 \pm 22$ & $0.12 \pm 0.05$ & Negligible \\
 &  & ZP & $4.8^{+1.6}_{-1.6}$\% & $157 \pm 102$ & $59 \pm 17$ & $1.03 \pm 0.05$ & Weak \\
\cline{2-8}
 & \multirow{3}{*}{Manticore} & $\mathbf{V}_{\rm ext}$ & $716^{+394}_{-409}$\,km/s & $117 \pm 83$ & $43 \pm 24$ & $-0.37 \pm 0.06$ & Disfavoured \\
 &  & $H_0$ & $< 2.3$\% & $182 \pm 89$ & $35 \pm 30$ & $-1.71 \pm 0.05$ & Disfavoured \\
 &  & ZP & $4.5^{+1.8}_{-1.7}$\% & $161 \pm 104$ & $57 \pm 18$ & $0.70 \pm 0.04$ & Negligible \\
\cline{2-8}
 & \multirow{3}{*}{None} & $\mathbf{V}_{\rm ext}$ & $936^{+373}_{-364}$\,km/s & $115 \pm 85$ & $50 \pm 20$ & $0.40 \pm 0.04$ & Negligible \\
 &  & $H_0$ & $5.7^{+2.0}_{-2.1}$\% & $152 \pm 99$ & $57 \pm 18$ & $1.69 \pm 0.03$ & Weak \\
 &  & ZP & $4.8^{+1.8}_{-1.8}$\% & $152 \pm 98$ & $57 \pm 18$ & $1.17 \pm 0.05$ & Weak \\
\hline\hline
\multirow{9}{*}{LTYT} & \multirow{3}{*}{C15} & $\mathbf{V}_{\rm ext}$ & $< 1276$\,km/s & $135 \pm 48$ & $19 \pm 24$ & $-0.93 \pm 0.09$ & Disfavoured \\
 &  & $H_0$ & $1.8^{+0.8}_{-0.8}$\% & $129 \pm 41$ & $28 \pm 20$ & $0.01 \pm 0.08$ & Negligible \\
 &  & ZP & $4.4^{+1.9}_{-2.0}$\% & $143 \pm 80$ & $49 \pm 21$ & $0.26 \pm 0.07$ & Negligible \\
\cline{2-8}
 & \multirow{3}{*}{Manticore} & $\mathbf{V}_{\rm ext}$ & $628^{+376}_{-369}$\,km/s & $138 \pm 44$ & $17 \pm 22$ & $-1.02 \pm 0.07$ & Disfavoured \\
 &  & $H_0$ & $< 2.1$\% & $166 \pm 78$ & $27 \pm 31$ & $-1.93 \pm 0.09$ & Disfavoured \\
 &  & ZP & $4.2^{+1.9}_{-2.0}$\% & $138 \pm 78$ & $46 \pm 23$ & $0.04 \pm 0.07$ & Negligible \\
\cline{2-8}
 & \multirow{3}{*}{None} & $\mathbf{V}_{\rm ext}$ & $785^{+440}_{-440}$\,km/s & $124 \pm 83$ & $41 \pm 26$ & $-0.55 \pm 0.07$ & Disfavoured \\
 &  & $H_0$ & $5.6^{+2.2}_{-2.2}$\% & $144 \pm 81$ & $54 \pm 18$ & $1.34 \pm 0.07$ & Weak \\
 &  & ZP & $4.7^{+1.6}_{-1.7}$\% & $147 \pm 83$ & $54 \pm 18$ & $0.90 \pm 0.08$ & Negligible \\
\hline
\end{tabular}
\label{tab:dipole_results_full}
\end{table*}

\begin{table*}
\centering
\caption{Radially varying $\Vext$ model with fixed direction for all scaling relations and reconstructions. Four radial knots are used, with amplitudes $V_0$--$V_3$ corresponding to increasing distance. A common direction $(\ell, b)$ is fit across all radial bins. The large uncertainties on individual knot amplitudes indicate the data do not strongly constrain radial variations in the bulk flow. All $\Delta\ln\mathcal{Z}$ values are negative, indicating no preference over the simpler dipole model.}
\begin{tabular}{|ll|cccc|cc|c|}
\hline\hline
Rel. & Recon & $V_0$ & $V_1$ & $V_2$ & $V_3$ & $\ell$ & $b$ & $\Delta\ln\mathcal{Z}$ \\
\hline\hline
LT & C15 & $1135^{+504}_{-541}$ & $< 1555$ & $2284^{+1697}_{-1517}$ & $2054^{+1847}_{-1473}$ & $147 \pm 35$ & $13 \pm 18$ & $-0.72 \pm 0.08$ \\
 & Manticore & $1227^{+552}_{-600}$ & $< 1536$ & $2257^{+1653}_{-1489}$ & $2009^{+1827}_{-1469}$ & $148 \pm 33$ & $13 \pm 17$ & $-0.88 \pm 0.07$ \\
 & No velocity field & $< 1620$ & $< 1984$ & $3278^{+1175}_{-1772}$ & $2449^{+1660}_{-1744}$ & $181 \pm 110$ & $45 \pm 23$ & $-1.44 \pm 0.05$ \\
\hline
YT & C15 & $< 1137$ & $1380^{+587}_{-580}$ & $3357^{+1121}_{-1492}$ & $2393^{+1615}_{-1677}$ & $162 \pm 95$ & $60 \pm 15$ & $0.73 \pm 0.04$ \\
 & Manticore & $< 1213$ & $1320^{+548}_{-564}$ & $3287^{+1137}_{-1663}$ & $2358^{+1628}_{-1626}$ & $151 \pm 93$ & $58 \pm 16$ & $0.27 \pm 0.05$ \\
 & No velocity field & $< 1491$ & $1388^{+571}_{-574}$ & $3404^{+1049}_{-1492}$ & $2406^{+1673}_{-1722}$ & $151 \pm 93$ & $61 \pm 15$ & $0.96 \pm 0.06$ \\
\hline
LTYT & C15 & $< 1685$ & $< 1939$ & $2839^{+1356}_{-1450}$ & $2141^{+1772}_{-1432}$ & $149 \pm 61$ & $34 \pm 22$ & $-1.30 \pm 0.12$ \\
 & Manticore & $< 1814$ & $< 1856$ & $2866^{+1354}_{-1576}$ & $2206^{+1912}_{-1592}$ & $147 \pm 50$ & $29 \pm 20$ & $-1.30 \pm 0.09$ \\
 & No velocity field & $< 1650$ & $1111^{+655}_{-627}$ & $3263^{+1103}_{-1448}$ & $2517^{+1612}_{-1671}$ & $158 \pm 96$ & $55 \pm 17$ & $-0.28 \pm 0.07$ \\
\hline
\end{tabular}
\label{tab:radial_params}
\end{table*}

\begin{table*}
\centering
\caption{Radially varying $\Vext$ model parameters with free direction per bin for the joint \LTYT\ relation. Unlike \cref{tab:radial_params}, each radial bin has an independent direction $(\ell_i, b_i)$ in addition to amplitude $V_i$. The \Manti\ reconstruction shows weak positive evidence ($\Delta\ln\mathcal{Z} = 1.25$), though this is sensitive to the binning choice (see text).}
\begin{tabular}{|ll|cccc|c|}
\hline\hline
Rel. & Recon & Bin 0 & Bin 1 & Bin 2 & Bin 3 & $\Delta\ln\mathcal{Z}$ \\
\hline\hline
LTYT & C15 & $1064^{+529}_{-530}$ & $1023^{+664}_{-652}$ & $3085^{+1314}_{-1596}$ & $2470^{+1772}_{-1698}$ & $0.36 \pm 0.66$ \\
 & & ($160 \pm 33$, $-1 \pm 19$) & ($139 \pm 134$, $38 \pm 22$) & ($213 \pm 84$, $41 \pm 27$) & ($206 \pm 107$, $6 \pm 37$) & \\
\hline
 & Manticore & $1129^{+535}_{-523}$ & $1008^{+687}_{-633}$ & $3198^{+1224}_{-1618}$ & $2576^{+1730}_{-1743}$ & $1.25 \pm 0.10$ \\
 & & ($161 \pm 28$, $-1 \pm 17$) & ($141 \pm 136$, $39 \pm 21$) & ($218 \pm 83$, $43 \pm 25$) & ($204 \pm 110$, $7 \pm 38$) & \\
\hline
 & No velocity field & $< 1739$ & $1345^{+625}_{-705}$ & $3331^{+1165}_{-1581}$ & $2707^{+1606}_{-1776}$ & $0.20 \pm 0.07$ \\
 & & ($162 \pm 101$, $13 \pm 34$) & ($110 \pm 91$, $36 \pm 22$) & ($221 \pm 83$, $47 \pm 23$) & ($216 \pm 108$, $10 \pm 36$) & \\
\hline
\end{tabular}
\label{tab:radial_free_dir_params}
\end{table*}

\begin{table*}
\centering
\caption{Pixelised model parameters for all scaling relations and reconstructions. We use 12 HEALPix pixels ($N_{\rm side}=1$); $p_i$ denotes the value in each pixel. For the $\Vext$ model, values are in $\kmsec$. For the ZP and $H_0$ models, values are expressed as fractional variations in percent, equivalent to $\delta H_0/H_0$ (the ZP magnitude is converted to this form via \cref{eq:Deltaa_DeltaH0_linear} for direct comparison). With reconstructions, individual pixel values are typically consistent with zero within their $\sim 3$--$5\%$ uncertainties, though some coherent structure may be present.}
\label{tab:pixel_params}
\makebox[\textwidth][c]{\resizebox{1.15\textwidth}{!}{\begin{tabular}{|c|c|l|cccccccccccc|c|}
\hline\hline
Rel. & Recon & Model & $p_{0}$ & $p_{1}$ & $p_{2}$ & $p_{3}$ & $p_{4}$ & $p_{5}$ & $p_{6}$ & $p_{7}$ & $p_{8}$ & $p_{9}$ & $p_{10}$ & $p_{11}$ & $\Delta\ln\mathcal{Z}$ \\
\hline\hline
\multirow{9}{*}{LT} & \multirow{3}{*}{C15} & $\mathbf{V}_{\rm ext}$ & $-466 \pm 574$ & $746 \pm 891$ & $-463 \pm 533$ & $1268 \pm 1338$ & $-751 \pm 632$ & $170 \pm 635$ & $-958 \pm 502$ & $4308 \pm 2600$ & $-1718 \pm 962$ & $-68 \pm 830$ & $-1365 \pm 637$ & $-703 \pm 603$ & $1.58 \pm 0.11$ \\
 &  & ZP & $4.0 \pm 5.0$\% & $0.0 \pm 5.0$\% & $4.0 \pm 4.0$\% & $-2.0 \pm 5.0$\% & $-1.0 \pm 4.0$\% & $1.0 \pm 5.0$\% & $1.0 \pm 4.0$\% & $2.0 \pm 5.0$\% & $-6.0 \pm 3.0$\% & $-1.0 \pm 5.0$\% & $1.0 \pm 5.0$\% & $1.0 \pm 4.0$\% & $-1.67 \pm 0.08$ \\
 &  & $H_0$ & $6.0 \pm 3.0$\% & $2.0 \pm 4.0$\% & $1.0 \pm 1.0$\% & $-2.0 \pm 5.0$\% & $-2.0 \pm 1.0$\% & $1.0 \pm 3.0$\% & $-0.0 \pm 2.0$\% & $2.0 \pm 5.0$\% & $-2.0 \pm 1.0$\% & $-1.0 \pm 3.0$\% & $-1.0 \pm 1.0$\% & $-1.0 \pm 2.0$\% & $-7.80 \pm 0.11$ \\
\cline{2-16}
 & \multirow{3}{*}{Manticore} & $\mathbf{V}_{\rm ext}$ & $-4031 \pm 1029$ & $-5512 \pm 1679$ & $-4946 \pm 1007$ & $-7217 \pm 1919$ & $-6017 \pm 1071$ & $-4681 \pm 1284$ & $-4837 \pm 953$ & $-326 \pm 2973$ & $-8295 \pm 1101$ & $-5123 \pm 1657$ & $-5377 \pm 1058$ & $-5673 \pm 1156$ & $15.08 \pm 0.09$ \\
 &  & ZP & $4.0 \pm 5.0$\% & $0.0 \pm 5.0$\% & $3.0 \pm 4.0$\% & $-2.0 \pm 5.0$\% & $-1.0 \pm 4.0$\% & $1.0 \pm 5.0$\% & $1.0 \pm 4.0$\% & $2.0 \pm 5.0$\% & $-6.0 \pm 3.0$\% & $-1.0 \pm 5.0$\% & $1.0 \pm 5.0$\% & $1.0 \pm 4.0$\% & $-1.59 \pm 0.09$ \\
 &  & $H_0$ & $6.0 \pm 3.0$\% & $2.0 \pm 5.0$\% & $0.0 \pm 1.0$\% & $-2.0 \pm 4.0$\% & $-2.0 \pm 1.0$\% & $1.0 \pm 3.0$\% & $-1.0 \pm 1.0$\% & $2.0 \pm 5.0$\% & $-1.0 \pm 1.0$\% & $-0.0 \pm 3.0$\% & $0.0 \pm 2.0$\% & $-1.0 \pm 1.0$\% & $-12.13 \pm 0.06$ \\
\cline{2-16}
 & \multirow{3}{*}{No velocity field} & $\mathbf{V}_{\rm ext}$ & $74 \pm 576$ & $294 \pm 695$ & $106 \pm 447$ & $126 \pm 804$ & $-418 \pm 567$ & $-60 \pm 610$ & $-219 \pm 525$ & $1097 \pm 1463$ & $-952 \pm 926$ & $70 \pm 699$ & $-231 \pm 553$ & $113 \pm 530$ & $-1.24 \pm 0.07$ \\
 &  & ZP & $4.0 \pm 4.0$\% & $0.0 \pm 5.0$\% & $3.0 \pm 4.0$\% & $-3.0 \pm 5.0$\% & $-2.0 \pm 4.0$\% & $-1.0 \pm 5.0$\% & $3.0 \pm 4.0$\% & $3.0 \pm 5.0$\% & $-7.0 \pm 3.0$\% & $-0.0 \pm 5.0$\% & $3.0 \pm 4.0$\% & $0.0 \pm 4.0$\% & $-1.01 \pm 0.12$ \\
 &  & $H_0$ & $4.0 \pm 5.0$\% & $0.0 \pm 5.0$\% & $3.0 \pm 4.0$\% & $-2.0 \pm 5.0$\% & $-2.0 \pm 4.0$\% & $-0.0 \pm 5.0$\% & $3.0 \pm 4.0$\% & $3.0 \pm 5.0$\% & $-7.0 \pm 3.0$\% & $-0.0 \pm 5.0$\% & $3.0 \pm 5.0$\% & $0.0 \pm 4.0$\% & $-0.69 \pm 0.03$ \\
\hline\hline
\multirow{9}{*}{YT} & \multirow{3}{*}{C15} & $\mathbf{V}_{\rm ext}$ & $162 \pm 546$ & $524 \pm 747$ & $373 \pm 447$ & $-77 \pm 956$ & $-634 \pm 517$ & $-212 \pm 646$ & $544 \pm 480$ & $1132 \pm 1259$ & $-1611 \pm 872$ & $-434 \pm 681$ & $-590 \pm 665$ & $824 \pm 590$ & $1.34 \pm 0.05$ \\
 &  & ZP & $4.0 \pm 4.0$\% & $2.0 \pm 4.0$\% & $6.0 \pm 3.0$\% & $-4.0 \pm 4.0$\% & $-3.0 \pm 3.0$\% & $-0.0 \pm 5.0$\% & $6.0 \pm 3.0$\% & $0.0 \pm 5.0$\% & $-7.0 \pm 3.0$\% & $-2.0 \pm 4.0$\% & $1.0 \pm 4.0$\% & $2.0 \pm 4.0$\% & $2.21 \pm 0.11$ \\
 &  & $H_0$ & $5.0 \pm 3.0$\% & $3.0 \pm 4.0$\% & $2.0 \pm 2.0$\% & $-4.0 \pm 5.0$\% & $-2.0 \pm 1.0$\% & $0.0 \pm 3.0$\% & $4.0 \pm 3.0$\% & $-0.0 \pm 5.0$\% & $-2.0 \pm 1.0$\% & $-2.0 \pm 3.0$\% & $-1.0 \pm 2.0$\% & $-1.0 \pm 2.0$\% & $-6.15 \pm 0.14$ \\
\cline{2-16}
 & \multirow{3}{*}{Manticore} & $\mathbf{V}_{\rm ext}$ & $339 \pm 552$ & $471 \pm 707$ & $314 \pm 418$ & $-149 \pm 900$ & $-648 \pm 506$ & $-263 \pm 638$ & $528 \pm 490$ & $1097 \pm 1213$ & $-1633 \pm 823$ & $-453 \pm 652$ & $-452 \pm 551$ & $851 \pm 591$ & $1.27 \pm 0.13$ \\
 &  & ZP & $4.0 \pm 4.0$\% & $2.0 \pm 4.0$\% & $6.0 \pm 3.0$\% & $-4.0 \pm 4.0$\% & $-3.0 \pm 3.0$\% & $-0.0 \pm 5.0$\% & $6.0 \pm 3.0$\% & $0.0 \pm 5.0$\% & $-7.0 \pm 3.0$\% & $-2.0 \pm 5.0$\% & $0.0 \pm 4.0$\% & $2.0 \pm 4.0$\% & $1.89 \pm 0.09$ \\
 &  & $H_0$ & $3.0 \pm 4.0$\% & $2.0 \pm 4.0$\% & $1.0 \pm 1.0$\% & $-3.0 \pm 4.0$\% & $-2.0 \pm 1.0$\% & $-0.0 \pm 3.0$\% & $1.0 \pm 2.0$\% & $-1.0 \pm 5.0$\% & $-1.0 \pm 1.0$\% & $-1.0 \pm 3.0$\% & $1.0 \pm 2.0$\% & $-1.0 \pm 1.0$\% & $-11.26 \pm 0.10$ \\
\cline{2-16}
 & \multirow{3}{*}{No velocity field} & $\mathbf{V}_{\rm ext}$ & $282 \pm 574$ & $490 \pm 753$ & $444 \pm 468$ & $-124 \pm 946$ & $-705 \pm 523$ & $-347 \pm 698$ & $813 \pm 588$ & $1108 \pm 1256$ & $-1677 \pm 834$ & $-652 \pm 747$ & $-381 \pm 574$ & $749 \pm 598$ & $1.61 \pm 0.04$ \\
 &  & ZP & $4.0 \pm 4.0$\% & $2.0 \pm 4.0$\% & $6.0 \pm 2.0$\% & $-4.0 \pm 4.0$\% & $-3.0 \pm 3.0$\% & $-1.0 \pm 5.0$\% & $6.0 \pm 3.0$\% & $0.0 \pm 5.0$\% & $-7.0 \pm 3.0$\% & $-3.0 \pm 5.0$\% & $0.0 \pm 4.0$\% & $2.0 \pm 3.0$\% & $2.20 \pm 0.09$ \\
 &  & $H_0$ & $4.0 \pm 5.0$\% & $2.0 \pm 5.0$\% & $6.0 \pm 3.0$\% & $-4.0 \pm 5.0$\% & $-4.0 \pm 3.0$\% & $-1.0 \pm 5.0$\% & $6.0 \pm 3.0$\% & $-0.0 \pm 5.0$\% & $-7.0 \pm 3.0$\% & $-3.0 \pm 5.0$\% & $-0.0 \pm 5.0$\% & $2.0 \pm 4.0$\% & $2.67 \pm 0.04$ \\
\hline\hline
\multirow{9}{*}{LTYT} & \multirow{3}{*}{C15} & $\mathbf{V}_{\rm ext}$ & $-568 \pm 524$ & $966 \pm 812$ & $-256 \pm 443$ & $1082 \pm 1233$ & $-1031 \pm 582$ & $66 \pm 624$ & $-726 \pm 439$ & $3960 \pm 2108$ & $-1831 \pm 848$ & $85 \pm 735$ & $-1479 \pm 574$ & $-269 \pm 503$ & $2.08 \pm 0.19$ \\
 &  & ZP & $3.0 \pm 5.0$\% & $3.0 \pm 4.0$\% & $7.0 \pm 2.0$\% & $-4.0 \pm 5.0$\% & $-3.0 \pm 3.0$\% & $1.0 \pm 5.0$\% & $4.0 \pm 3.0$\% & $0.0 \pm 5.0$\% & $-7.0 \pm 3.0$\% & $-1.0 \pm 5.0$\% & $-2.0 \pm 4.0$\% & $2.0 \pm 4.0$\% & $1.30 \pm 0.14$ \\
 &  & $H_0$ & $6.0 \pm 3.0$\% & $3.0 \pm 3.0$\% & $2.0 \pm 2.0$\% & $-4.0 \pm 5.0$\% & $-2.0 \pm 1.0$\% & $1.0 \pm 3.0$\% & $0.0 \pm 2.0$\% & $0.0 \pm 5.0$\% & $-2.0 \pm 1.0$\% & $-1.0 \pm 3.0$\% & $-1.0 \pm 1.0$\% & $-0.0 \pm 2.0$\% & $-7.75 \pm 0.15$ \\
\cline{2-16}
 & \multirow{3}{*}{Manticore} & $\mathbf{V}_{\rm ext}$ & $-247 \pm 507$ & $773 \pm 782$ & $-199 \pm 428$ & $700 \pm 1108$ & $-846 \pm 600$ & $80 \pm 584$ & $-528 \pm 420$ & $2916 \pm 2206$ & $-1516 \pm 924$ & $56 \pm 670$ & $-1040 \pm 633$ & $-150 \pm 477$ & $0.44 \pm 0.20$ \\
 &  & ZP & $3.0 \pm 4.0$\% & $3.0 \pm 4.0$\% & $6.0 \pm 3.0$\% & $-4.0 \pm 5.0$\% & $-3.0 \pm 3.0$\% & $0.0 \pm 5.0$\% & $4.0 \pm 4.0$\% & $0.0 \pm 5.0$\% & $-7.0 \pm 3.0$\% & $-0.0 \pm 5.0$\% & $-2.0 \pm 4.0$\% & $3.0 \pm 4.0$\% & $1.23 \pm 0.09$ \\
 &  & $H_0$ & $5.0 \pm 3.0$\% & $3.0 \pm 4.0$\% & $1.0 \pm 1.0$\% & $-3.0 \pm 4.0$\% & $-2.0 \pm 1.0$\% & $1.0 \pm 3.0$\% & $-0.0 \pm 1.0$\% & $-0.0 \pm 5.0$\% & $-1.0 \pm 1.0$\% & $-0.0 \pm 2.0$\% & $-0.0 \pm 1.0$\% & $-1.0 \pm 1.0$\% & $-11.55 \pm 0.09$ \\
\cline{2-16}
 & \multirow{3}{*}{No velocity field} & $\mathbf{V}_{\rm ext}$ & $-9 \pm 595$ & $566 \pm 785$ & $299 \pm 506$ & $152 \pm 923$ & $-739 \pm 604$ & $-115 \pm 638$ & $-27 \pm 509$ & $1295 \pm 1435$ & $-1318 \pm 937$ & $31 \pm 760$ & $-363 \pm 566$ & $227 \pm 542$ & $-0.23 \pm 0.07$ \\
 &  & ZP & $4.0 \pm 4.0$\% & $3.0 \pm 4.0$\% & $7.0 \pm 2.0$\% & $-4.0 \pm 4.0$\% & $-4.0 \pm 3.0$\% & $0.0 \pm 5.0$\% & $4.0 \pm 3.0$\% & $1.0 \pm 5.0$\% & $-7.0 \pm 3.0$\% & $0.0 \pm 5.0$\% & $-1.0 \pm 4.0$\% & $1.0 \pm 4.0$\% & $1.78 \pm 0.08$ \\
 &  & $H_0$ & $3.0 \pm 5.0$\% & $2.0 \pm 4.0$\% & $7.0 \pm 2.0$\% & $-4.0 \pm 5.0$\% & $-4.0 \pm 3.0$\% & $-0.0 \pm 5.0$\% & $4.0 \pm 4.0$\% & $0.0 \pm 5.0$\% & $-7.0 \pm 3.0$\% & $-0.0 \pm 5.0$\% & $-1.0 \pm 5.0$\% & $1.0 \pm 4.0$\% & $2.08 \pm 0.08$ \\
\hline\hline
\end{tabular}}}
\end{table*}

\begin{table*}
\centering
\caption{Quadrupole model parameters for all scaling relations and reconstructions. For each model, we report the dipole amplitude $A_{\rm dip}$, dipole direction $(\ell, b)$, quadrupole amplitude $A_{\rm quad}$, and the two quadrupole axis directions $(\ell_1, b_1)$ and $(\ell_2, b_2)$. Amplitudes are in $\kmsec$ for $\Vext$ models and in percent ($\equiv \delta H_0/H_0$) for ZP models. The $H_0$ quadrupole model uses the same parameterisation as ZP. All quadrupole amplitudes are consistent with zero and the models are disfavoured by the Bayesian evidence.}
\scriptsize
\begin{tabular}{|c|c|c|c|cc|c|cccc|c|}
\hline\hline
Rel. & Recon & Model & $A_{\rm dip}$ & $\ell$ [$^\circ$] & $b$ [$^\circ$] & $A_{\rm quad}$ & $\ell_1$ [$^\circ$] & $b_1$ [$^\circ$] & $\ell_2$ [$^\circ$] & $b_2$ [$^\circ$] & $\Delta\ln\mathcal{Z}$ \\
\hline\hline
\multirow{9}{*}{LT} & \multirow{3}{*}{C15} & $\mathbf{V}_{\rm ext}$ & $722^{+393}_{-365}$ & $130 \pm 38$ & $4 \pm 24$ & $< 489$ & $181 \pm 104$ & $0 \pm 37$ & $181 \pm 105$ & $-1 \pm 38$ & $-2.54 \pm 0.10$ \\
 &  & ZP & $< 7.0$ & $140 \pm 99$ & $20 \pm 34$ & $9.2^{+8.5}_{-6.3}$ & $192 \pm 98$ & $-1 \pm 30$ & $180 \pm 97$ & $-0 \pm 30$ & $-2.11 \pm 0.11$ \\
 &  & $H_0$ & $1.8^{+0.8}_{-0.8}$ & $125 \pm 44$ & $21 \pm 22$ & $< 4.5$ & $184 \pm 102$ & $0 \pm 36$ & $177 \pm 103$ & $-0 \pm 37$ & $-1.43 \pm 0.24$ \\
\cline{2-12}
 & \multirow{3}{*}{Manticore} & $\mathbf{V}_{\rm ext}$ & $741^{+375}_{-427}$ & $132 \pm 39$ & $3 \pm 23$ & $< 536$ & $185 \pm 101$ & $-0 \pm 37$ & $185 \pm 101$ & $-0 \pm 37$ & $-2.97 \pm 0.06$ \\
 &  & ZP & $< 6.6$ & $139 \pm 99$ & $20 \pm 34$ & $8.9^{+9.1}_{-6.3}$ & $188 \pm 102$ & $-1 \pm 30$ & $181 \pm 101$ & $1 \pm 30$ & $-2.16 \pm 0.11$ \\
 &  & $H_0$ & $< 1.9$ & $169 \pm 85$ & $23 \pm 33$ & $< 4.3$ & $182 \pm 103$ & $1 \pm 39$ & $178 \pm 103$ & $-1 \pm 39$ & $-4.01 \pm 0.15$ \\
\cline{2-12}
 & \multirow{3}{*}{No velocity field} & $\mathbf{V}_{\rm ext}$ & $< 1483$ & $125 \pm 85$ & $28 \pm 30$ & $2298^{+1371}_{-1343}$ & $180 \pm 109$ & $-1 \pm 30$ & $182 \pm 109$ & $-0 \pm 29$ & $-0.80 \pm 0.11$ \\
 &  & ZP & $< 7.2$ & $147 \pm 114$ & $32 \pm 32$ & $8.5^{+8.2}_{-5.8}$ & $181 \pm 101$ & $1 \pm 31$ & $184 \pm 100$ & $-1 \pm 32$ & $-1.66 \pm 0.18$ \\
 &  & $H_0$ & $4.1^{+2.9}_{-2.8}$ & $141 \pm 112$ & $34 \pm 29$ & $7.0^{+5.4}_{-4.7}$ & $183 \pm 105$ & $0 \pm 34$ & $175 \pm 104$ & $-0 \pm 34$ & $-0.43 \pm 0.12$ \\
\hline\hline
\multirow{9}{*}{YT} & \multirow{3}{*}{C15} & $\mathbf{V}_{\rm ext}$ & $797^{+362}_{-384}$ & $120 \pm 87$ & $48 \pm 22$ & $< 761$ & $181 \pm 104$ & $1 \pm 38$ & $180 \pm 103$ & $1 \pm 38$ & $-2.43 \pm 0.14$ \\
 &  & ZP & $4.7^{+1.8}_{-1.8}$ & $163 \pm 101$ & $56 \pm 17$ & $< 11.8$ & $183 \pm 102$ & $-2 \pm 37$ & $180 \pm 102$ & $-0 \pm 37$ & $-0.60 \pm 0.08$ \\
 &  & $H_0$ & $2.1^{+1.0}_{-0.9}$ & $151 \pm 59$ & $40 \pm 22$ & $< 5.2$ & $181 \pm 104$ & $-0 \pm 38$ & $180 \pm 103$ & $2 \pm 38$ & $-1.26 \pm 0.14$ \\
\cline{2-12}
 & \multirow{3}{*}{Manticore} & $\mathbf{V}_{\rm ext}$ & $742^{+373}_{-389}$ & $117 \pm 84$ & $44 \pm 25$ & $< 800$ & $185 \pm 100$ & $-1 \pm 36$ & $186 \pm 102$ & $-0 \pm 36$ & $-2.51 \pm 0.19$ \\
 &  & ZP & $4.6^{+1.8}_{-1.8}$ & $159 \pm 99$ & $54 \pm 19$ & $< 12.0$ & $174 \pm 106$ & $-2 \pm 38$ & $179 \pm 103$ & $-2 \pm 37$ & $-1.14 \pm 0.05$ \\
 &  & $H_0$ & $< 2.6$ & $186 \pm 89$ & $38 \pm 29$ & $< 4.3$ & $181 \pm 105$ & $1 \pm 39$ & $178 \pm 103$ & $1 \pm 38$ & $-3.32 \pm 0.21$ \\
\cline{2-12}
 & \multirow{3}{*}{No velocity field} & $\mathbf{V}_{\rm ext}$ & $1002^{+389}_{-401}$ & $113 \pm 79$ & $53 \pm 19$ & $2277^{+1230}_{-1231}$ & $180 \pm 102$ & $2 \pm 25$ & $180 \pm 103$ & $1 \pm 26$ & $0.82 \pm 0.24$ \\
 &  & ZP & $4.7^{+1.8}_{-1.7}$ & $157 \pm 98$ & $56 \pm 19$ & $< 11.8$ & $181 \pm 100$ & $0 \pm 36$ & $177 \pm 101$ & $0 \pm 38$ & $-0.59 \pm 0.08$ \\
 &  & $H_0$ & $5.8^{+2.2}_{-2.2}$ & $155 \pm 100$ & $57 \pm 19$ & $4.9^{+4.9}_{-3.4}$ & $180 \pm 98$ & $-1 \pm 37$ & $177 \pm 99$ & $-1 \pm 37$ & $1.27 \pm 0.04$ \\
\hline\hline
\multirow{9}{*}{LTYT} & \multirow{3}{*}{C15} & $\mathbf{V}_{\rm ext}$ & $646^{+390}_{-367}$ & $138 \pm 47$ & $21 \pm 23$ & $< 593$ & $182 \pm 103$ & $-1 \pm 39$ & $184 \pm 104$ & $-0 \pm 40$ & $-3.24 \pm 0.12$ \\
 &  & ZP & $4.5^{+1.9}_{-1.9}$ & $143 \pm 78$ & $50 \pm 20$ & $4.5^{+4.3}_{-2.9}$ & $179 \pm 102$ & $1 \pm 38$ & $185 \pm 101$ & $-0 \pm 39$ & $-1.36 \pm 0.14$ \\
 &  & $H_0$ & $2.0^{+0.8}_{-0.8}$ & $135 \pm 43$ & $32 \pm 21$ & $< 4.5$ & $185 \pm 104$ & $-1 \pm 38$ & $182 \pm 104$ & $-3 \pm 38$ & $-1.25 \pm 0.10$ \\
\cline{2-12}
 & \multirow{3}{*}{Manticore} & $\mathbf{V}_{\rm ext}$ & $676^{+368}_{-400}$ & $139 \pm 43$ & $19 \pm 22$ & $< 531$ & $183 \pm 102$ & $0 \pm 38$ & $185 \pm 102$ & $-0 \pm 37$ & $-3.68 \pm 0.09$ \\
 &  & ZP & $4.4^{+1.9}_{-2.0}$ & $144 \pm 78$ & $49 \pm 21$ & $4.7^{+4.3}_{-3.1}$ & $178 \pm 104$ & $0 \pm 38$ & $185 \pm 104$ & $-0 \pm 38$ & $-1.47 \pm 0.10$ \\
 &  & $H_0$ & $< 2.2$ & $164 \pm 78$ & $33 \pm 29$ & $< 4.4$ & $185 \pm 104$ & $-0 \pm 40$ & $184 \pm 102$ & $-0 \pm 40$ & $-3.58 \pm 0.14$ \\
\cline{2-12}
 & \multirow{3}{*}{No velocity field} & $\mathbf{V}_{\rm ext}$ & $1034^{+426}_{-428}$ & $123 \pm 67$ & $49 \pm 18$ & $2360^{+1108}_{-1202}$ & $177 \pm 102$ & $0 \pm 33$ & $183 \pm 104$ & $3 \pm 32$ & $0.54 \pm 0.10$ \\
 &  & ZP & $4.7^{+1.7}_{-1.7}$ & $144 \pm 82$ & $54 \pm 18$ & $< 11.0$ & $181 \pm 103$ & $1 \pm 36$ & $181 \pm 103$ & $-0 \pm 37$ & $-1.18 \pm 0.22$ \\
 &  & $H_0$ & $5.5^{+2.2}_{-2.0}$ & $144 \pm 80$ & $54 \pm 19$ & $4.5^{+4.6}_{-3.2}$ & $181 \pm 101$ & $-0 \pm 38$ & $178 \pm 102$ & $-1 \pm 37$ & $0.80 \pm 0.16$ \\
\hline\hline
\end{tabular}
\label{tab:quad_params}
\end{table*}

\begin{table*}
\centering
\caption{Mixed dipole model parameters (simultaneous $\Vext$ + ZP or $\Vext$ + $H_0$ dipoles) for all scaling relations and reconstructions. $A_V$ is the bulk flow amplitude in $\kmsec$, and $A_{\rm ZP}$ is the zeropoint dipole amplitude expressed as a fractional variation equivalent to $\delta H_0/H_0$ (see \cref{eq:Deltaa_DeltaH0_linear}). Directions are in Galactic coordinates. No combination of dipoles is favoured over the simpler single-dipole models.}
\scriptsize
\begin{tabular}{|c|c|l|c|cc|c|cc|c|}
\hline\hline
Rel. & Recon & Model & $A_{V}$ & $\ell$ [$^\circ$] & $b$ [$^\circ$] & $A_{\rm ZP}$ & $\ell$ [$^\circ$] & $b$ [$^\circ$] & $\Delta\ln\mathcal{Z}$ \\
\hline\hline
\multirow{6}{*}{LT} & \multirow{2}{*}{C15} & ZP + $\mathbf{V}_{\rm ext}$ & $911^{+386}_{-393}$ & $135 \pm 37$ & $-7 \pm 23$ & $4.9^{+3.6}_{-3.1}$ & $188 \pm 135$ & $26 \pm 30$ & $-0.02 \pm 0.14$ \\
 &  & $H_0$ + $\mathbf{V}_{\rm ext}$ & $648^{+370}_{-358}$ & $134 \pm 44$ & $1 \pm 24$ & $1.7^{+0.7}_{-0.6}$ & $121 \pm 36$ & $19 \pm 18$ & $0.06 \pm 0.09$ \\
\cline{2-10}
 & \multirow{2}{*}{Manticore} & ZP + $\mathbf{V}_{\rm ext}$ & $977^{+458}_{-408}$ & $145 \pm 31$ & $-5 \pm 22$ & $5.5^{+3.7}_{-3.4}$ & $189 \pm 139$ & $25 \pm 28$ & $-0.77 \pm 0.07$ \\
 &  & $H_0$ + $\mathbf{V}_{\rm ext}$ & $714^{+373}_{-403}$ & $135 \pm 41$ & $2 \pm 23$ & $< 1.7$ & $171 \pm 87$ & $19 \pm 35$ & $-2.84 \pm 0.08$ \\
\cline{2-10}
 & \multirow{2}{*}{No velocity field} & ZP + $\mathbf{V}_{\rm ext}$ & $< 1574$ & $137 \pm 90$ & $4 \pm 35$ & $3.8^{+3.0}_{-2.4}$ & $173 \pm 121$ & $30 \pm 32$ & $-2.25 \pm 0.09$ \\
 &  & $H_0$ + $\mathbf{V}_{\rm ext}$ & $< 1606$ & $132 \pm 88$ & $5 \pm 35$ & $4.5^{+3.6}_{-2.9}$ & $173 \pm 123$ & $32 \pm 31$ & $-1.78 \pm 0.10$ \\
\hline\hline
\multirow{6}{*}{YT} & \multirow{2}{*}{C15} & ZP + $\mathbf{V}_{\rm ext}$ & $< 1350$ & $134 \pm 99$ & $19 \pm 35$ & $4.2^{+2.1}_{-2.2}$ & $187 \pm 103$ & $46 \pm 27$ & $-0.36 \pm 0.11$ \\
 &  & $H_0$ + $\mathbf{V}_{\rm ext}$ & $666^{+384}_{-381}$ & $120 \pm 97$ & $41 \pm 26$ & $1.8^{+1.0}_{-0.9}$ & $153 \pm 63$ & $34 \pm 25$ & $-0.51 \pm 0.06$ \\
\cline{2-10}
 & \multirow{2}{*}{Manticore} & ZP + $\mathbf{V}_{\rm ext}$ & $< 1291$ & $127 \pm 95$ & $17 \pm 36$ & $4.1^{+2.2}_{-2.3}$ & $185 \pm 101$ & $45 \pm 27$ & $-0.77 \pm 0.10$ \\
 &  & $H_0$ + $\mathbf{V}_{\rm ext}$ & $700^{+360}_{-416}$ & $117 \pm 88$ & $40 \pm 27$ & $< 2.2$ & $188 \pm 88$ & $29 \pm 32$ & $-2.40 \pm 0.06$ \\
\cline{2-10}
 & \multirow{2}{*}{No velocity field} & ZP + $\mathbf{V}_{\rm ext}$ & $< 1442$ & $132 \pm 102$ & $21 \pm 36$ & $4.1^{+2.3}_{-2.2}$ & $179 \pm 94$ & $45 \pm 28$ & $-0.19 \pm 0.04$ \\
 &  & $H_0$ + $\mathbf{V}_{\rm ext}$ & $< 1503$ & $127 \pm 98$ & $24 \pm 34$ & $4.7^{+2.9}_{-2.9}$ & $182 \pm 96$ & $42 \pm 29$ & $0.35 \pm 0.04$ \\
\hline\hline
\multirow{6}{*}{LTYT} & \multirow{2}{*}{C15} & ZP + $\mathbf{V}_{\rm ext}$ & $< 1248$ & $139 \pm 55$ & $-5 \pm 29$ & $4.8^{+2.3}_{-2.3}$ & $179 \pm 112$ & $50 \pm 24$ & $-0.55 \pm 0.12$ \\
 &  & $H_0$ + $\mathbf{V}_{\rm ext}$ & $< 1157$ & $133 \pm 49$ & $13 \pm 25$ & $1.7^{+0.8}_{-0.8}$ & $128 \pm 45$ & $25 \pm 21$ & $-1.20 \pm 0.07$ \\
\cline{2-10}
 & \multirow{2}{*}{Manticore} & ZP + $\mathbf{V}_{\rm ext}$ & $608^{+394}_{-333}$ & $140 \pm 51$ & $-6 \pm 28$ & $4.6^{+2.2}_{-2.3}$ & $177 \pm 113$ & $49 \pm 23$ & $-0.63 \pm 0.10$ \\
 &  & $H_0$ + $\mathbf{V}_{\rm ext}$ & $< 1211$ & $139 \pm 48$ & $16 \pm 24$ & $< 1.9$ & $173 \pm 83$ & $24 \pm 32$ & $-3.32 \pm 0.07$ \\
\cline{2-10}
 & \multirow{2}{*}{No velocity field} & ZP + $\mathbf{V}_{\rm ext}$ & $< 1358$ & $140 \pm 102$ & $14 \pm 36$ & $4.3^{+2.1}_{-2.3}$ & $158 \pm 84$ & $44 \pm 27$ & $-0.66 \pm 0.08$ \\
 &  & $H_0$ + $\mathbf{V}_{\rm ext}$ & $< 1454$ & $137 \pm 102$ & $14 \pm 35$ & $5.2^{+2.8}_{-2.9}$ & $164 \pm 86$ & $45 \pm 27$ & $-0.05 \pm 0.08$ \\
\hline\hline
\end{tabular}
\label{tab:mixed_dipoles}
\end{table*}

\section{Low-redshift mapping between an $H_0$ dipole and a dipolar flow}
\label{app:H0_to_flow}

Starting from the redshift composition law (Eq.~\ref{eq:redshift_composition}),
and writing $\zpec\equiv V_{\rm pec}/c$, expansion gives
\begin{equation}\label{eq:app_zexpand}
z_{\rm pred} = \zcosmo + \frac{V_{\rm pec}}{c}\,(1+\zcosmo).
\end{equation}

A dipolar anisotropy in the distance--redshift relation,
$\delta H_0(\hat{\bm n})/H_0 = A(\hat{\bm d}\cdot\hat{\bm n})$,
modifies the cosmological redshift at fixed comoving distance as
$\zcosmo \rightarrow \zcosmo[1 + A(\hat{\bm d}\cdot\hat{\bm n})]$.
Requiring the same predicted redshift $\zpred$ in
Eq.~\eqref{eq:app_zexpand} yields
\begin{equation}\label{eq:app_Vpec_z}
V_{\rm pec}
=
c\,\frac{\zcosmo}{1+\zcosmo}\,
A(\hat{\bm d}\cdot\hat{\bm n}) .
\end{equation}

Using the low-redshift relation $c\,\zcosmo\simeq100\,r$ (with $r$ in $\Mpch$),
this becomes
\begin{equation}\label{eq:app_Vpec_r}
V_{\rm pec}(r,\hat{\bm n})
\simeq
\frac{100\,r}{1+100r/c}\,
A(\hat{\bm d}\cdot\hat{\bm n}),
\end{equation}
which reduces for $r\ll c/100$ to
\begin{equation}\label{eq:app_Vpec_linear}
V_{\rm pec}(r,\hat{\bm n})
\simeq
100\,A\,r\,(\hat{\bm d}\cdot\hat{\bm n}) .
\end{equation}
\bsp
\label{lastpage}







\end{document}